\documentclass{lmcs}
\pdfoutput=1

\usepackage{lastpage}
\lmcsdoi{22}{4}{10}
\lmcsheading{}{\pageref{LastPage}}{}{}%
{Nov.~12,~2025}{Oct.~27,~2026}{}

\usepackage[noadjust,nocompress]{cite}
\usepackage{graphicx}
\usepackage{amsmath}
\usepackage{mathpartir}
\usepackage{color}
\usepackage[utf8]{inputenc}

\DeclareFontEncoding{LS1}{}{}
\DeclareFontSubstitution{LS1}{stix2}{m}{n}
\DeclareSymbolFont{stixsymbols2}{LS1}{stix2frak}{m}{n}

\DeclareFontEncoding{LS2}{}{}
\DeclareFontSubstitution{LS2}{stix2}{m}{n}

\DeclareSymbolFont{stixlargesymbols}  {LS2}{stix2ex}   {m} {n}
\DeclareMathSymbol{\lblkbrbrak}{\mathopen} {stixsymbols2}{"36}
\DeclareMathSymbol{\rblkbrbrak}{\mathclose}{stixsymbols2}{"37}

\DeclareMathDelimiter{\lbrbrak}{\mathopen}{stixlargesymbols}{"EE}{stixlargesymbols}{"14}
\DeclareMathDelimiter{\rbrbrak}{\mathclose}{stixlargesymbols}{"EF}{stixlargesymbols}{"15}
\DeclareMathDelimiter{\Lbrbrak}{\mathopen}{stixlargesymbols}{"DE}{stixlargesymbols}{"14}
\DeclareMathDelimiter{\Rbrbrak}{\mathclose}{stixlargesymbols}{"DF}{stixlargesymbols}{"15}

\DeclareUnicodeCharacter{2997}{\lblkbrbrak}
\DeclareUnicodeCharacter{2998}{\rblkbrbrak}

\DeclareUnicodeCharacter{2772}{\lbrbrak}
\DeclareUnicodeCharacter{2773}{\rbrbrak}

\usepackage{url}

\usepackage{stmaryrd}
\usepackage{amsfonts}
\usepackage{amssymb}
\usepackage{amsthm}
\usepackage{cmll}
\usepackage[all]{xy}
\usepackage{hyperref}
\usepackage{mathrsfs}
\newcommand{\Rel}{\mathbf{Rel}}

\newcommand{\ES}{\mathbf{ES}}
\newcommand{\CG}{\mathbf{CG}}
\newcommand{\TCG}{\mathbf{TCG}}

\newcommand{\ESS}{\mathbf{ESS}}

\newcommand{\Dsinn}{\mathbf{DSInn}}
\newcommand{\ScottL}{\mathbf{ScottL}}
\newcommand{\ScottFun}{\mathbf{ScottFun}}
\newcommand{\ScottP}{\mathbf{ScottP}}
\newcommand{\Scott}{\mathbf{Scott}}
\newcommand{\Ar}{\mathbf{Ar}}
\newcommand{\E}{\mathcal{E}}

\newcommand{\intr}[1]{\llbracket #1 \rrbracket}
\newcommand{\lin}{\multimap}
\newcommand{\Mf}{\mathscr{M}_f}
\newcommand{\iso}{\cong}
\newcommand{\bij}{\simeq}

\newcommand{\C}{\mathcal{C}}
\newcommand{\D}{\mathcal{D}}

\newcommand{\op}{\mathrm{op}}
\newcommand{\ot}{\leftarrow}
\newcommand{\id}{\mathrm{id}}

\newcommand{\Fam}{\mathsf{Fam}}
\newcommand{\N}{\mathbb{N}}

\newcommand{\tensor}{\otimes}

\newcommand{\der}{\mathrm{der}}
\newcommand{\dig}{\mathrm{dig}}

\DeclarePairedDelimiter\tuple\langle\rangle

\newcommand{\sym}{\cong}

\newcommand{\tto}{\Rightarrow}

\renewcommand{\cc}{\mathbf{\hspace{.5ex}c\!\!\!\!c\,}}
\newcommand{\cleq}{\vartriangleleft}
\newcommand{\CC}{\mathbf{CC}}

\renewcommand{\tilde}[1]{\mathscr{S}(#1)}
\newcommand{\ntilde}[1]{\mathscr{S}_-(#1)}

\newcommand{\ptilde}[1]{\mathscr{S}_+(#1)}
\newcommand{\tildep}[1]{\mathscr{S}^+(#1)}
\newcommand{\dom}{\mathrm{dom}}
\newcommand{\cod}{\mathrm{cod}}

\newcommand{\inj}{\mathsf{i}}
\newcommand{\q}{\mathsf{q}}

\newcommand{\confn}[1]{\mathscr{C}^{\neq \emptyset}(#1)}
\newcommand{\tildn}[1]{\mathscr{S}^{\neq \emptyset}(#1)}
\newcommand{\ntilden}[1]{\mathscr{S}^{\neq \emptyset}_-(#1)}
\newcommand{\ptilden}[1]{\mathscr{S}^{\neq \emptyset}_+(#1)}
\newcommand{\refines}{\vartriangleleft}

\newcommand{\bevm}{\mathsf{ev}}

\newcommand{\mon}{\mathsf{see}}
\newcommand{\coll}[1]{\mathfrak{R}(#1)}
\newcommand{\colloc}[1]{\mathfrak{R}_\oc(#1)}
\newcommand{\scoll}[1]{\mathfrak{S}(#1)}
\newcommand{\ccoll}[2]{\mathfrak{R}^{#2}(#1)}
\newcommand{\ccolloc}[2]{\mathfrak{R}^{#2}_{\oc}(#1)}
\newcommand{\cscoll}[2]{\mathfrak{S}^{#2}(#1)}
\newcommand{\stopc}[1]{\lbrbrak #1 \rbrbrak}
\newcommand{\supp}{\mathsf{supp}}
\newcommand{\fun}{\mathsf{fun}}
\newcommand{\tr}{\mathsf{tr}}
\newcommand{\Fun}{\mathsf{Fun}}
\newcommand{\Tr}{\mathsf{Tr}}
\newcommand{\col}{{\mathsf{col}}}
\newcommand{\inter}{\circledast}
\renewcommand{\ell}{\mathsf{l}}
\newcommand{\err}{\mathsf{r}}
\newcommand{\mar}[1]{\underline{#1}}
\newcommand{\lbl}{\mathsf{lbl}}
\newcommand{\ind}{\mathsf{ind}}
\newcommand{\Ind}{\mathsf{Ind}}
\newcommand{\pred}{\mathsf{pred}}
\newcommand{\strto}{\leadsto}
\newcommand{\strot}{\raisebox{5.4pt}{$\rotatebox{180}{$\leadsto$}$}}
\newcommand{\pstrto}{%
  \mathrel{\vbox{\offinterlineskip\ialign{%
    \hfil##\hfil\cr
    $\scriptscriptstyle +$\cr
    \noalign{\kern-0ex}
    $\leadsto$\cr
}}}}
\newcommand{\pstrot}{%
  \mathrel{\vbox{\offinterlineskip\ialign{%
    \hfil##\hfil\cr
    $\scriptscriptstyle +$\cr
    \noalign{\kern-0.3ex}
    $\raisebox{5.4pt}{$\rotatebox{180}{$\leadsto$}$}$\cr
}}}}
\newcommand{\nstrto}{%
  \mathrel{\vbox{\offinterlineskip\ialign{%
    \hfil##\hfil\cr
    $\scriptscriptstyle -$\cr
    \noalign{\kern+0ex}
    $\leadsto$\cr
}}}}
\newcommand{\nstrot}{%
  \mathrel{\vbox{\offinterlineskip\ialign{%
    \hfil##\hfil\cr
    $\scriptscriptstyle -$\cr
    \noalign{\kern-0.3ex}
    $\raisebox{5.4pt}{$\rotatebox{180}{$\leadsto$}$}$\cr
}}}}
\newcommand{\carmor}{%
  \mathrel{\vbox{\offinterlineskip\ialign{%
    \hfil##\hfil\cr
    $\scriptscriptstyle {-+}$\cr
    \noalign{\kern-0.3ex}
    $\raisebox{5.47pt}{$\rotatebox{180}{$\leadsto$}$}\hspace{-8.6pt}\leadsto$\cr
}}}}
\newcommand{\depth}{\mathrm{depth}}
\newcommand{\ppstrto}{%
  \mathrel{\vbox{\offinterlineskip\ialign{%
    \hfil##\hfil\cr
    $\scriptscriptstyle {+p}$\cr
    \noalign{\kern0ex}
    $\leadsto$\cr
}}}}
\newcommand{\ppstrot}{%
  \mathrel{\vbox{\offinterlineskip\ialign{%
    \hfil##\hfil\cr
    $\scriptscriptstyle {+p}$\cr
    \noalign{\kern0ex}
    $\raisebox{5.4pt}{$\rotatebox{180}{$\leadsto$}$}$\cr
}}}}
\newcommand{\pnstrto}{%
  \mathrel{\vbox{\offinterlineskip\ialign{%
    \hfil##\hfil\cr
    $\scriptscriptstyle {-p}$\cr
    \noalign{\kern-0.1ex}
    $\leadsto$\cr
}}}}
\newcommand{\pnstrot}{%
  \mathrel{\vbox{\offinterlineskip\ialign{%
    \hfil##\hfil\cr
    $\scriptscriptstyle {-p}$\cr
    \noalign{\kern-0.1ex}
    $\raisebox{5.4pt}{$\rotatebox{180}{$\leadsto$}$}$\cr
}}}}
\newcommand{\pre}{%
  \mathrel{\vbox{\offinterlineskip\ialign{%
    \hfil##\hfil\cr
    $\scriptscriptstyle {+-}$\cr
    \noalign{\kern-0.2ex}
    $\raisebox{5.4pt}{$\rotatebox{180}{$\leadsto$}$}\hspace{-8.5pt}\leadsto$\cr
}}}}
\newcommand{\parstrto}{%
  \mathrel{\vbox{\offinterlineskip\ialign{%
    \hfil##\hfil\cr
    $\scriptscriptstyle p$\cr
    \noalign{\kern-0.2ex}
    $\leadsto$\cr
}}}}
\newcommand{\restrict}{\upharpoonright}

\renewcommand{\P}{\mathscr{P}}

\newcommand{\specialraise}[1]{\text{\raisebox{-2.5pt}{$#1$}}}
\newcommand{\pview}[1]{\raisebox{2.5pt}{$\ulcorner$}\!#1\!\raisebox{2.5pt}{$\urcorner$}}
\newcommand{\oview}[1]{\specialraise{\llcorner}\!#1\!\specialraise{\lrcorner}}
\newcommand{\famc}[1]{\lfloor#1\rfloor}

\newcommand{\x}{\mathsf{x}}
\newcommand{\y}{\mathsf{y}}
\newcommand{\z}{\mathsf{z}}

\definecolor{grey}{rgb}{.5,.5,.5}
\newcommand{\grey}[1]{{\color{grey}#1}}
\definecolor{Red}{cmyk}{0,1,2,0}

\definecolor{Green}{rgb}{0,1,0}

\definecolor{Cyan}{cmyk}{1,0,0,0}

\definecolor{purple}{rgb}{.7, 0,1}

\newcommand{\ev}[1]{|#1|}
\newcommand{\conflict}{\mathrel{\#}}
\newcommand{\conf}[1]{\mathscr{C}(#1)}
\newcommand{\cconf}[2]{{\mathscr{C}}_{#2}(#1)}
\newcommand{\cconfp}[2]{{\mathscr{C}}_{#2}^+(#1)}
\newcommand{\confne}[1]{\mathscr{C}^{\neq \emptyset}(#1)}
\newcommand{\confp}[1]{\mathscr{C}^+(#1)}
\newcommand{\pol}{\mathrm{pol}}
\renewcommand{\qu}{\q}
\newcommand{\pr}{\partial}
\newcommand{\enb}{\vdash}
\newcommand{\imc}{\rightarrowtriangle}
\newcommand{\done}{\checkmark}
\newcommand{\gcc}{\mathsf{gcc}}
\newcommand{\mconflict}{\!\!\xymatrix@C=15pt{\, \ar@{~}[r]&\,}\!\!\!\!\!}
\newcommand{\nconf}[1]{\mathscr{C}^0(#1)}
\newcommand{\etcg}{\emptyset}
\newcommand{\one}{\mathbf{1}}

\newdir{|>}{%
!/4.5pt/@{|}*:(1,-.2)@{>}*:(1,+.2)@_{>}}
\newdir{pb}{:(1,-1)@^{|-}}
\def\pb#1{\save[]+<16 pt,0 pt>:a(#1)\ar@{pb{}}[]\restore}

\newcommand{\tbool}{\mathbb{B}}
\newcommand{\tnat}{\mathbb{N}}
\newcommand{\tunit}{\mathbb{U}}

\newcommand*{\seq}[1]{\langle #1 \rangle}

\newcommand{\ttrue}{\mathbf{t\!t}}
\newcommand{\tfalse}{\mathbf{f\!f}}

\begin{document}
\title{The Qualitative Collapse of Concurrent Games (Extended Version)}

\author[P. Clairambault]{Pierre Clairambault\lmcsorcid{0000-0002-3285-6028}}

\address{CNRS, Aix Marseille Univ, LIS, Marseille, France}
\email{Pierre.Clairambault@cnrs.fr}

\begin{abstract}
In this paper, we construct an interpretation-preserving functor from a
category of concurrent games to the category of Scott domains and
Scott-continuous functions. We give a concrete description of this
functor, extending earlier results on the \emph{relational} collapse
of game semantics. The crux is an intricate combinatorial
lemma allowing us to synchronize states of strategies which reach the
same resources, but with different multiplicity. 

Putting this together with the previously established relational
collapse, this provides a new proof of the
\emph{qualitative}-\emph{quantitative} correspondence first established
by Ehrhard in his celebrated \emph{extensional collapse} theorem.
Whereas Ehrhard's proof is indirect and rests on an abstract
realizability construction, our result gives a concrete, combinatorial
description of the extraction of quantitative information from a
qualitative model.
\end{abstract}

\maketitle

\section{Introduction}

The heart of denotational semantics is certainly \emph{domain theory},
where types are interpreted as certain partially ordered sets, and
programs as (continuous) functions between those. This 
idea, originally pioneered by Scott and Strachey
\cite{scott1971toward}, has spread wide and far, and underlies much of
the modern theory of programming languages. In the terminology of this
paper, this functional semantics is \emph{qualitative}: it tracks the
amount of information about the input need to compute a given part of
the output, but not \emph{how many times} that information is needed,
or how many times the argument of a function is evaluated.

Another deeply influential discovery, in that field of research, is
Girard's invention of \emph{linear logic}
\cite{DBLP:journals/tcs/Girard87}. Linear logic is a logic of
\emph{resources}; it gives a special status to those functions that are
\emph{linear} in the sense that they evaluate their argument
\emph{exactly once}. Starting with the interpretation of
$\lambda$-terms as normal functors \cite{DBLP:journals/apal/Girard88},
linear logic prompted the development of denotational models that are
sensitive to resources, in the sense that they also record the
multiplicity of resource usage: in the terminology of this paper, they
are \emph{quantitative}. Quantitative models have been under active
development in the following three decades, with a number of remarkable
achievements. For instance, quantitative models (and their
type-theoretic presentations as \emph{non-idempotent intersection
types}) provide a semantic characterization of execution time
\cite{carv:ex}. Their resource-sensitivity lets them track numerous
quantitative aspects of computation \cite{DBLP:conf/lics/LairdMMP13}, or
provide models of properly quantitative computational effects, such as
probabilistic choice \cite{DBLP:journals/jacm/EhrhardPT18} or quantum
effects \cite{DBLP:conf/popl/PaganiSV14}, for which they give fully
abstract models
\cite{DBLP:journals/jacm/EhrhardPT18,DBLP:journals/pacmpl/ClairambaultV20}.

The drawback of this quantitative aspect, however, is that they are
infinitary. Even for the simply-typed $\lambda$-calculus with a
finite interpretation for ground types, they give infinitary semantics,
because they rely on finite multisets to represent the arrow type. A
``proof'' that a certain point is in the quantitative semantics of a
term (which can often be represented as a derivation in a
non-idempotent intersection type system), is really a de-temporised,
``static'' representation of the full execution. In contrast, the
functional models as in domain theory, and their syntactic
presentations as idempotent intersection type systems, remain finitary:
for instance, they give a finite interpretation to simply-typed
programs with finite ground types. In this way, they talk by finitary
means of an infinitary object: the execution -- this is the key to
their algorithmic use in \emph{e.g.} higher-order model-checking
\cite{DBLP:conf/csl/Aehlig06,DBLP:conf/lics/KobayashiO09}. 

There is a fascinating scientific tension between these
\emph{qualitative} and \emph{quantitative} models. On the one hand,
they are remarkably similar: with the right presentation, the only
significant difference in their construction is whether the exponential
modality should be based on finite sets or finite multisets. On the
other hand, the associated proof methods are very different:
quantitative models are infinitary, but their connection with the
execution is simple logically (though it can still be subtle
combinatorially), allowing them to provide useful program approximants
\cite{DBLP:journals/pacmpl/BarbarossaM20}; qualitative models are
finitary, but linking them with execution requires tools with
considerable logical complexity, such as logical relations.
Surprisingly, this tension has been somewhat little studied, perhaps
also because the two families of models correspond to different
communities. However, there is one important paper that strikes right
at that tension: Ehrhard's result that the linear Scott model of the
simply-typed $\lambda$-calculus is the \emph{extensional collapse} of
its relational model \cite{DBLP:journals/tcs/Ehrhard12}.  Ehrhard's
result entails, in particular, that a point $a$ in the qualitative
model is in the semantics of a program $M$ iff it has a
``quantitativation'' $a'$ in the \emph{quantitative} semantics of $M$.
At the core of this result is the construction of a model that is
somewhat hybrid between qualitative and quantitative; of quantitative
relations which behave well with respect to a preorder relation
rearranging resources. But this hybrid model is obtained by formulating
and maintaining an invariant (by \emph{biorthogonality}) implying this
quantitativation, it gives us no combinatorial understanding of that
process, and no way to compute it in concrete cases.

Here, we provide a combinatorial understanding of this
quantitativation process, using \emph{game semantics}. Game semantics
is another quantitative denotational model, originally developed to
attack the famous full abstraction problem for PCF
\cite{DBLP:journals/iandc/AbramskyJM00,DBLP:journals/iandc/HylandO00}.
Game semantics enriches the relational model with \emph{time} or
\emph{causality}, presenting interactive executions of a program with
its runtime environment as \emph{plays} on a game whose rules are
determined by the type. Despite its clear intellectual affiliation with
quantitative semantics, the precise relationship between games and
relational models has been the center of a longstanding line of
research
\cite{DBLP:conf/csl/BaillotDER97,DBLP:journals/apal/Ehrhard96,DBLP:journals/tcs/Mellies06,DBLP:conf/lics/Mellies05,DBLP:conf/tlca/Boudes09}.
In the modern dresses of \emph{thin concurrent games}
\cite{DBLP:journals/lmcs/CastellanCW19,hdr}, building on
\emph{concurrent games on event structures}
\cite{DBLP:conf/lics/RideauW11} and in the footsteps of Melliès's
insightful work on \emph{asynchronous games}
\cite{DBLP:conf/lics/Mellies05}, this relationship now appears as a
simple forgetful interpretation-preserving functor to the relational
model, erasing the ``dynamic'' causal dependency coming from the
program, keeping only the ``static'' causal dependency from the type --
this is summed up in \cite{hdr}, see also
\cite{DBLP:conf/lics/CastellanCPW18,DBLP:journals/pacmpl/ClairambaultV20}
for extensions with quantitative features and
\cite{DBLP:conf/lics/ClairambaultOP23} for a bicategorical version.  

In this paper, we complement this ``relational collapse'' with a
related interpretation-preserving functor to the \emph{linear Scott
model}, a linear decomposition of a (full subcategory of) Scott domains
due independently to Huth \cite{DBLP:conf/mfps/Huth93} and Winskel
\cite{DBLP:conf/amast/Winskel98}. To construct a Scott domain from a
game, we equip the latter with adequate notions of morphisms,
\emph{cartesian morphisms}, which allow the rearrangement (contraction
and weakening) of resources. The crux of the issue is then to show that
this collapse operation to the linear Scott model preserves
composition: this rests on a crucial proposition (Proposition
\ref{prop:main_link}) showing that if innocent
strategies can synchronize up to cartesian morphisms, then one can find
adequate expansions of the strategies making them synchronize on the
nose. This forms the core of our combinatorial account of Ehrhard's
quantitativation result: because our games model has
interpretation-preserving functors to both the relational model and the
Scott model, we also obtain a precise connection between the two
(Theorem \ref{thm:main}). This is also a contribution to the line of
work connecting game semantics to ``static'' semantics, targeting from
the first time a qualitative semantics, spanning across communities:
Scott domains.

After this scientific introduction, we include in the next section a
more technical introduction, setting the scene and giving the main
intuitions, and exposing the outline.

\paragraph{Note.} This is an extended version of an earlier paper
\cite{DBLP:conf/lics/Clairambault25} presented at the LICS conference,
providing all technical details and proofs.

\section{The Qualitative and the Quantitative}

\subsection{The Relational Model and Quantitative Semantics}

Our starting point, in this discussion, will be the \emph{relational
model} of linear logic.

\subsubsection{The relational model}\label{subsec:rel_intro}
At its heart, it is the interpretation of the
$\lambda$-calculus into the category $\Rel$ of sets and relations, with
which we assume that the reader is familiar (see \emph{e.g.}
\cite{DBLP:journals/tcs/Ehrhard12} for a reference).  $\Rel$ is a Seely
category \cite{panorama}: its monoidal product is the cartesian product
of sets, its cartesian product is given by the disjoint union, and its
exponential modality $\oc$ sends a set $A$ to the set $\Mf(A)$ of
\emph{finite multisets} of elements of $A$. We adopt standard
conventions for multisets: we adopt a list notation $[a_1, \dots, a_n]$
possibly with repetitions, with the empty multiset written $[]$.
Multiset union is written with a sum $+$.

As $\Rel$ is a Seely category, one can consider the Kleisli category
for the exponential comonad $\oc$, which is cartesian closed. One can
then interpret the simply-typed $\lambda$-calculus following the
standard lines of its interpretation into a cartesian closed category:
this sends any type $A$ to a set $\intr{A}$, following the
straightforward inductive definition\footnote{The interpretation is
parametrised by the choice of an interpretation for the base type. For
now we use a singleton type, which is restrictive but simplifies the
relationship with game semantics. In Section~\ref{subsec:colors}, we
will see how to extend that if the base type is interpreted with an
arbitrary set.} 
\[
\begin{array}{rcl}
\intr{o} &=& \{\star\}\\
\intr{A\to B} &=& \Mf(\intr{A}) \times \intr{B}\,,
\end{array}
\]
this set $\intr{A}$ is often referred to as the \textbf{web} of $A$.
Likewise, any well-typed term $\vdash M : A$ is sent to $\intr{M}
\subseteq \intr{A}$ a subset of the web. It is a reasonable intuition
to think of elements of $\intr{A}$ as sort of \emph{detemporalized
execution traces}, and indeed it is central in this paper that they do
correspond to plays in the game semantics sense where time has been
suppressed.

Let us illustrate this with an example. We have
\begin{eqnarray}
([([\star], \star), ([\star, \star], \star)],([\star,\star], \star))
\quad\in\quad
\intr{\lambda fx.\,f\,(f\,x) : (o \to o) \to o \to o}
\label{eqn:ex_rel}
\end{eqnarray}
which may be interpreted as an execution of the term (the Church
integer for $2$) which calls $f$ twice. For one of these calls, $f$
calls its argument once; the other time, twice -- so the term ends up
using $x$ twice.  We will revisit this example later armed with better
notations.

\subsubsection{Non-idempotent intersection types} The relational model
is a denotational semantics -- in fact, it is the core of numerous
\emph{quantitative} denotational models including coherence spaces,
probabilistic coherence spaces, and many others. But one of its striking
features is that it can be presented purely syntactically, via an
intersection type system known as \emph{non-idempotent intersection
types}, or \emph{multi-types}.

Non-idempotent intersection types come in two shapes: on the one hand,
\emph{single} types are of the form $\star$ or $\bar{\alpha} \lin
\beta$, where $\beta$ is a single type and $\bar{\alpha}$ is a
multi-type. On the other hand, a \emph{multi-type} $\bar{\alpha}$ is a
finite multiset $[\alpha_1, \ldots, \alpha_n]$ of single types.
Because we refer here to the relational semantics of the simply-typed
$\lambda$-calculus, we shall only consider those non-idempotent
intersection types that refine -- \emph{i.e.}, follow the
structure of -- a simple type:
\begin{gather*}
  \inferrule
	{ }
	{\star \refines o}
  \qquad\qquad
  \inferrule
	{\bar{\alpha} \refines A \\
         \beta \refines B}
  	{\bar{\alpha} \lin \beta \refines A \to B}
  \qquad\qquad
  \inferrule
	{\forall i \in I\qquad
	 \alpha_i \refines A }
	{ [\alpha_i \mid i \in I] \refines A }
\end{gather*}

\noindent
It looks like there is no intersection in this syntax for
non-idempotent intersection types, but a finite multiset $[\alpha_1,
\dots, \alpha_n]$ can be read as a formal intersection
\[
\alpha_1 \wedge \ldots \wedge \alpha_n
\]
with $\wedge$ an associative, commutative operation -- but crucially,
not idempotent. 

It is direct to see by induction on types that refinements of a simple
type $A$ are in one-to-one correspondence with elements of $\intr{A}$,
provided we interpret the base type as $\intr{o} = \{\star\}$.
Accordingly, we shall from now on identify the two, and see
non-idempotent intersection types as a syntax for elements of the
relational model. Though we shall not rely on it in this paper, this
also extends to typing rules for terms: the relational
interpretation of $\vdash M : A$ is precisely the set of $\alpha
\refines A$ such that $\vdash M : \alpha$ is derivable -- for instance,
\[
\lambda fx.\,f\,(f\,x) : [[\star] \lin \star, [\star, \star] \lin
\star] \lin [\star, \star] \lin \star
\]
is the typing judgment corresponding to \eqref{eqn:ex_rel}.

\subsubsection{Plays and refinement types.} These objects,
points of the web in the relational model or non-idempotent
intersection types, are at the heart of many quantitative models.
Importantly for this paper, this includes (some presentations of) 
\emph{game semantics}.

\emph{Game semantics} present computation as an exchange of moves
between two players: Player $(+)$, who plays for the program under
scrutiny, and Opponent $(-)$, who plays for the execution environment.
In this setting, an execution is traditionally represented as a
\emph{play}, a chronological sequence of moves linked with so-called
\emph{pointers} indicating their hierarchical relationships. 
\begin{figure}
\begin{minipage}{.49\linewidth}
\[
\xymatrix@R=-1pt@C=0pt{
\lambda fx.\,f\,(f\,x) &:& 
(o &\to& o)& \to&o&\to&o\\
&&&&&&&&\qu^-\\
&&&&\qu^+	
	\ar@{.}@/^/[urrrr]\\
&&\qu^-	\ar@{.}@/^/[urr]\\
&&&&\qu^+
	\ar@{.}@/^/[uuurrrr]\\
&&\qu^-	\ar@{.}@/^/[urr]\\
&&&&&&\qu^+
	\ar@{.}@/^/[uuuuurr]\\
&&\qu^-	\ar@{.}@/^/[uuurr]\\
&&&&&&\qu^+
	\ar@{.}@/^/[uuuuuuurr]
}
\]
\caption{Example of a play}
\label{fig:ex_play}
\end{minipage}
\hfill
\begin{minipage}{.49\linewidth}
\bigskip
\[
(o_1 \to o_2) \to o_3 \to o_4
\]
\bigskip
\[
\xymatrix@R=20pt@C=10pt{
&&&\qu^-_4
	\ar@{-|>}[dlll]\\
\qu^+_2
	\ar@{-|>}[d]
	\ar@{.}@/^/[urrr]
&&\qu^+_2
	\ar@{.}@/^/[ur]
	\ar@{-|>}[dl]
	\ar@{-|>}[dr]
&&\qu^+_3
	\ar@{.}@/_/[ul]
&\qu^+_3
	\ar@{.}@/_/[ull]\\
\qu^-_1	\ar@{.}@/^.1pc/[u]
	\ar@{-|>}[urr]&
\qu^-_1	\ar@{.}@/^/[ur]
	\ar@{-|>}[urrr]
&&\qu^-_1
	\ar@{.}@/_/[ul]
	\ar@{-|>}[urr]
}
\]
\bigskip
\caption{Its position}
\label{fig:deseq}
\end{minipage}
\end{figure}
As an example, we show in Figure \ref{fig:ex_play} a play in (the
strategy for) $\lambda fx.\,f\,(f\,x)$. It is read from top to
bottom, and each move is placed under the corresponding type component.
Opponent starts computation, which prompts the evaluation of $f$ with
$\qu^+$. Then $f$ calls its argument, which prompts the evaluation of
the second occurrence of $f$. After that, $f$ calls its argument twice,
and $x$ gets evaluated twice. Moves are linked with so-called
\emph{justification pointers}, carrying the hierarchical
relationships between variable calls.

\emph{Time} is critical in traditional presentations of game semantics.
But it is nonetheless sensible to forget time, retaining only the tree
structure induced by moves and justification pointers -- as pictured in
Figure \ref{fig:deseq}, where the correspondence between moves and
atoms in the type is conveyed via subscripts rather than the horizontal
position of moves. In this time-forgetting operation lies the main
intuition behind the link between game semantics and relational
semantics: Figure \ref{fig:deseq} (ignoring solid arrows $\imc$) turns
out to be an alternative representation of  
the non-idempotent intersection refinement 
\[
[[\star] \lin \star, [\star, \star] \lin
\star] \lin [\star, \star] \lin \star
\quad\refines \quad
(o \to o) \to o \to o
\]
encountered earlier -- carrying the same information as in the play in
Figure \ref{fig:ex_play}
about the distinct variable calls and their hierarchical
dependencies\footnote{The formal link between plays in Hyland-Ong games
and points of the web in the relational model was first established by
Boudes \cite{DBLP:conf/tlca/Boudes09} -- at least under this specific
form: the correspondence between static and dynamic denotational models
was first explored in
\cite{DBLP:conf/csl/BaillotDER97,DBLP:journals/apal/Ehrhard96}. This
tension between static and dynamic models underlies Melliès's work on
asynchronous games \cite{DBLP:conf/lics/Mellies05}.}.
Modern presentations of game semantics
\cite{DBLP:conf/lics/Mellies05,DBLP:conf/lics/RideauW11,DBLP:journals/lmcs/CastellanCRW17}
reject time; instead, \emph{positions} as pictured in Figure
\ref{fig:deseq} are primitive. In \emph{concurrent games}, positions
are enriched instead with \emph{causal} wiring conveying the causal
dependencies from the term, pictured with solid arrows $\imc$ in Figure
\ref{fig:deseq}.

\subsubsection{Rigidity and symmetries.} There is a seemingly small,
but actually fundamental subtlety in the explanation above.
\emph{Positions}, those trees matching points in the relational model,
are \emph{unordered}: children of a same node which correspond to the
same type component can be permuted at will -- this corresponds to the
fact that elements of $\oc A$ in the relational model are
\emph{multisets} rather than merely lists. They are quotiented
structures.

In \emph{thin concurrent games}, strategies play not
on these quotiented structures, but rather on some choice of concrete
representatives; this is in particular required so that positions can
be sensibly enriched with causal information as in Figure
\ref{fig:deseq}. In thin concurrent games in particular, concrete
representatives of positions are called \emph{configurations}. In these
objects, distinct copies of moves are kept separate by attributing each
an identifier, an integer called its \emph{copy index}. Copy indices
are not unique to a move, but two moves sharing the same justifier and
corresponding to the same type component cannot have the same copy
index.
\begin{figure}
\[
\raisebox{40pt}{
\scalebox{.8}{
\xymatrix@R=20pt@C=0pt{
&&&\qu^-_{4,\grey{0}}\\
\qu^+_{2,\grey{0}}
        \ar@{.}@/^/[urrr]
&&\qu^+_{2,\grey{4}}
        \ar@{.}@/^/[ur]
&&\qu^+_{3,\grey{2}}
        \ar@{.}@/_/[ul]
&\qu^+_{3,\grey{6}}
        \ar@{.}@/_/[ull]\\
\qu^-_{1,\grey{12}} \ar@{.}@/^.1pc/[u]&
\qu^-_{1,\grey{4}} \ar@{.}@/^/[ur]
&&\qu^-_{1,\grey{2}}
        \ar@{.}@/_/[ul]
}}}
~\sym^-~
\raisebox{40pt}{
\scalebox{.8}{
\xymatrix@R=20pt@C=0pt{
&&&\qu^-_{4,\grey{1}}\\
\qu^+_{2,\grey{0}}
        \ar@{.}@/^/[urrr]
&&\qu^+_{2,\grey{4}}
        \ar@{.}@/^/[ur]
&&\qu^+_{3,\grey{2}}
        \ar@{.}@/_/[ul]
&\qu^+_{3,\grey{6}}
        \ar@{.}@/_/[ull]\\
\qu^-_{1,\grey{13}} \ar@{.}@/^.1pc/[u]&
\qu^-_{1,\grey{2}} \ar@{.}@/^/[ur]
&&\qu^-_{1,\grey{4}}
        \ar@{.}@/_/[ul]
}}}
~\sym^+~
\raisebox{40pt}{
\scalebox{.8}{
\xymatrix@R=20pt@C=0pt{
&&&\qu^-_{4,\grey{1}}\\
\qu^+_{2,\grey{1}}
        \ar@{.}@/^/[urrr]
&&\qu^+_{2,\grey{5}}
        \ar@{.}@/^/[ur]
&&\qu^+_{3,\grey{6}}
        \ar@{.}@/_/[ul]
&\qu^+_{3,\grey{2}}
        \ar@{.}@/_/[ull]\\
\qu^-_{1,\grey{13}} \ar@{.}@/^.1pc/[u]&
\qu^-_{1,\grey{2}} \ar@{.}@/^/[ur]
&&\qu^-_{1,\grey{4}}
        \ar@{.}@/_/[ul]
}}}
\]
\caption{Concrete configurations with copy indices}
\label{fig:config}
\end{figure}
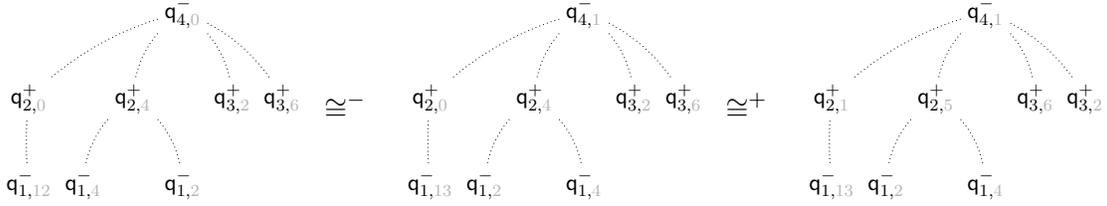
As an example, we draw in Figure \ref{fig:config} concrete
representatives for the position of Figure \ref{fig:deseq}. Copy
indices appear in \grey{grey}, to distinguish them from the subscript
for the type component. 

This feature of working with concrete representatives of positions is
not unique to thin concurrent games; it is in fact common in
\emph{categorifications} of the relational model, such as generalized
species of structure \cite{fiore2008cartesian}.  There, types are
interpreted not as sets comprising quotiented structures, but as
\emph{groupoids}: their objects are concrete representatives 
of
non-idempotent intersection types obtained as for the relational models
but with \emph{lists} 
\[
\seq{\alpha_1, \ldots, \alpha_n}
\]
instead of finite multisets. In quantitative
semantics, these concrete representatives of quotiented objects are
often referred to as \emph{rigid}. In this categorified situation, the
quotient is replaced with explicit morphisms generated by permutations
between elements of these lists.  

Just like generalized species, thin concurrent games are \emph{rigid};
and concrete configurations are related by so-called \emph{symmetries},
tree isomorphisms which can only change copy indices. In Figure
\ref{fig:config} we show two symmetries, which are the tree
isomorphisms respecting the topological position of nodes. In thin
concurrent games, the polarity of moves lets us set apart sub-groupoids
of \emph{polarized symmetries}\footnote{This fundamental distinction
was identified for the first time in Melliès's \emph{orbital games}
\cite{mellies2003asynchronous}; it is also at the core of the more
recent setting of \emph{thin spans of groupoids}
\cite{DBLP:conf/lics/ClairambaultF23}.}: some symmetries, dubbed
\emph{positive}, only change the copy index of positive moves; while
others, dubbed \emph{negative}, only reindex \emph{negative} moves. Not
every symmetry is negative or positive: the composite symmetry in
Figure \ref{fig:config} is neither. But every symmetry factors
as a composition of the two, as illustrated in Figure \ref{fig:config}.
We leave these structures on the side from now on, but they will play
a crucial role later on.

Note in passing that these similarities between thin concurrent games
and generalized species of structure are not superficial; the two
models are connected at the bicategorical level by an
interpretation-preserving (cartesian closed) pseudofunctor
\cite{DBLP:conf/lics/ClairambaultOP23}.

\subsection{The Qualitative} Quantitative models record the
multiplicity of resource usage. This is their defining feature, and a
significant part of their remarkable ability to handle quantitative
effects such as probabilistic choice
\cite{DBLP:journals/iandc/DanosE11}, quantum primitives
\cite{DBLP:conf/popl/PaganiSV14} and others
\cite{DBLP:conf/lics/LairdMMP13}. But most denotational models --
starting with Scott domains \cite{DBLP:journals/tcs/Plotkin77} -- do
\emph{not} record such quantitative information: they are
\emph{qualitative}; they record the \emph{presence} of resource calls
but not their multiplicity. They correspond to a different notion of
approximation according to which a function is greater than another if
it produces \emph{more} using \emph{less} information about the input,
regardless of how many times the input is evaluated.

\subsubsection{Idempotent intersection types} In terms of intersection
types, being \emph{qualitative} means 
\[
\alpha \wedge \alpha = \alpha\,,
\]
\emph{i.e.} that $\wedge$ is \emph{idempotent}: an expression $\alpha_1
\wedge \ldots \wedge \alpha_n$ no longer corresponds to the finite
multiset $[\alpha_1, \dots, \alpha_n]$, but to the \emph{set}
$\{\alpha_1, \dots, \alpha_n\}$.  But brutally enforcing $\alpha \wedge
\alpha = \alpha$ in this way fails: the finite powerset endofunctor on
$\Rel$ fails to be a comonad, as the candidate co-unit fails
naturality. For instance, with $A = \{a_1, a_2\}$, $B = \{b\}$,
with $R = \{(a_1, b), (a_2, b)\}$, 
\[
(\{a_1, a_2\}, b) \in \der_B \circ \oc R\,,
\qquad
\qquad
(\{a_1, a_2\}, b) \not \in R \circ \der_A\,.
\]

\noindent
Intuitively, the issue is that seeing $b$ may correspond to one
occurrence, or to many. If it is really just one, then we get it
through either $\{a_1\}$ or $\{a_2\}$ but not through $\{a_1, a_2\}$.
However, if there are actually several ``merged'' occurrences of $b$,
then some may arise through $a_1$ and others through $a_2$, so we
really do need $\{a_1, a_2\}$. To sort out this confusion, we must
relax our intuition about resources: while in
the relational model, $(a, b) \in \intr{M}$ means that there is an
execution where $M$ consumes \emph{exactly} $a$ to produce $b$, here we
must allow $M$ to produce $b$ \emph{or more} using $a$ \emph{or less}.
In idempotent intersection types, this means that it is unavoidable to
consider a \emph{partial order} on types, \emph{i.e.} a \emph{subtyping
relation}. By insisting that $R$ is adequately down-closed for this
subtyping relation, one reinstates naturality.

It is then fairly natural to consider ``idempotent'' intersection types
as simply the addition of a preorder on non-idempotent intersection
types, defined by the rules below: 
\begin{eqnarray}
  \inferrule
	{ }
	{\star \leq \star}
  \qquad
  \qquad
  \inferrule
	{\bar{\alpha_2} \leq \bar{\alpha_1} \\
         \beta_1 \leq \beta_2 }
  	{\bar{\alpha_1} \lin \beta_1 \leq \bar{\alpha_2} \lin \beta_2}
  \qquad
  \qquad
  \inferrule
	{\forall i \in I~\exists j \in J~\alpha_i \leq \beta_j }
	{ [\alpha_i \mid i \in I] \leq [\beta_j \mid j \in J] }
\label{eq:rules}
\end{eqnarray}
noting that the order is contravariant on the left hand side for the
arrow. Those may not look idempotent as they are still based on
multisets; but it follows that for all $\bar{\alpha}$, 
\[
\bar{\alpha} \wedge \bar{\alpha} \leq \bar{\alpha}\,,
\qquad
\qquad
[] \leq \bar{\alpha}
\]
which reminds us of the logical laws of \emph{contraction} and
\emph{weakening}. In particular, $\bar{\alpha} \leq \bar{\alpha} \wedge
[] \leq \bar{\alpha} \wedge \bar{\alpha}$. Hence, though idempotence is
not enforced primitively, it is derived as the equivalence relation
generated by the subtyping preorder\footnote{One can equivalently work
with non-idempotent intersection types equipped with the subtyping
preorder, or with idempotent intersection types equipped with the
induced partial order.}. This view on idempotent intersection
types is implicit in the \emph{linear Scott model}, the linear
decomposition of Scott domains discovered by Huth
\cite{DBLP:conf/mfps/Huth93} and Winskel
\cite{DBLP:conf/amast/Winskel98}; and in particular in Winskel's
presentation\footnote{The preordered set of intersection types is not
the order of Scott domains, which may be obtained by considering the
complete lattice of down-closed sets -- see Section~\ref{subsubsec:scott_domains}.}.

The heart of Ehrhard's celebrated \emph{extensional collapse} theorem
\cite{DBLP:journals/tcs/Ehrhard12} is then that in this language, the
interpretation of a simply-typed $\lambda$-term in the linear Scott
model is simply the \emph{down-closure} of its relational
interpretation, linking the qualitative and the quantitative.

\subsubsection{Cartesian maps} Now, as we discussed earlier, the set
of non-idempotent intersection types may be \emph{categorified} into a
groupoid of \emph{rigid intersection types}, as formalized in the
cartesian closed bicategory of generalized species of structure; or
into a \emph{groupoid} of configurations in the sense of thin
concurrent games. It is natural that the preorder $\leq$ should then be
refined into a \emph{category}: this is achieved, for instance,
through the cartesian closed bicategory of \emph{cartesian
distributors} \cite{ol:intdist}. There, given a category $A$ a morphism
in $\oc A$
\[
\seq{\alpha_i \mid i \in I} ~\to~ \seq{\beta_j \mid j \in J}
\]
consists in a \emph{function} $h: I \to J$, together with
a family $(f_i)_{i\in I}$ where $f_i : \alpha_i \to \beta_{f(i)}$ in
$A$ -- we call this a \emph{cartesian} morphism, as it can
\emph{contract} several copies together, and it can \emph{weaken}
copies on the right hand side by not giving them a pre-image.
A morphism from $\alpha_1 \lin \beta_1$ to $\alpha_2 \lin \beta_2$
consists in morphisms $f : \alpha_2 \to \alpha_1$ and $g : \beta_1 \to
\beta_2$, reflecting \eqref{eq:rules}, contravariantly on the left hand
side. 

In this paper, we achieve an analogous categorification in thin
concurrent games, turning the groupoid of configurations in thin
concurrent games into a \emph{category} of configurations. So, what
should be the adequate morphisms between configurations? Drawing
inspiration from symmetries and the contraction maps above, a natural
guess is that they should simply be forest morphisms which preserve the
type component. For instance, we could have
\[
\raisebox{35pt}{
\scalebox{.8}{
\xymatrix@R=20pt@C=0pt{
&&&\qu^-_{4,\grey{0}}\\
\qu^+_{2,\grey{0}}
        \ar@{.}@/^/[urrr]
&&\qu^+_{2,\grey{4}}
        \ar@{.}@/^/[ur]
&&\qu^+_{3,\grey{2}}
        \ar@{.}@/_/[ul]
&\qu^+_{3,\grey{6}}
        \ar@{.}@/_/[ull]\\
\qu^-_{1,\grey{12}} \ar@{.}@/^.1pc/[u]&
\qu^-_{1,\grey{4}} \ar@{.}@/^/[ur]
&&\qu^-_{1,\grey{2}}
        \ar@{.}@/_/[ul]
}}}
\qquad
\strto
\qquad
\scalebox{.8}{
\raisebox{40pt}{
\xymatrix@R=20pt@C=0pt{
&&\qu^-_{4,\grey{0}}\\
&\qu^+_{2,\grey{0}}
        \ar@{.}@/^/[ur]
&&\qu^+_{3,\grey{0}}
        \ar@{.}@/_/[ul]\\
&\qu^-_{1,\grey{0}} \ar@{.}[u]
}}}
\]
contracting all copies down to copy index $\grey{0}$. But this is not
right, because it does not take into account the contravariance on the
left hand side of arrows. The missing ingredient is to account for
\emph{polarity} -- \emph{negative} contraction maps can only
contract and weaken negative moves, while \emph{positive} contraction
maps can only contract and weaken positive moves:
\[
\raisebox{35pt}{
\scalebox{.8}{
\xymatrix@R=20pt@C=0pt{
&&&\qu^-_{4,\grey{0}}\\
\qu^+_{2,\grey{0}}
        \ar@{.}@/^/[urrr]
&&\qu^+_{2,\grey{1}}
        \ar@{.}@/^/[ur]
&&\qu^+_{3,\grey{0}}
        \ar@{.}@/_/[ul]\\
\qu^-_{1,\grey{12}} \ar@{.}@/^.1pc/[u]&
\qu^-_{1,\grey{4}} \ar@{.}@/^/[ur]
&&\qu^-_{1,\grey{2}}
        \ar@{.}@/_/[ul]
}}}
\quad
\pstrot
\quad
\raisebox{35pt}{
\scalebox{.8}{
\xymatrix@R=20pt@C=0pt{
&&&\qu^-_{4,\grey{0}}\\
\qu^+_{2,\grey{0}}
        \ar@{.}@/^/[urrr]
&&\qu^+_{2,\grey{4}}
        \ar@{.}@/^/[ur]
&&\qu^+_{3,\grey{2}}
        \ar@{.}@/_/[ul]
&\qu^+_{3,\grey{6}}
        \ar@{.}@/_/[ull]\\
\qu^-_{1,\grey{12}} \ar@{.}@/^.1pc/[u]&
\qu^-_{1,\grey{4}} \ar@{.}@/^/[ur]
&&\qu^-_{1,\grey{2}}
        \ar@{.}@/_/[ul]
}}}
\quad
\nstrto
\quad
\raisebox{35pt}{
\scalebox{.8}{
\xymatrix@R=20pt@C=0pt{
&&&\qu^-_{4,\grey{1}}\\
\qu^+_{2,\grey{0}}
        \ar@{.}@/^/[urrr]
&&\qu^+_{2,\grey{4}}
        \ar@{.}@/^/[ur]
&&\qu^+_{3,\grey{2}}
        \ar@{.}@/_/[ul]
&\qu^+_{3,\grey{6}}
        \ar@{.}@/_/[ull]\\
\qu^-_{1,\grey{1}}
	\ar@{.}[u]
&&\qu^-_{1,\grey{0}} \ar@{.}[u]
}}}
\]
and \emph{cartesian morphisms} between configurations are obtained as
relational compositions 
\[
\carmor \qquad=\qquad \pstrto ~ \nstrot ~ \pstrto ~ \nstrot ~ \ldots ~
\pstrto ~ \nstrot
\]
which are therefore no longer forest morphisms, but \emph{do} induce
the adequate preorder on configurations to match the linear Scott
model, as we shall demonstrate in this paper.

\subsubsection{Cartesian matching problems} The main contribution of
this paper is to extend this into a structure-preserving functor from a
category of thin concurrent games into the linear Scott model. This
builds on earlier results on the relational collapse of thin concurrent
games: informally, a strategy $\sigma : A \vdash B$ from $A$ to $B$ is
an aggregate of \emph{states} $x^\sigma$, with
\[
x^\sigma_A 
\qquad
\raisebox{5pt}{$
\rotatebox{180}{$\mapsto$}
$}
\qquad
x^\sigma
\qquad
\mapsto
\qquad
x^\sigma_B
\]
\emph{projections} obtained -- ignoring the issue of taking symmetry
classes -- by simply forgetting the causal arrows $\imc$ displayed
\emph{e.g.} in Figure \ref{fig:deseq}. The \emph{relational collapse}
of $\sigma$ then gathers all pairs $(x^\sigma_A, x^\sigma_B)$. But to
reach the linear Scott model, we need to build a relation that is
\emph{down-closed}! We shall achieve this by sending $\sigma$ to all
pairs $(y_A, y_B)$ such that we have
\[
y_A
\quad \pre \quad
x^\sigma_A 
\quad
\raisebox{5pt}{$
\rotatebox{180}{$\mapsto$}
$}
\quad
x^\sigma
\quad
\mapsto
\quad
x^\sigma_B
\quad\carmor \quad
y_B\,,
\]
\emph{i.e.} simply the down-closure with respect to $\pre$.
This does
yield a valid morphism in the linear Scott model. But this leaves us with
the demanding task to show that this down-closure remains functorial.
And this means that for $\sigma : A \vdash B$ and $\tau : B \vdash C$,
given
\[
x^\sigma 
\quad\mapsto\quad 
x^\sigma_B 
\quad\carmor\quad
x_B
\quad\carmor\quad 
x^\tau_B 
\quad\raisebox{5pt}{$
\rotatebox{180}{$\mapsto$}
$}\quad
x^\tau\,,
\]
\emph{i.e.} $(x^\sigma, x^\sigma_B \carmor x^\tau_B, x^\tau)$, we
must find a synchronization $z^{\tau\odot \sigma}$ in $\tau \odot
\sigma$ whose (down-closure of the) projection on $A, C$ is the same.
An analogous property is necessary with respect to symmetries to
construct thin concurrent games \cite[Proposition 7.4.4]{hdr}. But
here, the situation is significantly more complex: both $x^\sigma$ and
$x^\tau$ are trying to duplicate and erase each other, and we must find
a satisfactory state where all these duplications and erasures are
satisfied -- we call this a \emph{cartesian matching problem}. 

In game semantics we approach the question concretely, and provide a
combinatorial argument to resolve such matchings. This is the crux of
this paper; once this is solved it is not hard to provide an
interpretation-preserving functor from to the
linear Scott model. 

\subsection{Outline.} This sums up the main thrust of the paper; of
course there are various technical subtleties that come blur
the waters. First, we must introduce thin concurrent
games, along with their relational collapse. This already comes with a
significant technical set-up on top of thin concurrent games:
typically, those configurations that match points in the relational
model must be identified via a \emph{payoff} mechanism. One must also
introduce the slightly unorthodox concept of a \emph{relative Seely
($\sim$-)categories}, a weakening of Seely categories, as the structure
of plain Seely categories is not preserved by the relational collapse.
This content is well-covered in other sources
\cite{hdr,DBLP:conf/lics/ClairambaultOP23}, which our presentation
follows.

We must then give concrete definitions for cartesian morphisms.
Unfortunately, we can only do that relying on a fairly concrete
description of the games considered, referring explicitly to copy
indices. For this we must import from \cite{hdr} the rather clunky
notion of \emph{mixed board}.
Though this paper focuses on the semantic structures, we will apply
those to obtain results on the simply-typed $\lambda$-calculus. We
consider the $\lambda$-calculus with one base type $o$ -- our results
of course apply in the presence of many base types, but for this paper
we estimate that the additional notational burden is not worth the
extra generality.

Concretely, the paper is organized as follows. In
Section~\ref{sec:tcg}, we recall the main definitions of thin
concurrent games. In Section~\ref{sec:rel_collapse} we describe the
\emph{relational collapse} of thin concurrent games -- the material is
mainly taken from \cite{hdr}, with the extra development required in
order to allow interpreting the atom as an arbitrary set. The main
thrust of the contributions start in Section~\ref{sec:games_scottl}:
there, we refine our games to allow the collapse to the linear Scott
model. In particular, we introduce the notion of \emph{cartesian
morphisms} on a mixed board, and develop some of their basic
properties.  Finally, in Section~\ref{sec:qualcoll} we show how to
solve cartesian matching problems, and derive our main results.

\section{Thin Concurrent Games}
\label{sec:tcg}

The goal of this section is to give an introduction to thin concurrent
games, geared towards its relational collapse: we wish to present the
situation in which this collapse is the most natural, which we may then
specialize in later parts of this paper.

Though those are well-understood notions, setting up all the required
infrastructure for the collapse is demanding -- the reader can find a
more detailed introduction in \cite{hdr}.

\subsection{Basic Concurrent Games}
\label{subsec:basic-cg}

The framework of \emph{concurrent games}
\cite{DBLP:conf/concur/MelliesM07,DBLP:conf/tlca/FaggianP09,DBLP:conf/lics/RideauW11}
is not merely a game semantics for concurrency -- though it can serve
that purpose -- but a deep reworking of the basic mechanisms of game
semantics using causal ``truly concurrent'' structures from concurrency
theory \cite{DBLP:conf/scc/NielsenPW79}, which we must first introduce.

\subsubsection{Event structures} Concurrent games and
strategies are based on event structures. An event structure
represents the behaviour of a system
as a set of possible computational events equipped with
dependency and incompatibility constraints.

\begin{defi}\label{def:es}
An \textbf{event structure (es)} is $E = (\ev{E}, {\leq_E},
{\conflict_E})$,
where $\ev{E}$ is a (countable) set of \textbf{events}, $\leq_E$ is a
partial order called \textbf{causal dependency} and $\conflict_E$ is an
irreflexive symmetric binary relation on $\ev{E}$ called
\textbf{conflict}, satisfying:
\[
\begin{array}{rl}
\text{\emph{finite causes:}}& \forall e\in \ev{E},~[e]_E = \{e'\in
\ev{E} \mid
e'\leq_E e\}~\text{is finite,}\\
\text{\emph{vendetta:}}&
\forall e_1 \conflict_E e_2,~\forall e_2 \leq_E e'_2,~e_1 \conflict_E
e'_2\,.
\end{array}
\]
\end{defi}
\noindent
Operationally, an event can occur if \emph{all} its
dependencies are met, and \emph{no} conflicting events
have occurred. A finite set $x \subseteq_f \ev{E}$
down-closed for $\leq_E$ and comprising no conflicting pair
is called a \textbf{configuration} -- we write $\conf{E}$ for the set
of configurations on $E$, naturally ordered by inclusion. If $x \in
\conf{E}$ and $e \in \ev{E}$ is such that $e \not \in x$ but $x \cup
\{e\} \in \conf{E}$, we say that $e$ is \textbf{enabled} by $x$ and
write $x \enb_E e$. For $e_1, e_2 \in \ev{E}$ we write $e_1 \imc_E e_2$
for the \textbf{immediate causal dependency}, \emph{i.e.} $e_1 <_E
e_2$ with no event strictly in between. Finally, two events $e_1, e_2
\in \ev{E}$ are in \textbf{immediate conflict}, written $e_1\!\mconflict_E\,e_2$, if $e_1 \conflict_E
e_2$, and this conflict is not inherited: if $e'_1 < e_1$ then $\neg
(e_1 \conflict_E e_2)$, and likewise on the other side.

There is an accompanying notion of \emph{map}: a \textbf{map of event
structures} from $E$ to $F$ is a function $f : \ev{E} \to \ev{F}$ such
that: \emph{(1)} for all $x \in \conf{E}$, the direct image $f x \in
\conf{F}$; and \emph{(2)} for all $x \in \conf{E}$ and $e, e' \in x$,
if $f e = f e'$ then $e = e'$. There is a category $\ES$ of event
structures and maps.

\subsubsection{Games and strategies} Throughout this paper, we will
gradually refine our notion of game. For now, a \textbf{plain game} is
simply an event structure $A$ together with a \textbf{polarity}
function $\pol_A : \ev{A} \to \{-, +\}$ which specifies, for each event
$a \in A$, whether it is \textbf{positive} (\emph{i.e.} due to Player /
the program) or \textbf{negative} (\emph{i.e.} due to Opponent / the
environment). Events are often called \textbf{moves}, and annotated
with their polarity.

A strategy is an event structure with a projection map to $A$:
\begin{defi}\label{def:plain_strategy}
Consider $A$ a plain game. A \textbf{strategy} on $A$, written $\sigma
: A$, is an event structure $\sigma$ together with a map $\pr_\sigma :
\sigma
\to A$ called the \textbf{display map}, satisfying:
\[
\begin{array}{rl}
\text{\emph{(1)}} &
\text{for all $x\in \conf{\sigma}$ and $\pr_\sigma x \enb_A a^-$,
there is a unique $x \enb_\sigma s$ such that $\pr_\sigma s =
a$.}\\
\text{\emph{(2)}} &
\text{for all $s_1 \imc_\sigma s_2$, if $\pol_A(\pr_\sigma(s_1)) = +$
or $\pol_A(\pr_\sigma(s_2)) = -$, then $\pr_\sigma(s_1) \imc_A
\pr_\sigma(s_2)$.}
\end{array}
\]
\end{defi}

\noindent
There two conditions (called \emph{receptivity} and
\emph{courtesy})
ensure that the strategy does not constrain the behaviour of Opponent
any more than the game does. They are essential
for the compositional structure we describe below, but they do not
play a major role in this paper (their use is encapsulated
in technical lemmas and propositions proved elsewhere). Note also that
though a strategy does not come with a polarity function for the moves
in $\sigma$, they do inherit a polarity through $\pr_\sigma$. This is
used implicitly from now on.

As a simple example, the usual game $\tbool$ for booleans in
call-by-name is
\[
\xymatrix@R=10pt@C=10pt{
&\qu^-	\ar@{.}[dl]
	\ar@{.}[dr]\\
\ttrue^+\ar@{~}[rr]&&
\tfalse^+
}
\]
drawn from top to bottom: Opponent initiates computation with the first
move $\qu$, to which Player can react with either $\ttrue$ or
$\tfalse$. 

Strategies give a ``proof-relevant'' account of
execution, in the sense that moves and configurations of the game can
have multiple witnesses in the strategy. For example, on the left
below, $b$ and $c$ are both mapped to the same move $\ttrue$:
\[
\raisebox{20pt}{
\xymatrix@R=5pt@C=5pt{
&\sigma&
	\ar[rr]^{\pr_\sigma}&&&\tbool\\
&a    \ar@{-|>}[dl]
      \ar@{|->}@/^/[rrrr]
      \ar@{-|>}[dr]&&\qquad&&
\qu^- \ar@{.}[dl]
      \ar@{.}[dr]\\
b     \ar@{~}[rr]
      \ar@{|->}@/_1pc/[rrrr]&&
c     \ar@{|->}[rr]&&
\ttrue^+\ar@{~}[rr]&&
\tfalse^+
}}
\qquad =: \qquad
\raisebox{20pt}{
\xymatrix@R=5pt@C=0pt{
&\tbool\\
&\qu^-        \ar@{-|>}[dl]
      \ar@{-|>}[dr]\\
\ttrue^+\ar@{~}[rr]
      \ar@/^/@{.}[ur]&&
\ttrue^+
      \ar@/_/@{.}[ul]
}}
\]

\medskip

\noindent
We denote immediate causality by $\imc$ in strategies, and by dotted
lines for games -- this lets us represent the strategy in a single
diagram, as on the right above. Similar diagrams may represent not
entire games and strategies but \emph{configurations} of games and
strategies, which implicitly inherit a partial order from the ambiant
event structure. 

\subsubsection{Morphisms between strategies}
\label{subsubsec:mor_strat}
For $\sigma$ and $\tau$ two strategies on $A$, a \textbf{morphism} from
$\sigma$ to
$\tau$, written $f : \sigma \Rightarrow \tau$, is a map of event
structures $f : \sigma \to \tau$ preserving the dependency relation
$\leq$
(we say it is \textbf{rigid})
and such that $\pr_\tau \circ f = \pr_\sigma$.

\subsubsection{+-covered configurations}
We now describe a useful technical tool: it turns out that a strategy
is completely characterized by a subset of its configurations, called
$+$-covered. 

For a strategy $\sigma$ on a game $A$, a configuration $x \in
\conf{\sigma}$ is \textbf{$+$-covered} if all its
maximal events are positive, so every Opponent move in $x$ has at least
one Player successor. We write $\confp{\sigma}$ for the partially
ordered set (by inclusion) of $+$-covered configurations of $\sigma$. 

\begin{lem}\label{lem:iso_confp}
 Consider a plain game $A$, and strategies $\sigma, \tau : A$.

If $f : \confp{\sigma} \iso \confp{\tau}$ is an order-isomorphism such
that $\pr_\tau \circ f = \pr_\sigma$, then there is a unique
isomorphism of strategies $\hat{f} :
\sigma \iso \tau$ such that for all $x \in
\confp{\sigma}$, $\hat{f}(x) = f(x)$.
\end{lem}
\begin{proof}
Immediate consequence of \cite[Lemma 6.3.4]{hdr}.
\end{proof}

\noindent
This is the first hint of a methodology that is central to this paper:
in concurrent games, we rarely reason at the level of individual
events, preferring whenever possible to reason at the level of
configurations, especially when linking with relational-like models.

\subsection{\texorpdfstring{A $\sim$-category}{A ~-category} of concurrent games and strategies}

We now show how games and strategies are organized into a
$\sim$-category -- that is, a bicategory where $2$-cells are
degenerated so that each hom-set forms a setoid, a set with an
equivalence relation.

\subsubsection{Strategies between games}
\label{subsec:strat-betw-games}

If $A$ is a plain game, its
\textbf{dual} $A^\perp$ has the same components as $A$ except for
the reversed polarity. In particular $\conf{A} = \conf{A^\perp}$. The
\textbf{parallel composition} $A \parallel B$ of $A$ and $B$ is simply
$A$ and $B$ side by side, with no interaction -- its events are the
tagged disjoint union $\ev{A \parallel B} = \ev{A} + \ev{B} = \{1\}
\times \ev{A} \uplus \{2\} \times \ev{B}$, and other components are
inherited. 
Likewise, the \textbf{hom} $A\vdash B$ is simply defined as $A^\perp
\parallel B$.  We write $x_A \parallel x_B$ for the configuration of $A
\tensor B$ that has $x_A \in \conf{A}$ on the left and $x_B \in
\conf{B}$ on the right, and likewise for $x_A \vdash x_B \in \conf{A
\vdash B}$, informing order-isomorphisms\footnote{Throughout this
paper, we write $\bij$ for mere bijections, and $\iso$ for isomorphisms
also preserving structure.}
\begin{eqnarray}
- \parallel - \quad:\quad \conf{A} \times \conf{B} &\iso&
  \conf{A\parallel B}\,,
\label{eq:isopar}\\
- \vdash - \quad:\quad \conf{A} \times \conf{B} &\iso& \conf{A\vdash
  B}\,.
\label{eq:isohom}
\end{eqnarray}

\noindent
A \textbf{strategy from $A$ to $B$} is a strategy on the game $A \vdash
B$. Note that if $\sigma : A \vdash B$ and $x^\sigma \in
\conf{\sigma}$, by convention we write $\pr_\sigma(x^\sigma) =
x^\sigma_A \vdash x^\sigma_B \in \conf{A\vdash B}$. 

Our first example of a strategy between games is \textbf{copycat}
$\cc_A : A \vdash A$, the identity morphism on $A$ in our
$\sim$-category. Concretely, copycat on $A$ has the same events as $A
\vdash A$, but adds immediate causal links between copies of the same
move across components.
By Lemma~\ref{lem:iso_confp}, the following characterizes copycat up to
isomorphism.
\begin{prop}\label{prop:confp_cc}
If $A$ is a game, there is an order-isomorphism
\[
\cc_{(-)} \quad:\quad \conf{A} ~\iso~ \confp{\cc_A}
\]
such that for all $x \in \conf{A}$, $\pr_{\cc_A}(\cc_x) = x \vdash
x$.
\end{prop}
\begin{proof}
Follows from \cite[Lemma 6.4.4]{hdr}.
\end{proof}

\noindent
This shows that the copycat strategy is essentially the diagonal
relation, which is the first hint of the connection between concurrent
games and the relational model.

\subsubsection{Composition} \label{subsubsec:cg_comp}
Consider $\sigma : A \vdash B$ and $\tau :
B \vdash C$. We define their composition $\tau \odot \sigma :
A \vdash C$. Concurrent games are a dynamic model, and to successfully
synchronize, $\sigma$ and $\tau$ must agree to play the same events
\emph{in the same order}; this is defined in two steps. 

We say that configurations $x^\sigma \in \conf{\sigma}$ and $x^\tau
\in \conf{\tau}$ are \textbf{matching} if they reach the same
configuration on $B$, \emph{i.e.} $x^\sigma_B =
x^\tau_B = x_B$. If that is the case, it induces a synchronization (and
we may then ask if that synchronization induces a deadlock).
If all events of
$x^\sigma$ and $x^\tau$ were in $B$, this synchronization would take
the form of a
bijection $x^\sigma \bij x^\tau$. But some moves of $x^\sigma$ are in
$A$ and some moves of $x^\tau$ are in $C$, so instead we form the
bijection
\[
\varphi[x^\sigma, x^\tau] : x^\sigma \parallel x^\tau_C 
\stackrel{\pr_\sigma \parallel x^\tau_C}{\bij}
x^\sigma_A \parallel x_B \parallel x^\tau_C 
\stackrel{x^\sigma_A \parallel \pr_\tau^{-1}}{\bij}
x^\sigma_A \parallel x^\tau
\]
where $x\parallel y$ is the tagged disjoint union.
This uses the fact that from the conditions on maps of event
structures,
$\pr_\sigma : x^\sigma \bij x^\sigma_A \vdash x^\sigma_B$ is a
bijection and likewise for $\pr_\tau$.

\begin{figure}
\[
\raisebox{35pt}{$
\scalebox{.8}{$
\xymatrix@R=5pt@C=20pt{
\tunit  \ar@{}[r]|\lin& \tunit\\
&\qu^-  \ar@{-|>}[dl]\\
\qu^+   \ar@{.}@/^/[ur]
\ar@{-|>}[d]\\
\done^- \ar@{.}@/^/[u]
\ar@{-|>}[dr]\\
&\done^+\ar@{.}@/_/[uuu]
}$}$}
\quad
\text{vs}
\quad
\raisebox{35pt}{$
\scalebox{.8}{$
\xymatrix@R=5pt@C=20pt{
(\tunit \ar@{}[r]|\lin&\tunit)\ar@{}[r]|\vdash&\tnat\\
&&\qu^-  \ar@{-|>}[dl]\\
&\qu^+   \ar@{-|>}[dl]
\ar@{-|>}[d]\\
\qu^-    \ar@{-|>}[d]
\ar@{.}@/^/[ur]&
\done^- \ar@{-|>}[dr]
\ar@{.}@/^/[u]
\ar@{-|>}[dl]\\
\done^+ \ar@{.}@/^/[u]&&
0^+\ar@{.}@/_/[uuu]
}$}$}
\]
\caption{An example of matching but causally incompatible
  configurations, in the composition of $\sigma :
  \tunit \lin \tunit$ and $\tau : \tunit \lin \tunit \vdash
  \tnat$. The underlying games are left undefined, but can be recovered
  by removing the arrows $\imc$. The configurations
  are matching on $\tunit \lin \tunit$,  but the arrows $\imc$ impose
  incompatible orders (i.e.~a cycle) between the two occurrences of
$\done$.
}
\label{fig:deadlock}
\end{figure}
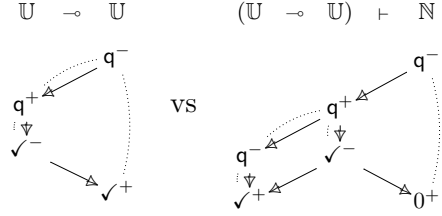

Next, we import the causal constraints of $\sigma$ and $\tau$ to (the
graph of) $\varphi[x^\sigma, x^\tau]$, via: 
\[
\begin{array}{rcl}
(m, n) \cleq_\sigma (m', n') &\Leftrightarrow& 
m <_{\sigma \parallel C} m'\\
(m, n) \cleq_\tau (m', n') &\Leftrightarrow& 
n <_{A \parallel \tau} n'
\end{array}
\]
letting us finally say that matching $x^\sigma$ and $x^\tau$ are
\textbf{causally compatible} if ${\cleq} = {\cleq_\sigma \cup
  \cleq_\tau}$ on (the graph of) $\varphi[x^\sigma, x^\tau]$ is
acyclic.
In particular,  $x^\sigma$ and $x^\tau$ in Figure \ref{fig:deadlock}
are
\emph{not} causally compatible, the synchronization induces a
\emph{deadlock}.

\noindent
The \textbf{composition} of $\sigma$ and $\tau$ is the unique (up to
iso) strategy whose $+$-covered configurations are essentially causally
compatible pairs of $+$-covered configurations. Writing $\CC(\sigma,
\tau)$ for causally compatible pairs $(x^\sigma, x^\tau)  \in
\confp{\sigma} \times \confp{\tau}$ (ordered componentwise):

\begin{propC}[{\cite[Proposition 6.2.1]{hdr}}]\label{prop:char_comp}
Consider strategies $\sigma : A \vdash B$ and $\tau : B \vdash C$.

There is a strategy $\tau \odot \sigma : A \vdash C$, unique up
to isomorphism, with an order-isomorphism
\[
\begin{array}{rcrcl}
- \odot -
&:&
\CC(\sigma, \tau) ~\iso~ \confp{\tau \odot \sigma}
\end{array}
\]
s.t. for all $x^\sigma \in \confp{\sigma}$ and $x^\tau \in
\confp{\tau}$ causally compatible,
$\pr_{\tau\odot \sigma}(x^\tau \odot x^\sigma) = x^\sigma_A \vdash
x^\tau_C$.
\end{propC}

\noindent
This description of composition emphasizes the conceptual difference
between a static model, in which composition is based merely on
matching pairs, and a dynamic model, based on causal compatibility and
sensitive to deadlocks.
We get:

\begin{thmC}[{\cite[Theorem 6.4.11]{hdr}}]
There is a $\sim$-category $\CG$ with: \emph{(1)} objects, plain games;
\emph{(2)} morphisms from $A$ to $B$, strategies $\sigma : A \vdash B$;
and equivalence relation, isomorphism of strategies.
\end{thmC}

\subsection{Adding Symmetry} The ambient $\sim$-category in which this
paper takes place is not quite $\CG$, but a refinement sensitive to
\emph{symmetry} -- this is necessary so that the model supports an
exponential modality.
We now go from $\CG$ to $\TCG$ by replacing the set of
configurations  $\conf{A}$ with a groupoid of configurations
$\tilde{A}$ whose morphisms are chosen bijections called
\emph{symmetries}, that behave well with respect to the causal order.

\subsubsection{Event structures with symmetry}
\label{subsubsec:ess}
Our starting point is to replace event structures with \emph{event
structures with symmetry}, due to Winskel
\cite{DBLP:journals/entcs/Winskel07}: 
\begin{defi}\label{def:isofam}%
An \textbf{isomorphism family} on es $E$ is a groupoid $\tilde{E}$
having as objects all configurations, and as morphisms certain
bijections between configurations, satisfying:
\[
\begin{array}{rl}
\text{\emph{restriction:}}&
\text{for all~$\theta : x \bij y \in \tilde{E}$ and $x \supseteq x' \in
\conf{E}$,}\\
&\text{there is $\theta \supseteq \theta' \in \tilde{E}$
such that $\theta' : x' \bij y'$.}\\
\text{\emph{extension:}}&\text{for all $\theta : x \bij y \in
\tilde{E}$, $x
\subseteq x' \in \conf{E}$,}\\
&\text{there is $\theta \subseteq \theta' \in
\tilde{E}$ such that $\theta' : x' \bij y'$.}
\end{array}
\]

\noindent
We call $(E, \tilde E)$ an \textbf{event structure with symmetry
(\emph{ess})}.
\end{defi}
\noindent
We refer to morphisms in $\tilde{E}$ as \textbf{symmetries}, and write
$\theta : x \sym_E y$ if $\theta : x \bij y$ with $\theta \in
\tilde{E}$. The \textbf{domain} $\dom(\theta)$ of $\theta : x \sym_E y$
is $x$, and likewise its \textbf{codomain} $\cod(\theta)$ is $y$.
A \textbf{map of ess} $E \to F$ is a map of event
structures that preserves symmetry: the bijection
\[
f \theta \qquad \overset{\mathrm{def}}= \qquad 
f x 
\quad \stackrel{f^{-1}}{\bij} \quad
x 
\quad \stackrel{\theta}{\bij} \quad 
y
\quad \stackrel{f}{\bij} \quad
f y,
\]
is in $\tilde{F}$ for
every $\theta : x \sym_E y$ (recall that $f$ restricted to any
configuration is bijective).  This exactly amounts to making $f :
\tilde{E} \to \tilde{F}$ a functor of groupoids.  

We can define a 2-category $\ESS$ of ess, maps of ess, and natural
transformations between the induced functors. For $f, g : E \to F$ such
a natural transformation is necessarily unique
\cite{DBLP:journals/entcs/Winskel07}, and corresponds to the fact that
for every $x \in \conf{E}$ the composite bijection
\[
f\,x 
\quad \stackrel{f^{-1}}{\bij} \quad
x
\quad \stackrel{g}{\bij} \quad
g\,x
\]
via local injectivity of $f$ and $g$, is in $\tilde{F}$.
So this is an equivalence, denoted  $f \sim g$.

\subsubsection{Thin games} We define games with symmetry. To
match the polarized structure, a game is an ess with two
sub-symmetries, one for each player (see
e.g.~\cite{mellies2003asynchronous,DBLP:journals/lmcs/CastellanCW19,mfps22}).
\begin{defi}\label{def:tcg}
A \textbf{thin concurrent game (tcg)} is a game $A$ with
isomorphism families $\tilde{A}, \ptilde{A}, \ntilde{A}$ s.t.
$\ptilde{A}, \ntilde{A} \subseteq
\tilde{A}$, symmetries preserve polarity, and
\[
\begin{array}{rl}
\text{\emph{(1)}} & \text{if $\theta \in \ptilde{A} \cap \ntilde{A}$,
then $\theta = \id_x$ for $x \in \conf{A}$,}\\
\text{\emph{(2)}} & \text{if $\theta \in \ntilde{A}$,
$\theta \subseteq^- \theta' \in \tilde{A}$, then $\theta' \in
\ntilde{A}$,}\\
\text{\emph{(3)}} & \text{if $\theta \in \ptilde{A}$,
$\theta \subseteq^+ \theta' \in \tilde{A}$, then $\theta' \in
\ptilde{A}$,}
\end{array}
\]
where $\theta \subseteq^p \theta'$ is $\theta \subseteq
\theta'$ with (pairs of) events of polarity $p$.
\end{defi}
\noindent
Elements of $\ptilde{A}$ (resp. $\ntilde{A}$) are called
\textbf{positive} (resp. \textbf{negative}); they intuitively
correspond to symmetries carried by positive (resp. negative) moves,
introduced by Player (resp. Opponent). We write $\theta : x
\sym_A^- y$ (resp. $\theta : x \sym_A^+ y$) if $\theta \in \ntilde{A}$
(resp. $\theta \in \ptilde{A}$).

Each symmetry has a unique positive-negative factorization:
\begin{lemC}[{\cite[Lemma 7.1.18]{hdr}}]\label{lem:sym_factor}
Consider $A$ a tcg and $\theta : x \sym_A z$ a symmetry.

Then, there are unique $y \in \conf{A}$, $\theta_- : x \sym_A^- y$ and
$\theta_+ : y \sym_A^+ z$ s.t. $\theta = \theta_+ \circ \theta_-$. 
\end{lemC}

\noindent
We extend with symmetry the basic constructions on games:
the \textbf{dual} $A^\perp$ has the same symmetries as $A$, but
$\ptilde{A^\perp} = \ntilde{A}$ and $\ntilde{A^\perp} = \ptilde{A}$;
the \textbf{parallel composition} $A_1 \parallel A_2$ has symmetries
those $\theta_1 \parallel \theta_2 : x_1 \parallel x_2 \sym_{A_1
\parallel A_2} y_1 \parallel y_2$,   where each $\theta_i : x_i
\sym_{A_i} y_i$, and similarly for positive and negative symmetries;
the \textbf{hom} $A \vdash B$ is $A^\perp \parallel B$. 

\subsubsection{Thin strategies} We now define strategies on thin
concurrent games:

\begin{defi}\label{def:thin}
Consider $A$ a tcg.

A \textbf{strategy} on $A$, written $\sigma :
A$, is an ess $\sigma$ equipped with a morphism of ess $\pr_\sigma :
\sigma \to A$ forming a strategy in the sense of Definition
\ref{def:plain_strategy}, and such that:
\[
\begin{array}{rl}
\text{\emph{(1)}} &
\text{if $\theta \in \tilde{\sigma}, \pr_\sigma \theta
\enb_A (a^-, b^-)$, there are unique $\theta \vdash_\sigma (s, t)$ s.t.
$\pr_\sigma s = a$ and
$\pr_\sigma t = b$.}\\
\text{\emph{(2)}} & 
\text{if $\theta : x \sym_\sigma y$ is such that $\pr_\sigma \theta
\in\ptilde{A}$, then $x = y$ and $\theta = \id_x$.} 
\end{array}
\]

\noindent
As before, a \textbf{strategy from $A$ to $B$} is a strategy on $\sigma
: A
\vdash B$.
\end{defi}
\noindent
The first condition forces $\sigma$ to acknowledge Opponent symmetries
in $A$; the notation $\theta \vdash_A (a, b)$
means $(a, b) \not \in \theta$ and $\theta \cup \{(a, b)\} \in
\tilde{A}$. The second condition is \textbf{thinness}: it means that
any
non-identity symmetry in the strategy must originate from Opponent.

\subsubsection{\texorpdfstring{A $\sim$-category}{A ~-category}}
The composition of thin
strategies $\sigma : A \vdash B$ and $\tau : B \vdash C$ is obtained by
equipping $\tau \odot \sigma$ (Proposition
\ref{prop:char_comp}) with an adequate isomorphism family. 

If
$\tildep{\sigma}$ is the restriction of $\tilde{\sigma}$ to
$+$-covered configurations, then we can write $\CC(\tildep{\sigma},
\tildep{\tau})$
for the pairs $(\varphi^\sigma, \varphi^\tau)$ of symmetries
which are matching, i.e. $\varphi^\sigma_B = \varphi^\tau_B$ and whose
domain (or equivalently, codomain) are causally compatible.

\begin{prop}\label{prop:pcomp}
Consider $\sigma : A \vdash B$ and $\tau : B \vdash C$
thin strategies.

There is a unique symmetry on $\tau \odot \sigma$ with a bijection
commuting with $\dom$ and $\cod$
\[
(- \odot -) : \CC(\tildep{\sigma}, \tildep{\tau}) \bij \tildep{\tau
\odot \sigma}
\]
and compatible with
display maps, \emph{i.e.}
$(\varphi^\tau \odot \varphi^\sigma)_A = \varphi^\sigma_A$ and
$(\varphi^\tau \odot \varphi^\sigma)_C = \varphi^\tau_C$.
\end{prop}
\begin{proof}
This follows from \cite[Proposition 7.3.1]{hdr}.
\end{proof}

\noindent
It can be checked that this makes $\tau \odot \sigma : A \vdash C$ a
thin strategy as required. In order to form a $\sim$-category, it is
necessary to give the adequate equivalence relation between thin
strategies. For this, recall first the
2-dimensional structure in $\ESS$, given by the equivalence relation
$\sim$ on morphisms (Section~\ref{subsubsec:ess}). For two maps $f, g :
E \to A$ into a tcg, we write $f \sim^+ g$ if $f \sim g$ and for every
$x \in \conf{E}$ the symmetry obtained as the composition 
\[
f\,x 
\quad \stackrel{f^{-1}}{\bij} \quad
x
\quad \stackrel{g}{\bij} \quad
g\,x\,,
\]
witnessing $f \sim g$ for $x$, is positive. This lets us give the next
definition:

\begin{defi}
Let $\sigma, \tau : A \vdash B$ be thin strategies.
A \textbf{positive morphism of strategies} from $\sigma$ to $\tau$ is a
rigid map of ess $f : \sigma \to \tau$ such that $\pr_\tau \circ f
\sim^+ \pr_\sigma$. We write $f : \sigma \tto \tau$ to mean that $f$ is
a positive morphism from $\sigma$ to $\tau$.

A \textbf{positive isomorphism} $f : \sigma \iso \tau$ is an invertible
(on the nose) positive morphism.
\end{defi}
  \noindent
  As a convention, if $f$ is a 2-cell as above, for
  $x^\sigma \in \conf{\sigma}$ we write
\[
f[x^\sigma] \quad:\quad \pr_\sigma x^\sigma ~\sym_{A\vdash B}^+~
\pr_{\tau}(f\,x^\sigma)
\]
for the positive symmetry witnessing this, which may be decomposed into
two symmetries on the two sides, 
$f[x^\sigma]_A : x^\sigma_A \sym_A^- (f\,x^\sigma)_A$ and
$f[x^\sigma]_B : x^\sigma_B \sym_B^+ (f\,x^\sigma)_B$.

Positive isomorphism will provide the equivalence relation for the
$\sim$-categorical structure of thin concurrent games. A crucial
challenge in constructing this $\sim$-category is then to ensure that
positive isomorphism is preserved under composition. This demands in
particular, given $f : \sigma \tto \sigma' : A \vdash B$ and $g : \tau
\tto \tau' : B \vdash C$, to form a \emph{horizontal composition}
\[
g\odot f 
\quad:\quad 
\tau \odot \sigma ~\tto~ \tau' \odot \sigma' 
\quad:\quad A \vdash C\,,
\]
which requires us to transport $x^\tau \odot x^\sigma \in \confp{\tau
\odot \sigma}$ to $\confp{\tau'\odot \sigma'}$ via $f$ and $g$.
However, the issue is that $f(x^\sigma)$ and $g(x^\tau)$ may not be
matching: the hypotheses at hand only yield
\[
g[x^\tau]_B \circ f[x^\sigma]_B^{-1} 
\quad:\quad 
f(x^\sigma)_B ~\sym_B~ g(x^\tau)_B
\]
a mediating symmetry -- hence to achieve our goals, we use:

\begin{propC}[{\cite[Proposition 7.4.4]{hdr}}]\label{prop:sync_sym}
Let $x^\sigma \in \confp{\sigma}, \theta_B : x^\sigma_B \sym_B
x^\tau_B, x^\tau \in \confp{\tau}$ be \textbf{causally compatible},
\emph{i.e.}
the relation $\cleq$ induced on the graph of the composite bijection
\[
\scalebox{.90}{$
x^\sigma \parallel x^\tau_C 
 \quad\stackrel{\pr_\sigma \parallel x^\tau_C}{\bij}\quad
x^\sigma_A \parallel x^\sigma_B \parallel x^\tau_C
 \quad\stackrel{x^\sigma_A \parallel \theta \parallel
x^\tau_C}{\bij}\quad
x^\sigma_A \parallel x^\tau_B \parallel x^\tau_C
 \quad\stackrel{x^\sigma_A \parallel \pr_\tau^{-1}}{\bij}\quad
x^\sigma_A \parallel x^\tau
$}
\]
by $<_{\sigma \parallel C}$ and $<_{A\parallel \tau}$
as in \S\ref{subsubsec:cg_comp}, is acyclic -- we also say the
composite bijection is \textbf{secured}.

Then, there are unique $y^\tau \odot y^\sigma \in \confp{\tau
\odot \sigma}$ with symmetries
$\varphi^\sigma : x^\sigma \sym_\sigma y^\sigma$ and $\varphi^\tau :
x^\tau \sym_\tau y^\tau$, such that $\varphi^\sigma_A \in \ntilde{A}$
and $\varphi^\tau_C \in \ptilde{C}$, and $\varphi^\tau_B \circ \theta =
\varphi^\sigma_B$.
\end{propC}

\noindent
Altogether, this allows us to construct the desired $\sim$-category:

\begin{thmC}[{\cite[Theorem 7.4.13]{hdr}}]
There is a $\sim$-category $\TCG$ with: \emph{(1)} objects, thin
concurrent games; \emph{(2)} morphisms, strategies $\sigma : A \vdash
B$; \emph{(3)} equivalence, positive isomorphism.
\end{thmC}

\subsection{Boards}
In this line of work connecting concurrent games with relational-like
models, a difficulty is that \emph{points} in the sense of the
relational model are not \emph{all} configurations, but only some of
them. This means that following the approach first outlined by Melliès
\cite{DBLP:conf/lics/Mellies05} and adapted to concurrent games in
earlier work (see \emph{e.g.} \cite{hdr}), we must enrich tcgs with
structure allowing us to identify those configurations that are
\emph{stopping}, in the sense that they correspond to points in the
relational model.

We now introduce boards along with useful constructions on them.

\begin{defi}\label{def:board}
A \textbf{board} is a tcg $A$ along with
$\kappa_A : \conf{A} \to \{-1, 0, +1\}$
a \textbf{payoff function}, such that this data satisfies the following
conditions:
\[
\begin{array}{rl}
\text{\emph{invariant:}} & \text{for all $\theta : x \sym_A y$, we have
$\kappa_A(x) = \kappa_A(y)$,}\\
\text{\emph{race-free:}} & \text{for all $a \mconflict_A\,a'$, we have
$\pol_A(a) = \pol_A(a')$.}\\
\text{\emph{forestial:}} & \text{for all $a_1, a_2, a \in {A}$, if
$a_1, a_2 \leq_A a$, then $a_1 \leq_A a_2$ or $a_2 \leq_A
a_1$,}\\
\text{\emph{alternating:}} &\text{for all $a_1, a_2 \in {A}$, if
$a_1 \imc_A a_2$, then $\pol_A(a_1) \neq \pol_A(a_2)$,}
\end{array}
\]
\indent
A \textbf{$-$-board} must also satisfy the following two additional
conditions:
\[
\begin{array}{rl}
\text{\emph{negative:}} & \text{for all $a$ minimal in $A$, $\pol_A(a)
= -$,}\\
\text{\emph{initialized:}} & \kappa_A(\emptyset) \geq 0\,.
\end{array}
\]
\indent
Finally, a $-$-board $A$ is \textbf{strict} if
$\kappa_A(\emptyset) = 1$ and all its initial moves are in pairwise
conflict. It is \textbf{well-opened} if it is strict
with exactly one initial move.
\end{defi}

\noindent
The payoff function $\kappa_A$ assigns a value to each configuration.
Configurations $x$ with payoff $0$ are called \textbf{complete},
written $x \in \nconf{A}$: those correspond to points in the
relational model.  Otherwise, $\kappa_A$ assigns a responsibility for
why a configuration is non-complete: if $\kappa_A(x) = -1$ then Player
is responsible, otherwise it is Opponent.

We recall a few constructions on boards. The objects of our forthcoming
category will be $-$-boards. The first basic $-$-boards are the units.
In the presence of the payoff function the empty tcg $\etcg$ splits
into two units, reflecting the units of multiplicative and additive
conjunctions in linear logic: the \textbf{top} $\top$ has
$\kappa_\top(\emptyset) = 1$, while the \textbf{one} $\one$ has
$\kappa_\one(\emptyset) = 0$.
To interpret the
base type we shall use a strict board, also written $o$, with only one
move $\qu$, which is negative. Its payoff function is given by
$\kappa_o(\emptyset) = 1$ and $\kappa_o(\{\qu\}) = 0$. 

\subsubsection{Dual, tensor and par.} First, the \textbf{dual} extends
with payoff via $\kappa_{A^\perp}(x) = -\kappa_A(x)$ as expected. Of
course, the dual does not preserve $-$-boards.  \emph{Parallel
composition} splits into:

\begin{figure}
\[
\begin{array}{c|ccc}
\tensor & -1 & 0 & +1\\
\hline
-1 & -1 & -1 & -1\\
0 & -1 & 0 & +1 \\
+1 & -1 & +1 & +1
\end{array}
\qquad\qquad\qquad\qquad
\begin{array}{c|ccc}
\parr & -1 & 0 & +1\\
\hline
-1 & -1 & -1 & +1\\
0 & -1 & 0 & +1 \\
+1 & +1 & +1 & +1
\end{array}
\]
\caption{Payoff for $\tensor$ and $\parr$}
\label{fig:op_payoff}
\end{figure}

\begin{defi}\label{def:board_tensor}
Consider $A$ and $B$ two boards.\\
\indent
Their \textbf{tensor} $A\tensor B$ and their
\textbf{par} $A \parr B$ are $A\parallel B$ enriched with:
\[
\kappa_{A\tensor B}(x_A \parallel x_B) = \kappa_A(x_A) \tensor
\kappa_B(x_B)\,,
\qquad
\kappa_{A\parr B}(x_A \parallel x_B) = \kappa_A(x_A) \parr
\kappa_B(x_B)
\]
with the operations $\tensor$ and $\parr$ defined on $\{-1, 0, +1\}$ in
Figure \ref{fig:op_payoff}.
\end{defi}

\noindent
The tensor of two $-$-boards is still a $-$-board, though tensor does
not
preserve \emph{strict} $-$-boards. The par also preserves $-$-boards,
but
we shall not use it on $-$-boards: if $A$ and $B$ are $-$-boards, let
us
use $A \vdash B$ to denote the board $A^\perp \parr B$ used to define
the strategies \emph{from $A$ to $B$}. Observe that even if $A$ and $B$
are $-$-boards, $A \vdash B$ is not.

By definition of payoff, the order-isomorphisms of
\eqref{eq:isopar} and \eqref{eq:isohom} refine to bijections: 
\begin{eqnarray}
- \tensor - \quad:\quad \nconf{A} \times \nconf{B} &\iso&
  \nconf{A\tensor B}\,,\\ \label{eq:comptensor}
- \parr - \quad:\quad \nconf{A} \times \nconf{B} &\iso& 
  \nconf{A \parr B}\,. \label{eq:comppar}
\end{eqnarray}

\subsubsection{The with.} We only consider the additive conjunction of
linear logic: the with. But in order to define it, we must first define
a new operation on ess and tcgs.

\begin{defi}
Let $A_1$ and $A_2$ be two tcgs.\\
\indent
Then, we define their \textbf{sum} $A_1 + A_2$
as comprising the components:
\[
\begin{array}{rrcl}
\text{\emph{events:}} & \ev{A_1\parallel A_2} &=& \ev{A_1} + \ev{A_2}\\
\text{\emph{causality:}} & (i, a) \leq_{A_1 \parallel A_2} (j, a')
&\Leftrightarrow&
i=j~\&~a \leq_{A_i} a'\\
\text{\emph{conflict:}} & (i, a) \conflict_{A_1 \parallel A_2} (j, a')
&\Leftrightarrow& i\neq j ~\vee~ a \conflict_{E_i} a'\,,\\
\text{\emph{symmetry:}} & \theta \in \tilde{A_1 \parallel A_2}
&\Leftrightarrow& \exists \theta_i \in \tilde{A_i}, \theta = \theta_1
\parallel \theta_2\,,\\
\text{\emph{positive symmetries:}} & 
\theta_1 \parallel \theta_2 \in \ptilde{A_1 + A_2} &\Leftrightarrow&
\theta_1 \in \ptilde{A_1}\,\&\,\theta_2 \in \ptilde{A_2}\\
\text{\emph{negative symmetries:}} &
\theta_1 \parallel \theta_2 \in \ntilde{A_1 + A_2} &\Leftrightarrow& 
\theta_1 \in \ntilde{A_1}\,\&\,\theta_2 \in \ntilde{A_2}\,.
\end{array}
\]
where, necessarily, one of $\theta_1$ or $\theta_2$ must be empty.
\end{defi}

\noindent
Ignoring the positive and negative symmetries, this also yields an
operation $+$ on plain event structures with symmetry that we shall use
later on.

If $A, B$ are tcgs and $x_A \in \conf{A}$, we write $(1, x_A) \in
\conf{A + B}$ as a shorthand for $\{1\} \times x_A$ and likewise for
$(2, x_B) = \{2\} \times x_B \in \conf{A + B}$ for $x_B \in \conf{B}$.
Note that all configurations of $A+B$ have the form $(1, x_A)$ for $x_A
\in \conf{A}$ or $(2, x_B)$ for $x_B \in \conf{B}$. For non-empty
configurations, this decomposition is \emph{unique}. We shall also use
the corresponding notations for symmetries, with \emph{e.g.} $(1,
\theta_A) : (1, x_A) \sym_{A+B} (1, y_A)$ for $\theta_A : x_A \sym_A
y_A$ comprising all $((1, a), (1, a'))$ for $(a, a') \in \theta_A$.
This sum operation yields the \emph{with} operation on strict boards:

\begin{defi}
Consider $S$ and $T$ two strict $-$-boards.\\
\indent
Then, their \textbf{with} $S \with T$ is the strict $-$-board with tcg
the sum $S + T$ and
\[
\kappa_{S\with T}(1, x_S) = \kappa_S(x_S)\,,
\qquad
\kappa_{S\with T}(2, x_T) = \kappa_T(x_T)\,,
\]
for non-empty configurations and $\kappa_{S\with T}(\emptyset) = 1$.
\end{defi}

\noindent
As we will see, this construction will give a cartesian product in the
subcategory of strict $-$-boards.
It can also be applied to
non-strict $-$-boards, but then it is not a product: if in $A_1 \with
A_2$, one of the $A_i$ is not strict, then the
corresponding projection does not respect payoff (in the sense of
Definition \ref{def:strat_winning}), because we have set $\kappa_{A_1
\with A_2}(\emptyset) = 1$. On the other hand, setting $\kappa_{A_1
\with A_2}(\emptyset) = 0$ breaks the correspondence with the
relational model, since the empty configuration does not correspond in
a canonical way to one of the components. 

In the sequel, we shall use the obvious $n$-ary generalization of the
product. Observe that any strict $-$-board $S$ decomposes uniquely (up
to forest isomorphism) as $S \iso \with_{i \in I} S_i$, where each
$S_i$ is well-opened. We need notations for configurations of this
board.
Writing $\confn{E}$ (resp. $\tildn{E}$) for the non-empty
configurations (resp. symmetries) of $E$, we observe:

\begin{lem}\label{lem:with_confn}
Consider $(S_i)_{i\in I}$ a family of well-opened $-$-boards. Then
there
are
\[
\begin{array}{rcl}
\confn{\with_{i\in I} S_i} &\iso& \sum_{i\in I} \confn{S_i}\\
\tildn{\with_{i\in I} S_i} &\iso& \sum_{i\in I} \tildn{S_i}
\end{array}
\qquad
\begin{array}{rcl}
\ntilden{\with_{i\in I} S_i} &\iso& \sum_{i\in I} \ntilden{S_i}\\
\ptilden{\with_{i\in I} S_i} &\iso& \sum_{i\in I} \ptilden{S_i}
\end{array}
\]
order-isos commuting with $\dom$ and $\cod$.
\end{lem}

\subsubsection{Linear closure.} First, we define it in the case the rhs
board is well-opened:

\begin{defi}
Consider $A$ a $-$-board and $S$ a well-opened $-$-board.\\
\indent
Then, $A\lin S$ has tcg all components set as $A \vdash S$ except for:
\[
\begin{array}{rrcl}
\text{\emph{causality:}} &
\leq_{A\lin S} &=& {\leq_{A \vdash S}} \uplus \{((2, s_0),
(1, a)) \mid a \in A\}
\end{array}
\]
writing $\min(S) = \{s_0\}$, yielding a well-opened tcg. Its payoff
function is:
\[
\kappa_{A\lin S}(x_A \parallel x_S) = \kappa_{A\vdash S}(x_A
\parallel x_S) = \kappa_{A^\perp}(x_A) \parr \kappa_S(x_S)\,.
\]  
\end{defi}

\noindent
This corrects the non-negativity of $A \vdash S$, by forcing the
missing dependency. In the sequel, we shall need
$A \lin S$ not only when $S$ is well-opened (which has no particular
status in the definition of relative Seely categories), but when it is
strict. In that case, $A \lin S$ may be defined directly via the
decomposition into strict boards, as done in:

\begin{defi}
Consider $A$ a $-$-board, and $S$ a strict $-$-board, with $S
\iso \with_{i\in I} S_i$.

Then, we define $A \lin S = \with_{i\in I} (A\lin S_i)$.
\end{defi}

\noindent
The following lemma then follows from  Lemma \ref{lem:with_confn}:

\begin{lem}\label{lem:confn_lin}
Consider $A, S$ $-$-boards with $S$ strict. Then, there are:
\[
\begin{array}{rcl}
\confn{A\lin S} &\iso& \conf{A} \times \confn{S}\\
\tildn{A\lin S} &\iso& \tilde{A} \times \tildn{S}
\end{array}
\qquad
\begin{array}{rcl}
\ntilden{A\lin S} &\iso& \ptilde{A} \times \ntilden{S}\\
\ptilden{A\lin S} &\iso& \ntilde{A} \times \ptilden{S}
\end{array}
\]
order-isos commuting with $\dom$ and $\cod$.
\end{lem}

\noindent
This anticipates on the link with the relational model, where the
linear arrow is obtained with a cartesian product.
Following this, we adopt the convention that for each $x_A \in
\conf{A}$ and $x_S \in \confn{S}$, $x_A \lin x_S \in \confn{A\lin S}$
denotes the corresponding configuration.

\subsubsection{Exponential.} We start by defining the bang as a
construction on mere ess:

\begin{defi}
Consider $E$ an ess. Then, we define the \textbf{bang} $\oc E$ with:
\[
\begin{array}{rrcl}
\text{\emph{events:}} &
\ev{\oc A} &=& \mathbb{N} \times \ev{A}\\
\text{\emph{causality:}} &
(\grey{i}, a_1) \leq_{\oc A} (\grey{j}, a_2) &\Leftrightarrow&
\grey{i}=\grey{j} \,\wedge\, a_1 \leq_A a_2\\
\text{\emph{conflict:}} &
(\grey{i}, a_1) \conflict_{\oc A} (\grey{j} , a_2) &\Leftrightarrow&
\grey{i}=\grey{j} \,\wedge\, a_1 \conflict_A a_2\\
\text{\emph{symmetries:}} &
\theta \in \tilde{\oc A} &\Leftrightarrow& 
\exists \pi : \mathbb{N} \bij \mathbb{N},\quad
\exists (\theta_n)_{n \in \mathbb{N}} \in \tilde{A}^\mathbb{N}\,\\
&&&\forall (\grey{i}, a) \in \dom(\theta), \theta(\grey{i}, a) =
(\grey{\pi(i)}, \theta_{\grey{i}}(a))\,.
\end{array}
\]
\end{defi}

\noindent
In $(\grey{i}, e) \in \oc E$, we refer to $\grey{i}$ as
a \textbf{copy index}. This is in this definition of games that
symmetries really come into play: they are used to express that these
copy indices can be reindexed at will. It will be convenient
to characterize the shape of configurations on $\oc E$:

\begin{lem}\label{lem:dec_conf_oc}
For $E$ an ess, there is an order-isomorphism
\begin{eqnarray}
\,\famc{-} &:& \Fam\left(\confn{E}\right)  ~\bij~
\conf{\oc E}
\label{eq:bijfam}
\end{eqnarray}
with $\Fam(X)$ the set of families of elements of $X$ indexed
by finite subsets of $\mathbb{N}$.
\end{lem}
\begin{proof}
This associates to $(x_i)_{i\in I}$ the set 
$\parallel_{i\in I} x_i = \uplus_{i\in I} \{i\} \times x_i$.
\end{proof}

\noindent
To clarify the notation above: we mean that for any family $(x_i)_{i\in
I} \in \Fam(\confn{E})$, we write $\famc{x_i \mid i \in I} \in
\conf{\oc E}$ for the corresponding configuration of $\oc E$. In
addition, it is clear that if $X$ is a set, then $\Fam(X)$ quotiented
by permutations of indices is in bijection with $\Mf(X)$. Hence, the
order-isomorphism of Lemma \ref{lem:dec_conf_oc} immediately yields a
bijection
\begin{eqnarray}
\conf{\oc E}/\!\sym_{\oc E}
&\bij&
\Mf(\confne{E}/\!\sym_E)
\label{eq:bijsym}
\end{eqnarray}
which again suggests the forthcoming connection with the relational
model.

Now, we extend the bang construction to boards:

\begin{defi}\label{def:bang}
Consider $S$ a strict board. Then, the \textbf{bang} $\oc S$
additionally has:
\[
\begin{array}{rrcl}
\text{\emph{polarities:}} &
\pol_{\oc S}(\grey{i}, s) &=& \pol_S(s)\\
\text{\emph{pos. symmetries:}} &
\theta \in \ptilde{\oc S} &\Leftrightarrow&
\exists (\theta_n)_{n\in \mathbb{N}} \in \ptilde{S}^{\mathbb{N}}\,,\\
&&&\forall (\grey{i}, s) \in \dom(\theta), \theta(\grey{i}, s) =
(\grey{i}, \theta_{\grey{i}}(s))\\
\text{\emph{neg. symmetries:}} &
\theta \in \ntilde{\oc S} &\Leftrightarrow&
\exists \pi : \mathbb{N} \bij \mathbb{N},\,
\exists (\theta_n)_{n \in \mathbb{N}} \in \ntilde{S}^{\mathbb{N}}\,,\\
&&&\forall (\grey{i}, s) \in \dom(\theta), \theta(\grey{i}, s) =
(\grey{\pi(i)}, \theta_{\grey{i}}(s))
\end{array}
\]
with payoff given by
$\kappa_{\oc S}(\famc{x_i \mid i \in I}) = \bigotimes_{i\in I}
\kappa_S(x_i)$, and $\kappa_{\oc S}(\famc{}) = 0$.
\end{defi}

\noindent
If $S$ is a strict board, then $\oc S$ is still a $-$-board, but no
longer strict. Since the minimal events of $S$ are negative, an
exchange in the copy indices arising from this definition is viewed as
\emph{negative}. Hence, positive symmetries can not affect them. 
Also, because $S$ is strict, its complete configurations are non-empty;
hence \eqref{eq:bijfam} and \eqref{eq:bijsym} refine to:
\begin{eqnarray}
\famc{-} : \hspace{18pt}\Fam(\nconf{S}) &\iso& \nconf{\oc S}\\
\,[-] : \Mf(\nconf{S}/\!\sym_S) &\bij& \nconf{\oc S}/\!\sym_{\oc S}
\label{eq:coll_bang}
\end{eqnarray}
which again suggests the forthcoming relational collapse. Note that
this only holds if $S$ is a strict board. The bang $\oc S$ does work in
more generality \cite{DBLP:journals/lmcs/CastellanC24,hdr}, but not
in a way that is compatible with the relational model. Accordingly, we
shall focus not on the Seely category structure where $\oc$ is a
comonad, but in a variation called \emph{relative Seely category} where
$\oc$ is a \emph{relative comonad}; we will come back to that point
later on.

\subsection{The Relative Seely Category of Sequential Innocence}
As explained before, we will form a category whose objects are
$-$-boards. The morphisms will be certain strategies in the sense of
Definition \ref{def:thin}. But to ensure the existence of a functorial
collapse to the relational model, we must impose additional conditions
on strategies.

\subsubsection{Deterministic sequential innocence} We first introduce
our notion of strategies:

\begin{defi}\label{def:strat_winning}
Consider $A$ a board, and $\sigma : A$ a strategy. We define
conditions:
\[
\begin{array}{rl}
\text{\emph{negative:}} & 
\text{for all $s \in {\sigma}$ minimal for $\leq_\sigma$, we have
$\pol_A(\pr_\sigma\,s) = -$,}\\
\text{\emph{winning:}} &
\text{for all $x^\sigma \in \confp{\sigma}$,
$\kappa_A(\pr_\sigma\,x^\sigma) \geq 0$,}\\
\text{\emph{forestial:}} & \text{$\leq_\sigma$ is a forest.}\\
\text{\emph{deterministic:}} & \text{if $s^- \imc_\sigma s^+_1, s^+_2$
then $s^+_1 = s^+_2$.}
\end{array}
\]

\noindent
We say $\sigma$ is \textbf{deterministic sequential innocent (dsinn)}
if it satisfies all four.
\end{defi}

\noindent
\emph{Winning} ensures that strategies are well-behaved with respect to
payoff: in particular, a closed interaction between winning strategies
will always result in a complete position, which is essential for the
relational collapse. \emph{Forestial} and \emph{deterministic} make
$\sigma$ a negatively branching forest, which mimics a syntactic
tree\footnote{The relational collapse is possible for the more general
notion of \emph{visible} strategies, see \cite{hdr}. But the rest of
this paper will depend on this specific notion of deterministic
sequential strategy.}; this ensures that composition is deadlock-free
and therefore matches relational composition. 

Copycat strategies on $-$-boards are automatically deterministic
sequential innocent. Deterministic sequential innocence is also stable
under composition, which ensures:

\begin{thm}
There is a $\sim$-category $\Dsinn$ with: \emph{(1)} objects,
$-$-boards; \emph{(2)} morphisms, dsinn strategies; \emph{(3)}
equivalence, positive isomorphism.
\end{thm}

\subsubsection{Relative Seely categories} This category (we shall often
omit the ``$\sim$-'' from now on) has significant further structure.
In particular, it can be organized into a \emph{Seely category}, a
categorical model of intuitionistic linear logic \cite{panorama}; but
unfortunately that structure is not preserved by the relational
collapse. For instance, for a game $\oc \oc A$, there is only one empty
configuration, whereas in $\Mf(\Mf(\conf{A}))$ there are multiple ways
to be ``empty'': $[], [[]], [[],[]],\ldots$; even if the empty
configuration on $A$ is not deemed a valid position. In a sense, the
relational model counts how many times a program ``does nothing'',
which is meaningless if states in the relational model are to
correspond to sets of events. 

Fortunately, this mismatch arises outside of the translation of simple
types (or even standard linear/non-linear systems, where $\oc$ can only
occur on the left hand side of an arrow). The categorical structure
describing the structure that \emph{is} preserved is called a
\emph{relative Seely category}
\cite{DBLP:journals/corr/abs-2107-03155,hdr}; we now recall the
definition.

\begin{defi}\label{def:relativeseely}
A \textbf{relative Seely category} is a symmetric monoidal category
$(\C, \tensor, 1)$ equipped with a full subcategory $\C_s$ together
with the following data and axioms:

\begin{itemize}
\item $\C_s$ has finite products $(\with, \top)$ preserved by the
inclusion functor $J : \C_s \hookrightarrow \C$.
\item For every $B \in \C$ there is a functor $B \lin - : \C_s \to
\C_s$, such that there is
\[
\Lambda(-) : \C(A\tensor B, S) \bij \C(A, B \lin S).
\]
a bijection natural in $A \in \C$ and $S \in \C_s$.
\item There is a $J$-relative comonad $\oc : \C_s \to \C$. This means
that we have, for every $S \in \C_s$, an object $\oc S \in \C$ and a
\textbf{dereliction} morphism $\der_S : \oc S \to S$, and for every
$\sigma: \oc S \to T$, a \textbf{promotion} $\sigma^{\oc} : \oc S \to
\oc T$, subject to three axioms \cite{relativemonads}:
\[
\begin{array}{rclcl}
\der_T \circ \sigma^{\oc} &=& \sigma
&\qquad&(\sigma : \oc S \to T)\\
\der_S^{\oc} &=& \id_{\oc S} &&(S \in \C_s)\\
(\tau \,\circ\, \sigma^{\oc})^{\oc} &=& \tau^{\oc} \,\circ\,
\sigma^{\oc}
&&(\sigma : \oc S \to T, \tau : T\to U)\,,
\end{array}
\]
which make $\oc : \C_s \to \C$ a functor, via $\oc \sigma =
(\sigma \,\circ\, \der_S)^{\oc}$ for $\sigma : S \to T$.

\item The functor $\oc : \C_s \to \C$ is symmetric strong monoidal
$(\C_s, \with, \top) \to (\C, \tensor, 1)$, so
\[m_0 : 1 \to \oc \top \qquad \qquad m_{S, T} : \oc S \tensor \oc T
\to
\oc(S \with T)\]
are natural isos, additional compatible with promotion: the diagram
\[
\xymatrix@R=10pt@C=10pt{
\oc \Gamma
        \ar[rr]^{\tuple{f, g}^{\oc}}
        \ar[d]_{\tuple{\der, \der}^{\oc}}&&
\oc (S\with T)
        \ar[d]^{m^{-1}}\\
\oc (\Gamma \with \Gamma)
        \ar[dr]_{m^{-1}}&&
\oc S \tensor \oc T
        \ar@{<-}[dl]^{f^{\oc} \tensor g^{\oc}}\\
&\oc \Gamma \tensor \oc \Gamma
}
\]
commutes for all $\Gamma, S, T \in \C_s$, $f\in \C(\oc \Gamma, S), g
\in \C(\oc \Gamma, T)$.
\end{itemize}
\end{defi}

\noindent
Any Seely category is a relative Seely category with $\C =
\C_s$. For any relative Seely category, the Kleisli category associated
with $\oc$, denoted $\C_{\oc}$, is cartesian closed: it has objects
those of $\C_s$, morphisms $\C_{\oc}(S, T)  = \C(\oc S, T)$, products
$\with$, and function space $S \tto T = \oc S \lin T$.

There is an accompanying notion of morphism: a \textbf{relative Seely
functor} from $\C$ to $\D$ is a functor $F : \C \to \D$ together with
isomorphisms
\[
\begin{array}{rcrcl}
t^\tensor_{A,B} &:& FA \tensor FB &\iso& F(A\tensor B)\\
t^\with_{S,T} &:& FS \with FT &\iso& F(S\with T)\\
t^\lin_{A,S} &:& FA \lin FS &\iso& F(A\lin S)\\
\end{array}
\qquad\qquad
\begin{array}{rcrcl}
t^1 &:& 1 &\iso& F1\\
t^\top &:& \top &\iso& F\top\\
t^{\oc}_S &:& \oc FS &\iso& F\oc S
\end{array}
\]
satisfying appropriate naturality and coherence conditions \cite{hdr}.
This ensures in particular that $F$ lifts to a cartesian closed functor
$F_\oc : \C_\oc \to \D_\oc$.\footnote{These definitions must of course
be taken in $\sim$-categorical form: as before this means that all
operations preserve $\sim$, and all conditions hold up to $\sim$;
yielding  notions of \textbf{relative Seely $\sim$-category} and
\textbf{relative Seely $\sim$-functor}. We omit the straightforward
adaptation.} 

\subsubsection{\texorpdfstring{The relative Seely category $\Dsinn$}{The relative Seely category DSInn}.} We now spell out
the additional structure in $\Dsinn$. From now on, by \emph{strategy}
we mean a morphism in $\Dsinn$, \emph{i.e.} a sequential deterministic
strategy in the sense of Definition \ref{def:strat_winning}.
First, the symmetric monoidal structure is handled by:

\begin{propC}[{\cite[Propositions 8.1.1 and 8.2.16]{hdr}}]\label{prop:def_tensor}
Consider $A, B, C, D$ $-$-boards, and $\sigma : A \vdash B$, $\tau : C
\vdash D$ strategies.\\
\indent
Then, there is a strategy $\sigma \tensor \tau : A \tensor C
\vdash B \tensor D$, unique up to iso, s.t. there are
\[
\begin{array}{rcrcl}
(- \tensor -) &:& \confp{\sigma} \times \confp{\tau} &\simeq&
\confp{\sigma \tensor \tau}\\
(- \tensor -) &:& \tildep{\sigma} \times \tildep{\tau} &\simeq&
\tildep{\sigma \tensor \tau}
\end{array}
\]
order-isos commuting with $\dom, \cod$, and s.t. for all
$\theta^\sigma
\in
\tildep{\sigma}$ and $\theta^\tau \in \tildep{\tau}$,
\[
\pr_{\sigma \tensor \tau}(\theta^\sigma \tensor \theta^\tau) =
(\theta^\sigma_A \tensor \theta^\tau_C) \tensor (\theta^\sigma_B
\tensor \theta^\tau_D)\,.
\]
\indent
Moreover, $(- \tensor -)$ preserves positive isomorphism.
\end{propC}

\noindent
From this definition, Propositions \ref{prop:char_comp},
\ref{prop:pcomp} and a straightforward symmetry-aware extension of
Lemma \ref{lem:iso_confp}, it follows that $\tensor$ is a bifunctor.
In addition, there are associativity and unit natural isomorphisms
defined as the obvious copycat strategies, satisfying the required
conditions for a symmetric monoidal category \cite[Proposition
8.2.20]{hdr}.

Next, we move to the product, which is defined via:

\begin{propC}[{\cite[Proposition 8.2.22]{hdr}}]\label{prop:def_pairing}
For $-$-boards $\Gamma, S, T$, with $S, T$ strict, and strategies
$\sigma : \Gamma \vdash S, \tau : \Gamma \vdash T$, there is a strategy
$\tuple{\sigma, \tau} : \Gamma \vdash S \with T$, unique up to iso,
such that there are order-isos: 
\[
\begin{array}{rcl}
\confp{\sigma} + \confp{\tau}  &\simeq& \confp{\tuple{\sigma, \tau}}\\
\tildep{\sigma} + \tildep{\tau} & \simeq& \tildep{\tuple{\sigma, \tau}}
\end{array}
\]
commuting with $\dom, \cod$, and such that for all $\theta^\sigma \in
\tildep{\sigma}$ and $\theta^\tau \in \tildep{\tau}$, we have
\[
\pr_{\tuple{\sigma, \tau}}(\inj_\sigma(\theta^\sigma)) =
\theta^\sigma_\Gamma \vdash \inj_S(\theta^\sigma_S)\,,
\qquad
\pr_{\tuple{\sigma, \tau}}(\inj_\tau(\theta^\tau)) =
\theta^\tau_\Gamma \vdash \inj_T(\theta^\tau_T)
\]
with $\inj_\sigma : \confp{\sigma} \to \confp{\tuple{\sigma, \tau}}$
and $\inj_\tau : \confp{\tau} \to \confp{\tuple{\sigma, \tau}}$ the
induced injections.\\
\indent
Moreover, $\tuple{-, -}$ preserves positive isomorphism.
\end{propC}

\noindent
There are also projections, obtained again as the obvious copycat
strategies
\[
\pi_S : S\with T \vdash S\,,
\qquad
\qquad
\pi_T : S\with T \vdash T
\]
turning $S \with T$ into a categorical product \cite[Proposition
8.2.24]{hdr} for $S$ and $T$ strict, and additionally $\top$ is
terminal. Finally, for the arrow type, we have:

\begin{propC}[{\cite[Proposition 8.2.25]{hdr}}]\label{prop:def_curry}
Consider $\Gamma, A, S$ $-$-boards with $S$ strict.\\
\indent
For $\sigma : \Gamma \tensor A \vdash S$, there is
$\Lambda(\sigma) : \Gamma \vdash A \lin S$, unique up to iso, s.t.
there are
\[
\begin{array}{rcrcl}
\Lambda(-) &:& \confp{\sigma} &\iso& \confp{\Lambda(\sigma)}\\
\Lambda(-) &:& \tildep{\sigma} &\iso& \tildep{\Lambda(\sigma)}
\end{array}
\]
order-isos commuting with $\dom, \cod$, and such that for all
$\theta^\sigma$ non-empty,
\begin{eqnarray}
\pr_{\Lambda(\sigma)}(\Lambda(\theta^\sigma)) &=&
\theta^\sigma_\Gamma \vdash
\theta^\sigma_A \lin \theta^\sigma_S\label{eq:pr_cur}
\end{eqnarray}
whenever $\pr_\sigma(\theta^\sigma) = \theta^\sigma_\Gamma \parallel
\theta^\sigma_A \vdash \theta^\sigma_S$.\\
\indent
Moreover, $\Lambda(-)$ preserves positive isomorphism.
\end{propC}

\noindent
This construction, \textbf{currying}, is easily shown to be invertible
up to isomorphism. As usual, the \textbf{evaluation} is defined as
$\bevm_{A, S} = $ 
$\Lambda^{-1}(\cc_{A\lin S})$, the uncurrying of the
identity, for $A$, $S$ two $-$-boards with $S$ strict; altogether this
makes $A\lin S$ an arrow of $A$ and $S$.

The last outstanding construction is the \emph{promotion}, relative to
the exponential:

\begin{prop}\label{prop:def_bang}
Consider $S, T$ strict $-$-boards, and $\sigma : \oc S \vdash T$ a
strategy.

Then, the ess $\oc \sigma$ may be equipped with a display map
$\pr_{\sigma^\dagger}$ such that
\[
\pr_{\sigma^\dagger}(\famc{x^{\sigma,i} \mid i \in I}) =
\famc{x^{\sigma,i}_{A,j} \mid \tuple{i,j} \in \Sigma_{i\in I} J_i}
\vdash \famc{x^{\sigma,i}_B \mid i \in I}
\]
where $\pr_{\sigma}(x^{\sigma,i}) = \famc{x^{\sigma,i}_{A,j} \mid j \in
J_i} \vdash x^{\sigma,i}_B$, writing $\Sigma_{i\in I} J_i$ for the set
of encodings $\tuple{i,j} \in \N$ of all pairs of $i\in I$ and $j \in
J_i$, where $\tuple{-,-} : \N^2 \bij \N$ is an arbitrary bijection.

This makes $\sigma^\dagger : \oc S \vdash \oc T$ a strategy, and 
the construction $(-)^\dagger$ preserves positive iso.
\end{prop}

\noindent
The \textbf{dereliction} strategy $\der_A : \oc A \vdash A$
is defined as a copycat strategy, opening one copy with copy index
$\grey{0}$. Finally, the \emph{Seely isomorphisms} (up to positive iso)
are 
\[
\mon_{S, T} \quad:\quad \oc S \tensor \oc T ~\iso~ \oc (S \with T)\,,
\]
defined again by the obvious copycat strategy, and the obvious
isomorphism $\oc \top \iso \one$ between empty games. Altogether, these
provide all the components for the desired structure:

\begin{thm}
The components above make $\Dsinn$ a relative Seely category; where the
strict full subcategory $\Dsinn_s$ is restricted to strict $-$-boards.
\end{thm}

\noindent
In particular, the Kleisli category $\Dsinn_\oc$ is cartesian closed.

\section{Relational Collapse}
\label{sec:rel_collapse}

Now that we have set up our ambient game semantics, we are in position
to resume the discussion in the introduction now resting on precise
definitions. The aim of this section is to recall the relational
collapse of thin concurrent games, for deterministic sequential
strategies. 

This takes the form of a relative Seely functor
\[
\coll{-} : \Dsinn \to \Rel
\]
which we describe in this section. Again, we borrow much of the
presentation from \cite{hdr}.

\subsection{General Idea} We start by giving the basic definition of
the collapse.

\subsubsection{Collapsing games.}
As argued before, \emph{boards} come equipped
with a notion of \textbf{stopping configurations}: namely, those
configurations whose payoff is null:
\[
\stopc{A} = \{x \in \conf{A} \mid \kappa_A(x) = 0\}\,,
\]
which as argued in the introduction, are designed to match notions of
rigid intersection types; or alternatively, the objects of the
interpretation of a type in generalized species
\cite{DBLP:conf/lics/ClairambaultOP23}.

In turn, points of the web in the sense of the
relational model -- or non-idempotent intersection types -- will
correspond to stopping configurations \emph{up to symmetry}
\begin{eqnarray}
\coll{A} &=& \stopc{A}/\!\sym_A
\label{eq:def_pos}
\end{eqnarray}
called the \textbf{positions} of $A$. Here, we use symbols $\x, \y,
\z\ldots$ to range over \emph{symmetry classes of configurations} --
note the different font than for configurations. 

Now, the idea is simply to send a board $A$ to its set of positions.
This is well-behaved, in the sense that (by design!) there are
relatively straightforward bijections presented in Figure
\ref{fig:str_coll_bij}
\begin{figure}
\begin{minipage}{.48\linewidth}
\[
\scalebox{.9}{$
\begin{array}{rcrcl}
s^\vdash_{A, B} &:& \coll{A} \times \coll{B} &\bij& \coll{A\vdash B}\\
s^\tensor_{A, B} &:& \coll{A} \times \coll{B} &\bij& \coll{A \tensor
B}\\
s^\one &:& \one &\bij& \coll{\one}\\
s^{\oc}_S &:& \Mf(\coll{S}) &\bij& \coll{\oc S}\\
s^\top &:& \emptyset &\bij& \coll{\top}\\
s^\with_{S, T} &:& \coll{S} + \coll{T} &\bij& \coll{S\with T}\\
s^\lin_{A, S} &:& \coll{A} \times \coll{S} &\bij& \coll{A\lin S}
\end{array}
$}
\]
\caption{Structural collapse bijections}
\label{fig:str_coll_bij}
\end{minipage}
\hfill
\begin{minipage}{.48\linewidth}
\[
\scalebox{.9}{$
\begin{array}{rcrcl}
s^\vdash_{A, B} &:& \ccoll{A}{\C} \times \ccoll{B}{\C} &\bij&
\ccoll{A\vdash B}{\C}\\
s^\tensor_{A, B} &:& \ccoll{A}{\C} \times \ccoll{B}{\C} &\bij& \ccoll{A
\tensor B}{\C}\\
s^\one &:& \one &\bij& \ccoll{\one}{\C}\\
s^{\oc}_S &:& \Mf(\ccoll{S}{\C}) &\bij& \ccoll{\oc S}{\C}\\
s^\top &:& \emptyset &\bij& \ccoll{\top}{\C}\\
s^\with_{S, T} &:& \ccoll{S}{\C} + \ccoll{T}{\C} &\bij& \ccoll{S\with
T}{\C}\\
s^\lin_{A, S} &:& \ccoll{A}{\C} \times \ccoll{S}{\C} &\bij&
\ccoll{A\lin S}{\C}
\end{array}
$}
\]
\caption{Colored collapse bijections}
\label{fig:col_str_coll_bij}
\end{minipage}
\end{figure}
where $A, B$ are any $-$-boards and $S, T$ are strict. For $\one$ and
$\top$ this is clear. For the tensor and hom-game this follows from
\eqref{eq:comptensor} and \eqref{eq:comppar}. For the with this comes
from Lemma \ref{lem:with_confn}, for the linear arrow from Lemma
\ref{lem:confn_lin}, and for the bang, from \eqref{eq:coll_bang}.

From the above, as an immediate corollary we get a bijection, for every
simple type $A$:
\begin{eqnarray}
s_A 
\quad:\quad
\intr{A}_{\Rel} 
&\bij&
\coll{\intr{A}_{\Dsinn}}
\label{eq:bij_posrel}
\end{eqnarray}
obtained simply by induction on $A$ -- and similarly for a simply-typed
context $\Gamma$. So as announced, we are able to identify the points
in the relational interpretation of a type $A$ with certain (symmetry
classes of) configurations of the game semantics interpretation of $A$.

\subsubsection{Collapsing strategies}
Now, we must extend this to \emph{strategies}. The idea is rather
simple: we intend to simply send a strategy to its set of reached
positions. In fact, the slightly better behaved definition consists in
sending a strategy $\sigma : A \vdash B$ to 
\begin{eqnarray}
\coll{\sigma} 
&=& 
\{ (\x_A, \x_B) \in \coll{A} \times \coll{B} \mid \exists x^\sigma \in \confp{\sigma},~
x^\sigma_A \in \x_A,~x^\sigma_B \in \x_B \}
\label{eq:coll_rel_strat}
\end{eqnarray}
those positions reached by \emph{$+$-covered} configurations
only\footnote{In fact, in the case of deterministic sequential innocent
strategies, the configurations that reach positions are always
$+$-covered, under the mild assumption that stopping
configurations have as many Opponent as Player events. We stick with
the $+$-covered definition, which works without sequential innocence
\cite{hdr}.}.
\begin{figure}
\[
\raisebox{40pt}{
\xymatrix@R=20pt@C=10pt{
&&&\qu^-_4
        \ar@{-|>}[dlll]\\
\qu^+_{2,\grey{0}}
        \ar@{-|>}[d]
        \ar@{.}@/^/[urrr]
&&\qu^+_{2,\grey{1}}
        \ar@{.}@/^/[ur]
        \ar@{-|>}[dl]
        \ar@{-|>}[dr]
&&\qu^+_{3,\grey{1}}
        \ar@{.}@/_/[ul]
&\qu^+_{3,\grey{3}}
        \ar@{.}@/_/[ull]\\
\qu^-_{1,\grey{0}} 
	\ar@{.}@/^.1pc/[u]
        \ar@{-|>}[urr]&
\qu^-_{1,\grey{0}} 
	\ar@{.}@/^/[ur]
        \ar@{-|>}[urrr]
&&\qu^-_{1,\grey{1}}
        \ar@{.}@/_/[ul]
        \ar@{-|>}[urr]
}}
\qquad
\leadsto
\qquad
\raisebox{40pt}{
\xymatrix@R=20pt@C=10pt{
&&&\qu^-_4\\
\qu^+_2
        \ar@{.}@/^/[urrr]
&&\qu^+_2
        \ar@{.}@/^/[ur]
&&\qu^+_3
        \ar@{.}@/_/[ul]
&\qu^+_3
        \ar@{.}@/_/[ull]\\
\qu^-_1 \ar@{.}@/^.1pc/[u]&
\qu^-_1 \ar@{.}@/^/[ur]
&&\qu^-_1
        \ar@{.}@/_/[ul]
}}
\]
\caption{The relational collapse : forgetting the dynamic order}
\label{fig:coll_dyn}
\end{figure}
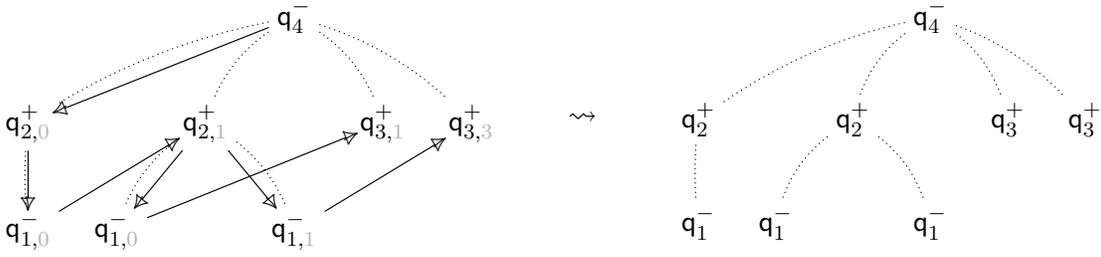
This is illustrated in Figure \ref{fig:coll_dyn}, with one $+$-covered
configuration arising from the interpretation of $\lambda
fx.\,f\,(f\,x) : (o_1 \to o_2) \to o_3 \to o_4$, labelling the
occurrences of the base type to match the moves in the diagram. There
are two phenomena at play here: firstly, the \emph{dynamic causal
links} $\imc$ are forgotten, leaving only the underlying configuration.
Secondly, the \emph{copy indices} are forgotten by taking the symmetry
class. Note that by doing only the first, one may collapse not merely
to the relational model, but to its categorification, generalized
species of structure \cite{DBLP:conf/lics/ClairambaultOP23}.

\subsection{Composition} \label{subsec:pres_comp}
Hopefully, the above conveys the idea of the
relational collapse. There are however a few conceptual subtleties that
have to do with preservation of composition.

For this section, fix two strategies $\sigma : A \vdash B$ and $\tau :
B \vdash C$ in $\Dsinn$.

\subsubsection{Oplax preservation} We start with the easy inclusion,
namely:
\[
\coll{\tau \odot \sigma} 
\quad \subseteq \quad 
\coll{\tau} \circ \coll{\sigma}\,.
\]

Consider $(\x_A, \x_C) \in \coll{\tau \odot \sigma}$. From the
definition, this means that there is some $x^{\tau\odot \sigma} \in
\confp{\tau \odot \sigma}$ such that $x^{\tau \odot \sigma}_A \in \x_A$
and $x^{\tau \odot \sigma}_C \in \x_C$. By Proposition
\ref{prop:char_comp}, $+$-covered configurations of $\tau \odot \sigma$
are in one-to-one correspondence with pairs $(x^\sigma, x^\tau)$ of
causally compatible $x^\sigma \in \confp{\sigma}$ and $x^\tau \in
\confp{\tau}$. Here, \emph{causally compatible} means that $x^\sigma_B
= x^\tau_B$, and that the induced synchronization of $x^\sigma$ and
$x^\tau$ causes no deadlock -- recall that we then write $x^{\tau \odot
\sigma} = x^\tau \odot x^\sigma \in \confp{\tau \odot \sigma}$. 

To show $(\x_A, \x_C) \in \coll{\tau} \circ \coll{\sigma}$, we must
exhibit $\x_B \in \coll{B}$ such that $(\x_A, \x_B) \in \coll{\sigma}$
and $(\x_B, \x_C) \in \coll{C}$. This seems simple: we have
$x^\sigma_B = x^\tau_B$, so we may simply take $\x_B$ their symmetry
class. We must however ensure that $\kappa_A(\x_B) = 0$ -- and this
follows easily from $\kappa_A(\x_A) = \kappa_C(\x_C) = 0$ using the
fact that $\sigma$ and $\tau$ are winning \cite[Lemma 10.4.6]{hdr}.

\subsubsection{Lax preservation} \label{subsubsec:lax_rel}
We now focus on the other inclusion:
\begin{eqnarray}
\coll{\tau} \circ \coll{\sigma} 
&\subseteq&
\coll{\tau \odot \sigma}\,.
\label{eq:lax_rel}
\end{eqnarray}

\noindent
In this direction, the situation is more subtle. Consider $(\x_A,
\x_B) \in \coll{\sigma}$ and $(\x_B, \x_C) \in \coll{\tau}$. By
definition, those are witnessed by $x^\sigma \in \confp{\sigma}$ and
$x^\tau \in \confp{\tau}$ such that 
\[
x^\sigma_A \in \x_A,\qquad
x^\sigma_B \in \x_B,\qquad
x^\tau_B \in \x_B,\qquad
x^\tau_C \in \x_C
\]
and ideally, we would like to form $x^\tau \odot x^\sigma \in
\confp{\tau \odot \sigma}$ using Proposition \ref{prop:char_comp}. But
there are two issues: firstly, we may not have $x^\sigma_B = x^\tau_B$,
in general all we have is $x^\sigma_B, x^\tau_B \in \x_B$ so that there
must exists some unspecified symmetry $\theta : x^\sigma_B \sym_B
x^\tau_B$. Secondly, even if we had $x^\sigma_B = x^\tau_B$, it is not
clear why the induced synchronization would be deadlock-free. 

Fortunately, it is a general fact that \emph{sequential innocent
strategies cannot deadlock}:

\begin{lem}\label{lem:deadlock_free}
Consider $A, B, C$ $-$-boards, $\sigma : A \vdash B$ and $\tau : B
\vdash C$ deterministic sequential innocent strategies, $x^\sigma
\in \conf{\sigma}$ and $x^\tau \in \conf{\tau}$ with a symmetry $\theta
: x^\sigma_B \sym_B x^\tau_B$.\\ 
\indent
Then, the composite bijection
\[
\varphi
~~
:
~~
x^\sigma \parallel x^\tau_C 
~~
\stackrel{\pr_\sigma \parallel x^\tau_C}{\simeq} 
~~
x^\sigma_A \parallel x^\sigma_B \parallel x^\tau_C 
~~
\stackrel{x^\sigma_A \parallel \theta \parallel x^\tau_C}{\simeq}
~~
x^\sigma_A \parallel x^\tau_B \parallel x^\tau_C
~~
\stackrel{x^\sigma_A \parallel \pr_\tau^{-1}}{\simeq}
~~
x^\sigma_A \parallel x^\tau\,,
\]
is secured, in the sense of Proposition \ref{prop:sync_sym}.
\end{lem}

\noindent
This \emph{deadlock-free lemma} was first proved in the context of
\cite{lics15} -- the reader is rather directed to the more recent and
detailed presentation in \cite[Lemma 10.4.8]{hdr}, where it is
proved with the more general hypothesis that $\sigma$ and $\tau$ should
be \emph{visible}. This crucial lemma bridges the main conceptual
difference between game semantics and relational semantics: the former
is sensitive to \emph{deadlocks}, whereas the latter is not.

Applying Lemma \ref{lem:deadlock_free} to the data at hand, the
obtained securedness hypothesis lets us \emph{reindex} $x^\sigma$ and
$x^\tau$ by Proposition \ref{prop:sync_sym} to obtain $y^\tau \odot
y^\sigma \in \confp{\tau \odot \sigma}$ such that $y^\sigma_A \sym
x^\sigma_A$ and $y^\tau_C \sym_C x^\tau_C$, so that $y^\tau \odot
y^\sigma$ witnesses $(\x_A, \x_C) \in \coll{\tau \odot \sigma}$ as
required.

\subsection{Further structure} \label{subsec:pres_further}
From the above, $\coll{\tau \odot
\sigma} = \coll{\tau} \circ \coll{\sigma}$. From Proposition
\ref{prop:confp_cc}, it is immediate that $\coll{\cc_A}$ is the
identity relation on $\coll{A}$. 
All other constructions on strategies are preserved in the appropriate
sense of a relative Seely functor, with respect to the isomorphisms in
$\Rel$ induced with the bijections of Figure \ref{fig:str_coll_bij}; as
follows via routine verifications from Propositions
\ref{prop:def_tensor}, \ref{prop:def_pairing}, \ref{prop:def_curry} and
\ref{prop:def_bang}. Altogether:

\begin{thm}
The above provide the components for a relative Seely functor:
\[
\coll{-} : \Dsinn \to \Rel\,.
\]
\end{thm}

\noindent
See \cite[Corollary 10.4.15]{hdr} for more details. It follows in
particular that we also get
\[
\colloc{-} : \Dsinn_\oc \to \Rel_\oc
\]
a cartesian closed functor between the induced cartesian closed
categories, so that:

\begin{cor} \label{cor:mainrelcoll}
Consider $\Gamma \vdash M : A$ a simply-typed $\lambda$-term.

Then, the following diagram commutes in $\Rel$:
\[
\xymatrix@C=60pt{
\oc \intr{\Gamma}_{\Dsinn_\oc}
	\ar[r]^{\coll{\intr{M}_{\Dsinn_\oc}}}
	\ar[d]_{\oc s_\Gamma}&
\intr{A}_{\Dsinn_\oc}
	\ar[d]^{s_A}\\
\oc \intr{\Gamma}_{\Rel_{\oc}}
	\ar[r]_{\intr{M}_{\Rel_\oc}}&
\intr{A}_{\Rel_\oc}
}
\]
\end{cor}

\subsection{Relational Collapse in Colors.}\label{subsec:colors} The
above theorem holds with respect to the relational interpretation of
simple types fixed in Section~\ref{subsec:rel_intro}, which specified
in particular that $\intr{o} = \{\star\}$ some singleton set. This
comes into play in the existence of the unique bijection between
singleton sets $s_o : \intr{o}_\Rel \bij \coll{\intr{o}_\Dsinn}$. Here,
we show how Corollary \ref{cor:mainrelcoll} must be adapted if we
instead have $\intr{o}_{\Rel} = \C$ some arbitrary set of
\emph{colors}. For disambiguation, from now on we shall explicitly
specify the interpretation of the base type in the relational
interpretation with $\intr{-}_\Rel^\C$.

Though the relational interpretation of types changes, the
interpretation of types as games remains the same; thus the connection
between games and positions must be adjusted to a connection between
games and positions \emph{in colors}. This mechanism is new in the
context of concurrent games (it does not appear in earlier published
works involving the relational collapse of concurrent games), but a
similar idea already appears in Tsukada and Ong's account of the
relationship between game semantics and the relational model
\cite{DBLP:conf/lics/TsukadaO16}.   

\subsubsection{Positions in colors} We start by adjusting
configurations and positions: 

\begin{defi}
Consider $A$ a board, and $x \in \conf{A}$ a configuration.

A \textbf{$\C$-coloring} (or just coloring, when $\C$ is clear from the
context) of $x$ is a function $\lambda : x \to \C$. We write
$\col(x)$ for the set of $\C$-colorings of $x$, and $\cconf{A}{\C}$
for the set of \textbf{configurations in colors}, \emph{i.e.} pairs
$(x, \lambda)$ of a configuration equipped with a coloring. 
\end{defi}

\noindent
Though a configuration with colors is a pair $(x, \lambda) \in
\cconf{A}{\C}$, we shall sometimes just write $x \in \cconf{A}{\C}$ and
refer to the coloring as $\lambda_x \in \col(x)$. 

If $A$ and $B$ are boards, $x_A \in \conf{A}$ and $x_B \in \conf{B}$,
then every pair of colorings $\lambda_A \in \col(x_A)$ and $\lambda_B
\in \col(x_B)$ induce a coloring $\lambda_A \tensor \lambda_B \in
\col(x_A \tensor x_B)$ simply by co-pairing, informing a bijection
$\col(x_A \tensor x_B) \bij \col(x_A) \times \col(x_B)$. Together with
the bijection $- \tensor - : \conf{A} \times \conf{B} \bij \conf{A
\tensor B}$, this yields a bijection $- \tensor - : \cconf{A}{\C}
\times \cconf{B}{\C} \bij \cconf{A\tensor B}{\C}$. We have similar
bijections for $\vdash, \with, \lin$ and $\oc$, defined in the obvious
way.

\begin{defi}
Two configurations with colors $x, y \in \cconf{A}{\C}$ are
\textbf{symmetric} if there is some $\theta : x \sym_A y$ that
preserves colors. This is an equivalence relation, and a
\textbf{position with colors} is a symmetry class of configurations
with colors of null payoff, written $\x \in \ccoll{A}{\C}$.
\end{defi}

\noindent
All the above bijections between configurations in colors are
compatible with symmetry, and yield the bijections of Figure
\ref{fig:col_str_coll_bij}. For the base type $o$, a coloring consists
simply in the choice of a color for the unique move $\qu$, so that we
indeed have $\ccoll{\intr{o}_\Dsinn}{\C} \bij \intr{o}^{\C}_{\Rel} =
\C$. Altogether, the bijection \eqref{eq:bij_posrel} extends in the
presence of colors to give
\[
s_A \quad:\quad \intr{A}^\C_{\Rel} ~\bij~ \ccoll{A}{\C}
\]
for every simple type $A$, extending our earlier situation in the
presence of colors.

\subsubsection{Experiments} Next, we must associate to any strategy a
set of positions in colors. This rests on the following notion of
\emph{experiment}, the name being inspired from the notion with the
same name in proof nets \cite{DBLP:journals/tcs/Girard87}. Intuitively,
an experiment is a coloring of a configuration of a strategy; except
that the axiom links must be preserved: a Player move must have the
same color as that given to its (necessarily unique) causal
predecessor.

\begin{defi}
Consider $A$ a board, $\sigma : A$ a strategy, and $x \in
\confp{\sigma}$.

A ($\C$-)\textbf{coloring} of $x$ is a function $\lambda :
x \to \C$ subject to:
\[
\begin{array}{rl}
\text{\emph{monochrome:}} &
\text{for all $s_1^- \imc_\sigma s_2^+$, we have
$\lambda(s_1^-) = \lambda(s_2^+)$.}
\end{array}
\]

\noindent
As above, we write $\col(x)$ for the set of colorings of
$x \in \confp{\sigma}$. 
An \textbf{experiment} on $\sigma$ is $x \in \confp{\sigma}$
together with a coloring, and we write 
$\cconfp{\sigma}{\C}$ for the set of experiments.
\end{defi}

\noindent
As for configurations in colors, experiments are pairs $(x, \lambda) \in \cconf{\sigma}{\C}$; nevertheless we shall often just write $x \in \cconf{\sigma}{\C}$ and refer to the coloring as $\lambda_{x} \in \col(x)$.

Given $x^\sigma \in \confp{\sigma}$, recall that the display map
$\pr_\sigma : \sigma \to A$ induces a bijection $\pr_\sigma : x^\sigma
\bij x^\sigma_A$. We use this bijection to transport any coloring
$\lambda^\sigma \in \col(x^\sigma)$ to $\lambda^\sigma_A =
\lambda^\sigma \circ \pr_\sigma^{-1} \in \col(x^\sigma_A)$. Likewise,
if $x^\sigma \in \cconf{\sigma}{\C}$, we write $x^\sigma_A \in
\cconf{A}{\C}$ for the corresponding configuration in colors.

\subsubsection{The colorful collapse}\label{subsubsec:color_coll}
With this we extend the collapse to the colored
setting:
\begin{eqnarray}
\ccoll{\sigma}{\C}
&=& 
\{ \x_A \vdash \x_B \in \ccoll{A \vdash B}{\C} \mid \exists x^\sigma
\in \cconfp{\sigma}{\C},~ x^\sigma_A \in \x_A,~x^\sigma_B \in \x_B
\}\,,
\label{eq:color_coll}
\end{eqnarray}
which we now aim to show still yields a relative Seely functor. 
As perhaps expected, the
noteworthy cases are preservation of composition, and of copycat.

We focus first on composition; fix $\sigma : A \vdash B$ and $\tau : B
\vdash C$, along with $x^\sigma \in \confp{\sigma}$ and $x^\tau \in
\confp{\tau}$ causally compatible. Given colorings $\lambda^\sigma \in
\col(x^\sigma)$ and $\lambda^\tau \in \col(x^\tau)$, we say that they
are \textbf{matching} if $\lambda^\sigma_B = \lambda^\tau_B$.
Preservation of composition rests on:

\begin{lem}\label{lem:col_pres_comp}
Fix $\sigma : A \vdash B, \tau : B \vdash C$ with $x^\sigma \in
\confp{\sigma}$ and $\confp{\tau}$ causally compatible. 

Then, there is a bijection:
\[
- \odot - 
\quad:\quad
\{(\lambda^\sigma, \lambda^\tau) \in \col(x^\sigma) \times \col(x^\tau)
\mid \text{matching} \}
\quad\bij\quad
\col(x^\tau \odot x^\sigma)\,,
\]
satisfying $(\lambda^\tau \odot \lambda^\sigma)_A = \lambda^\sigma_A$
and $(\lambda^\tau \odot \lambda^\sigma)_C = \lambda^\tau_C$.
\end{lem}
\begin{proof}
From left to right, fix $x^\tau \odot x^\sigma \in \confp{\tau
\odot \sigma}$ with matching colorings $\lambda^\sigma$ and
$\lambda^\tau$ -- as they are matching, they induce a coloring
$\lambda^\tau \inter \lambda^\sigma$ on $x^\tau \inter x^\sigma$,
which restricts to a coloring $\lambda^\tau \odot \lambda^\sigma$ for
$x^\tau\odot x^\sigma$; but we must check that it is \emph{monochrome}.
The main observation is that every immediate causality 
$p_1^- \imc_{\tau \odot \sigma} p_2^+$ in $x^\tau \odot x^\sigma$
arises from a sequence 
\[
p_1^- \imc_{\tau \inter \sigma}
q_1   \imc_{\tau \inter \sigma}
\ldots\imc_{\tau \inter \sigma}
q_n   \imc_{\tau \inter \sigma}
p_2^+
\]
where all the $q_i$s occur in $B$. In that case, all those causal links
arise from immediate causal links in $\sigma$ and $\tau$ \cite[Lemma
6.2.14]{hdr}, so the coloring $\lambda^\tau \inter \lambda^\sigma$ is
preserved along the chain and $(\lambda^\tau \inter
\lambda^\sigma)(p_1^-) = (\lambda^\tau \inter
\lambda^\sigma)(p_2^+)$, hence $(\lambda^\tau \odot
\lambda^\sigma)(p_1^-) = (\lambda^\tau \odot
\lambda^\sigma)(p_2^+)$ as well.

From right to left, fix a coloring $\lambda$ on $x^\tau \odot
x^\sigma$. The main observation we make is that every $q \in x^\tau
\inter x^\sigma$ is uniquely sandwiched between visible events. More
precisely, there are \emph{unique} $p_1^-, p_2^+$ visible, necessarily
with the polarities indicated, and a unique causal path  
\[
p_1^- ~=~ q_0 
~\imc_{\tau \inter \sigma}~
\ldots
~\imc_{\tau \inter \sigma}~
q_{n+1} ~=~ p_2^+
\]
such that $q$ appears in the $q_i$s, and every $q_1, \ldots, q_n$
occurs in $B$ (is \emph{hidden}) -- this follows easily from Lemmas
6.1.16 and 6.2.15 from \cite{hdr}, remarking that as $\sigma$ and
$\tau$ are \emph{forestial}, so is $\tau \inter \sigma$. This lets us
complete $\lambda$ to the interaction $x^\tau \inter x^\sigma$ by
setting $\lambda(q) = \lambda(p_1^-) = \lambda(p_2^+)$; this projects
to matching $\lambda^\sigma$ on $x^\sigma$ and $\lambda^\tau$ on
$x^\tau$. These projected colorings are \emph{monochrome}, because if
\emph{e.g.} $(p_1)^-_\sigma \imc_\sigma (p_2)_\sigma^+$, then a
straightforward case analysis shows that $p_1$ and $p_2$ must appear in
the same sandwich as above, and therefore receive the same color.

That these operations are inverses is an easy verification.
\end{proof}

\noindent
This immediately entails that the ``colorful collapse'' preserves
composition; we shall see the proof just later. For preservation of
copycat, the corresponding main lemma is:

\begin{lem}\label{lem:col_pres_cc}
For any $-$-board $A$ and any $x \in \conf{A}$, there is a bijection
\[
\cc_{-} 
\quad:\quad
\col(x)
\quad\bij\quad
\col(\cc_x)
\]
such that for all $\lambda \in \col(x)$, we have $\cc_\lambda =
[\lambda, \lambda]$ the co-pairing.
\end{lem}
\begin{proof}
Immediate causal links from negative moves to positive moves in $\cc_x$
have shape 
\[
(i, a) \quad \imc_{\cc_x} \quad (j,a)
\]
where $i \neq j$ (see \cite[Lemma 6.4.3]{hdr}). Thus, given $\lambda_A
\in \col(x)$, then the co-pairing $\lambda = [\lambda_A, \lambda_A] :
\cc_x \to \C$ satisfies \emph{monochrome}, therefore giving a valid
coloring on $\cc_x$. Reciprocally, for all $a \in x$, we always have
either $(1, a) \imc_{\cc_x} (2, a)$ or $(2, a) \imc_{\cc_x} (1, a)$
depending on the polarity of $a$ (see also \cite[Lemma 6.4.3]{hdr}).
Thus, if $\lambda \in \col(\cc_x)$, we must have $\lambda(i,a) =
\lambda(j,a)$ by \emph{monochrome}, so that $\lambda = [\lambda_A,
\lambda_A]$ for some $\lambda_A \in \col(x)$. 
\end{proof}

\noindent
Putting these two lemmas together, we have:

\begin{prop}\label{prop:color_rel_fun}
The colorful collapse defined above yields a functor:
\[
\ccoll{-}{\C} \quad:\quad 
\Dsinn 
\quad\to\quad
\Rel
\]
with a bijection $s_o : \ccoll{\intr{o}_{\Dsinn}}{\C} \bij \C$.
\end{prop}
\begin{proof}
This proposition builds on the functoriality of the earlier (colorless)
collapse. In this proof, we leave implicit the material already covered
in Sections~\ref{subsec:pres_comp} and \ref{subsec:pres_further}, and
focus on the preservation of colorings.

For preservation of copycat, consider $(\x_A, \y_A) \in
\ccoll{\cc_A}{\C}$. By definition, it is witnessed by an experiment in
$\cc_A$, \emph{i.e.} a pair $(\cc_x, \lambda) \in \cconfp{\cc_x}{\C}$
such that $(\cc_x, \lambda)_l \in \x_A$ and $(\cc_x, \lambda)_r \in
\y_A$. By Lemma \ref{lem:col_pres_cc}, $\lambda = [\lambda_A,
\lambda_A]$ for $\lambda_A \in \col(x)$, so that $(\cc_x, \lambda)_l =
(x, \lambda_A) \in \cconf{A}{\C}$ and $(\cc_x, \lambda)_r = (x,
\lambda_A)$ as well. Hence we have $(x, \lambda_A)$ in both symmetry
classes $\x_A$ and $\y_A$, which must therefore be equal.
Reciprocally, given $\x_A \in \ccoll{A}{\C}$, taking $(x, \lambda) \in
\cconf{A}{\C}$ a representative, it is immediate by Lemma
\ref{lem:col_pres_cc} again that $(\cc_x, \cc_\lambda) \in
\cconfp{\cc_A}{\C}$ provides an experiment which projects on the left
and right to $(x, \lambda) \in \x_A$ as required. 

For preservation of composition, consider first $(\x_A, \x_C) \in
\ccoll{\tau \odot \sigma}{\C}$. By definition, it is witnessed by an
experiment in $\tau \odot \sigma$, which by Lemma
\ref{lem:col_pres_comp} must have the form $(x^\tau \odot x^\sigma,
\lambda^\tau \odot \lambda^\sigma)$, where $(x^\sigma, \lambda^\sigma)
\in \cconfp{\sigma}{\C}$ and $(x^\tau, \lambda^\tau) \in
\cconfp{\tau}{\C}$ are compatible experiments, \emph{i.e.} $x^\sigma,
x^\tau$ are causally compatible and $\lambda^\sigma, \lambda^\tau$ are
matching. Writing $x_B = x^\sigma_B = x^\tau_B$ and $\lambda_B =
\lambda^\sigma_B = \lambda^\tau_B$, we form $(x_B, \lambda_B) \in
\cconf{B}{\C}$. Its symmetry class $\x_B$ is then in $\ccoll{B}{\C}$,
and $(\x_A, \x_B) \in \ccoll{\sigma}{\C}$ is witnessed by the
experiment $(x^\sigma, \lambda^\sigma)$ while $(\x_B, \x_C) \in
\ccoll{\tau}{\C}$ is witnessed by the experiment $(x^\tau,
\lambda^\tau)$. Hence $(\x_A, \x_C) \in \ccoll{\tau}{\C} \circ
\ccoll{\sigma}{\C}$ as required.

Reciprocally consider $(\x_A, \x_C) \in \ccoll{\tau}{\C} \circ
\ccoll{\sigma}{\C}$, so there is $\x_B \in
\ccoll{B}{\C}$ such that $(\x_A, \x_B) \in \ccoll{\sigma}{\C}$ and
$(\x_B, \x_C) \in \ccoll{\tau}{\C}$, respectively witnessed by
$(x^\sigma, \lambda^\sigma) \in \cconf{\sigma}{\C}$ and
$(x^\tau, \lambda^\tau) \in \cconf{\tau}{\C}$. Now, we must have
$(x^\sigma_B, \lambda^\sigma_B) \in \x_B$ and $(x^\tau_B,
\lambda^\tau_B) \in \x_B$, hence there must be $\theta_B :
x^\sigma_B \sym_B x^\tau_B$ compatible with the coloring, \emph{i.e.}
$\lambda^\tau_B \circ \theta_B = \lambda^\sigma_B$. Now, by Proposition
\ref{prop:sync_sym} (along with Lemma \ref{lem:deadlock_free}), there
are $y^\tau \odot y^\sigma \in \confp{\tau \odot \sigma}$ and
$\varphi^\sigma : x^\sigma \sym_\sigma y^\sigma$ and $\varphi^\tau :
x^\tau \sym_\tau y^\tau$ such that $\varphi^\tau_B \circ \theta_B =
\varphi^\sigma_B$. Now the idea is to equip $y^\sigma$ and $y^\tau$
with adequate colorings, to turn them into experiments: we simply set
$\mu^\sigma = \lambda^\sigma \circ (\varphi^\sigma)^{-1}$ and $\mu^\tau
= \lambda^\tau \circ (\varphi^\tau)^{-1}$. As $\varphi^\sigma$ and
$\varphi^\tau$ are order-isomorphisms preserving polarity, it is clear
that those are \emph{monochrome} so that $(y^\sigma, \mu^\sigma) \in
\cconfp{\sigma}{\C}$ and $(y^\tau, \mu^\tau)\in \cconfp{\tau}{\C}$ are
valid experiments. Additionally, by construction,
\[
\mu^\sigma_B = \lambda^\sigma_B \circ (\varphi^\sigma_B)^{-1} =
\lambda^\tau_B \circ \theta_B \circ (\varphi^\sigma_B)^{-1} =
\lambda^\tau_B \circ (\varphi^\tau_B)^{-1} = \mu^\tau_B
\]
ensuring that $\mu^\sigma$ and $\mu^\tau$ are matching, so that we can
form the experiment $\mathcal{E} = (y^\tau \odot y^\sigma, \mu^\tau
\odot \mu^\sigma)$. Finally, this experiment indeed witnesses that
$(\x_A, \x_C) \in \ccoll{\tau\odot \sigma}{\C}$ since 
\[
\varphi^\sigma_A : (x^\sigma, \lambda^\sigma)_A \sym_A
(y^\sigma, \mu^\sigma)_A = \mathcal{E}_A\,, 
\qquad
\varphi^\tau_C : (x^\tau, \lambda^\tau)_C \sym_C (y^\tau, \mu^\tau)_C =
\mathcal{E}_C
\]
so that $\mathcal{E}_A \in \x_A$ and $\mathcal{E}_C \in \x_C$ as
required.
\end{proof}

\subsubsection{Further structure.} It remains to extend $\ccoll{-}{\C}$
to a relative Seely functor. All the structural isomorphisms for that
are the same as in the colorless case, colored as for copycat in the
obvious way. Preservation of tensor and promotion are handled by two
lemmas:

\begin{lem}
Fix $\sigma : A \vdash B$, $\tau : C \vdash D$ strategies, and
$x^\sigma \in \confp{\sigma}, x^\tau \in \confp{\tau}$.

Then, there is a bijection:
\[
- \tensor - 
\quad:\quad 
\col(x^\sigma) \times \col(x^\tau) 
\quad\bij\quad
\col(x^\sigma \tensor x^\tau)
\]
such that $(\lambda^\sigma \tensor \lambda^\tau)_{A\tensor C} =
\lambda^\sigma_A \tensor \lambda^\tau_C$ and $(\lambda^\sigma \tensor
\lambda^\tau)_{B \tensor D} = \lambda^\sigma_B \tensor \lambda^\tau_D$.
\end{lem}

\noindent
We omit the direct proof. Because of this lemma, it
is straightforward that $\ccoll{\sigma \tensor \tau}{\C}$ and
$\ccoll{\sigma}{\C} \tensor \ccoll{\tau}{\C}$ coincide up to
$s^\tensor$, as required by relative Seely functors.

Finally, we need a corresponding observation for the \emph{promotion}:

\begin{lem}
Fix $\sigma : \oc S \vdash A$, and $\famc{x^{\sigma,i}\mid i \in I} \in
\confp{\sigma^\dagger}$.
Then, there is a bijection:
\[
\famc{-}
\quad:\quad
\prod_{i\in I} \col(x^{\sigma,i})
\quad\bij\quad
\col(\famc{x^{\sigma, i} \mid i \in I})
\]
such that for all $(\lambda^{\sigma, i})_{i\in I} \in \prod_{i\in I}
\col(x^{\sigma,i})$, writing $\lambda^{\sigma, i}_{\oc S} =
\famc{\lambda^{\sigma, i}_j \mid j \in J_i}$, we have
\begin{eqnarray*}
\famc{\lambda^{\sigma, i} \mid i \in I}_{\oc S} &=& 
\famc{\lambda^{\sigma, i}_{j} \mid \tuple{i,j} \in \Sigma_{i\in I} J_i}\,,\\
\famc{\lambda^{\sigma, i} \mid i \in I}_{A} &=&
\famc{\lambda^{\sigma,i}_A \mid i \in I}\,.
\end{eqnarray*}
\end{lem}

\noindent
Again there is no subtlety here besides the heavy notation, the
coloring $\famc{\lambda^{\sigma, i} \mid i \in I}$ is defined via the
obvious co-pairing, and verifications are direct. Via this lemma, it
follows that $\ccoll{\sigma^{\dagger}}{\C}$ and
$\ccoll{\sigma}{\C}^{\dagger}$ coincide up to $s^\oc$, as required by
relative Seely functors. 

Together with a few additional verifications for structural
isomorphisms, this yields:

\begin{thm}
The above provide the components for a relative Seely functor:
\[
\ccoll{-}{\C} : \Dsinn \to \Rel\,.
\]
\end{thm}

\noindent
Again, it follows in particular that we also get
$\ccolloc{-}{\C} : \Dsinn_\oc \to \Rel_\oc$
a cartesian closed functor between the induced cartesian closed
categories, so that:

\begin{cor} \label{cor:mainrelcollcolor}
Consider $\Gamma \vdash M : A$ a simply-typed $\lambda$-term.

Then, the following diagram commutes in $\Rel$:
\[
\xymatrix@C=60pt{
\oc \intr{\Gamma}_{\Dsinn_\oc}
        \ar[r]^{\ccoll{\intr{M}_{\Dsinn_\oc}}{\C}}
        \ar[d]_{\oc s_\Gamma}&
\intr{A}_{\Dsinn_\oc}
        \ar[d]^{s_A}\\
\oc \intr{\Gamma}_{\Rel_{\oc}}^{\C}
        \ar[r]_{\intr{M}_{\Rel_\oc}^{\C}}&
\intr{A}_{\Rel_\oc}^{\C}
}
\]
\end{cor}

\noindent
This is the same statement as in the colorless case, except that this
fact we have fixed an arbitrary set $\C$ for the interpretation of the
base type. Note however that neither $\Dsinn$ nor its interpretation of
the base type has changed: it is only the collapse that we changed.

\section{From Games to the Linear Scott Model}
\label{sec:games_scottl}

Now, we have finally finished setting up the scene, and we can finally
carry on with the journey announced in the introduction. 
So far, we have introduced the relational model $\Rel$, the (relative)
Seely category $\Dsinn$ of thin concurrent games, and the relational
collapse 
\[
\coll{-} : \Dsinn \to \Rel
\]
that preserves the interpretation of the simply-typed
$\lambda$-calculus. This collapse is \emph{quantitative}: $\Rel$ still
records the multiplicity of resource consumption. In the rest of this
paper, we set to construct a corresponding \emph{qualitative} collapse,
targeting the \emph{linear Scott model}. 

In this section, we first recall the linear Scott model. Then, we 
construct \emph{cartesian morphisms}, these morphisms between
configurations that allow contraction and weakening of resources. We
establish a few important properties of cartesian morphisms. Then,
we will show that from a board equipped with cartesian morphisms we are
able to construct a preorder, in a way compatible with the
interpretation of types in the linear Scott model.

\subsection{The Linear Scott Model}

We first recall the linear Scott model, following
\cite{DBLP:journals/tcs/Ehrhard12}. 

\subsubsection{The basic category}

The linear Scott model can be presented in two different ways: as a
category of (linear) functions between certain complete lattices, or as
a category of relations between certain preordered sets. In this paper
we pick the latter presentation, because it is more homogeneous with the
relational model and facilitates the relationship; but we shall
also include a discussion about the domain-theoretic
presentation.

We work with the following category:

\begin{defi}
$\ScottL$ has: \emph{(1)} objects, preorders $(\ev{A},
\leq_A)$; \emph{(2)} morphisms from $A$ to $B$, relations $\alpha
\subseteq \ev{A} \times \ev{B}$ which are \emph{down-closed}: if $(a,
b) \in \alpha$ and $a \leq_A a', b' \leq_B b$, then $(a', b') \in
\alpha$. \emph{Composition} is relational composition, and 
identities $\id_A = \{(a, a') \mid a' \leq_A a\}$.  
\end{defi}

\noindent
By a slight abuse of notation, we often write only $A$ for the support
set. 
If $A$ is a preorder, we write $A^\op$ for the \textbf{opposite}
preorder, with same support but reversed preorder. The \textbf{product}
preorder has $(a, b) \leq_{A\times B} (a', b')$ iff $a \leq_A a'$ and
$b \leq_B b'$.

As explained in the introduction, $a \leq_A a'$ expresses that
the resources in $a$ can be ``contracted'' into the resources in $a'$;
and the resources in $a'$ can be also ``weakened'' by not appearing
in $a$. For instance, we shall see that if $A$ is discrete, then
\[
[a, a] \leq_{\oc A} [a, b]
\]
where both copies of $a$ are contracted into $a$ and $b$ is weakened.
However, that intuition is incomplete as the preorder is crucially
reversed in contravariant position. 

\subsubsection{Seely category} Firstly, $\ScottL$ has a symmetric
monoidal structure: if $A$ and $B$ are preorders, then $A \tensor B = A
\times B$. The monoidal unit is $1 = (\{\star\}, =)$, and the
functorial action of $\tensor$ is as in the relational model, while
structural morphisms are the \emph{down-closure} of their relational
counterparts: for instance, for associativity we have
\[
\begin{array}{rcrclcl}
\alpha^{\ScottL}_{A, B, C} &:& (A \tensor B) \tensor C &\iso& A \tensor
(B \tensor C)\\
&=& \{(((a, b), c)&,&(a', (b',c'))) & \mid & a' \leq_A a,~b'\leq_B
b,~c'\leq_C c\}
\end{array}
\]
and likewise for the other components
\cite{DBLP:journals/tcs/Ehrhard12}. Likewise, the cartesian structure
of $\Rel$ adapts to $\ScottL$ transparently, with $A \with B = A + B$
the disjoint union of the two preorders; $\top = (\emptyset,
\emptyset)$ is terminal. The pairing operation is the
same as in $\Rel$, and projections in $\ScottL$ are obtained as the
down-closure of those in $\Rel$. We additionally set $A \lin B = A^\op
\times B$ -- again, currying is as in $\Rel$, and evaluation in
$\ScottL$ is the down-closure of evaluation in $\Rel$. Altogether, this
makes $\ScottL$ a cartesian symmetric monoidal closed category. 

Now, we get to the more critical definition of the exponential. Given
$A$, we set
\[
\ev{\oc A} = \Mf(\ev{A}),\qquad
\mu \leq_{\oc A} \nu ~\Leftrightarrow~
\forall a \in \supp(\mu),~\exists a'\in \supp(\nu),~a \leq_A a'\,,
\]
where the \textbf{support} $\supp(\mu)$ of a finite multiset $\mu \in
\Mf(X)$ is simply the set of $x \in X$ with non-zero multiplicity.
Again, this preorder is built on the same set as the exponential for
the plain relational model. We specify the additional components of the
exponential with:
\[
\begin{array}{rcrcl}
\oc \alpha &=& \{(\mu, \nu) \in \oc A \times \oc B \mid \forall b \in
\supp(\nu),~\exists a \in \supp(\mu),~(a, b) \in \alpha\}\\
\der_A &=& \{(\mu, a) \mid \exists a' \in \supp(\mu),~a \leq_A a'\}\\
\dig_A &=& \{(\mu, [\mu_1, \dots, \mu_n]) \mid \forall i,~\mu_i
\leq_{\oc A} \mu\}\\
\mon_{A_1, A_2} &=& \{((\mu_1, \mu_2), \nu) \mid \forall (i, a') \in
\supp(\nu),~ \exists a \in \supp(\mu_i),~a'\leq_{A_i} a\} 
\end{array}
\]
and $\mon_1 : 1 \iso \oc \top$ the obvious isomorphism. Altogether, we
have: 

\begin{thm}
The above components make $\ScottL$ a Seely category.
\end{thm}

\noindent
Though the exponential is built from finite multisets, this model does
not actually record quantitative information as morphisms are
down-closed -- note that $\oc A$ is isomorphic (in $\ScottL$) to the
preorder obtained with the same definitions but built on finite sets
instead of finite multisets.
This Seely category $\ScottL$ will be the target of our qualitative
collapse.

\subsubsection{The Linear Scott Model and Scott Domains} 
\label{subsubsec:scott_domains}
Though the rest
of the paper will mostly focus on $\ScottL$ as presented above, we take
a small detour to present its relationship with the standard category
of Scott domains and continuous functions.

\paragraph{Morphisms in $\ScottL$ as functions.} If $A$ is a
preorder, we write $\D(A)$ for the set of \emph{down-closed subsets} of
$\ev{A}$; it is a complete lattice. If $a \in A$, we write $[a]_A =
\{a' \in A \mid a' \leq_A a\} \in \D(A)$; likewise if $X \subseteq
\ev{A}$, we write $[X]_A \in \D(A)$ for its down-closure.  

We shall now construct a new category, which is an alternative
presentation of $\ScottL$ in terms of functions. Its objects are the
same as $\ScottL$, \emph{i.e.} preorders. A morphism from $A$ to $B$ is
a \emph{linear map} from $\D(A)$ to $\D(B)$, \emph{i.e.} a function $f
: \D(A) \to \D(B)$ such that
\[
f(\bigcup_{i\in I} x_i) = \bigcup_{i\in I} (f(x_i))
\]
for any family $(x_i)_{i\in I}$ of $x_i \in \D(A)$ -- in particular,
$f$ is monotone and $f(\emptyset) = \emptyset$. We write $\ScottFun$
for the category with preorders as objects and linear maps.

The main observation here is that $\ScottL$ and $\ScottFun$ are
isomorphic categories:

\begin{prop}
The following constructions yield an isomorphism of categories:
\[
\begin{array}{rcrcl}
\fun_{A,B} &:& \ScottL[A, B] &\to& \ScottFun[A,B]\\
&&\alpha &\mapsto& x \in \D(A) \mapsto \alpha\,x\\\\
\tr_{A,B} &:& \ScottFun[A,B] &\to& \ScottL[A, B]\\
&& f &\mapsto & \{(a, b) \mid b \in f([a]_A)\}
\end{array}
\]
where $\alpha\,x$ refers to relational composition, and where $\tr$ is
called the \textbf{linear trace}.
\end{prop}

\paragraph{A full subcategory of Scott domains} Now, we redo the
analogous construction but for the Kleisli category $\ScottL_\oc$. We
noted above that for every preorder $A$, $\D(A)$ is a complete lattice;
in fact it is a Scott domain, whose compact elements are the finitely
generated elements of $\D(A)$, that is, those $x \in \D(A)$ such
that $x = [X]_A$ for some finite $X \subseteq \ev{A}$. Let us write 
$\ScottP$ for the category whose objects are preorders, and morphisms
from $A$ to $B$ are Scott-continuous functions from $\D(A)$ to $\D(B)$.
Writing also $\Scott$ for the usual category of Scott domains and Scott
continuous functions, this yields a full and faithful (identity on
morphisms) functor $K : \ScottP \hookrightarrow \Scott$.

Given $X \subseteq \ev{A}$ for a preorder $A$, we set $X^\oc =
\Mf(X)$. With this, we have:
\begin{prop}
The following constructions yield an isomorphism of categories:
\[
\begin{array}{rcrcl}
\Fun_{A,B} &:& \ScottL_{\oc}[A, B] &\to& \ScottP[A,B]\\
&&\alpha &\mapsto& x \in \D(A) \mapsto \alpha\,x^\oc\\\\
\Tr_{A,B} &:& \ScottP[A,B] &\to& \ScottL_{\oc}[A, B]\\
&& f &\mapsto & \{(\mu, b) \mid b \in f([\supp(\mu)]_A) \}\,.
\end{array}
\]
\end{prop}

\noindent
Altogether we get a full and faithful functor
\[
K \circ \Fun : \ScottL_\oc \to \Scott
\]
which is easily shown to preserve the cartesian closed structure; so
that the Kleisli category $\ScottL_\oc$ is indeed a category of Scott
domains and Scott-continuous functions.

\paragraph{Cartesian Morphisms} 
In Ehrhard's presentation of the
linear Scott model \cite{DBLP:journals/tcs/Ehrhard12}, each type is
interpreted as a preorder whose support set is nothing but the
standard relational interpretation of the type. Thus, we expect to
extract a preorder from a board $A$ by simply attaching to $\coll{A}$
(and, later on, to $\ccoll{A}{\C}$) an adequate preorder relation. As
argued in the introduction, this preorder relation will arise from a
notion of \emph{cartesian morphisms} between configurations,
maps allowing the contraction and weakening of resources. 

We now move back to games, aiming for the definition of cartesian
morphisms.

\subsection{Mixed Boards} 
In the introduction, we sketched cartesian morphisms
as certain forest morphisms leaving the identity of moves
(\emph{i.e.} the type component) unchanged, but that may alter
\emph{copy indices}. Unfortunately, we are not able to define such
morphisms in the setting of \emph{boards} as presented in Definition
\ref{def:board} -- we need to refine our notion of game.

\subsubsection{Arenas}
Those boards arising from the
interpretation of simple types have a specific shape -- namely,
they are themselves an \emph{expansion} of a simpler structure called
an \emph{arena}:

\begin{defi}\label{def:arena}
An \textbf{arena} comprises $A = (\ev{A}, \pol_A, \leq_A)$ with
$\ev{A}$
a countable set of \textbf{moves}, $\pol_A : \ev{A} \to \{-, +\}$ a
\textbf{polarity function}, $\leq_A$ a \textbf{causality} partial
order, such that:
\[
\begin{array}{rl}
\text{\emph{forestial:}} & \text{for all $a_1, a_2, a \in \ev{A}$, if
$a_1, a_2 \leq_A a$, then $a_1 \leq_A a_2$ or $a_2 \leq_A
a_1$,}\\
\text{\emph{well-founded:}} & \text{there is no infinite descending
$a_1 >_A a_2 >_A a_3 >_A \dots$,}\\
\text{\emph{negative:}} &\text{if $a \in \ev{A}$ is minimal for
$\leq_A$, then $\pol_A(a) = -$,}\\
\text{\emph{alternating:}} &\text{for all $a_1, a_2 \in \ev{A}$, if
$a_1
\imc_A a_2$, then $\pol_A(a_1) \neq \pol_A(a_2)$.}
\end{array}
\]

\noindent
Additionally we fix, for each move $a \in A$, a set $\Ind(a)$ which is
either $\mathbb{N}$ or $\{*\}$.
\end{defi}

\noindent
This is close to the usual notion of Hyland-Ong arenas, with a slight
change in presentation so as to remain close to boards. The new
component is the set $\Ind(a)$, which specifies which moves are
duplicable by specifying the admissible copy indices\footnote{This data
is not required for the interpretation of simple types, where all
non-initial moves are necessarily duplicable. We need this here because
we wish to phrase the collapse as a relative Seely functor, also
handling the linear structure; and hence some moves may not be
duplicable in the arena.}. 

We briefly present the main constructions on arenas. First, for the
atom, we write $\mar{o}$ for the arena with exactly one (negative) move
$\qu^-$, with $\Ind(\qu^-) = \grey{\{*\}}$: this move is not
duplicable. If $A$ is an arena, the \emph{exponential} $\oc A$ has the
same components as $A$ (we do not duplicate the moves), but we set
$\Ind_{\oc A}(a^-) = \N$ for every $a^-$ minimal: we set the initial
moves as duplicable. The \emph{parallel composition} $A \parallel B$ adapts
transparently to arenas, with the $\Ind(-)$ function inherited. Note
that any arena may be written as $A\iso~\parallel_{i\in I} A_i$ with
$A_i$ well-opened. If $A$ and $B$ are arenas with $B$ well-opened, then
$A \lin B$ is an arena (again with $\Ind$ inherited); this extends to
$B$ not well-opened with 
\[
A \lin (\parallel_{i\in I} B_i) 
\quad=\quad 
\parallel_{i\in I} A \lin B_i\,.
\]

\noindent
Altogether, this lets us interpret any simple type as an arena
with $\intr{o}_{\Ar} = \mar{o}$ and $\intr{A\to B} = \oc \intr{A}_{\Ar} \lin
\intr{B}_{\Ar}$: moves are not explicitly duplicated, but simply marked
with the admissible copy indices.
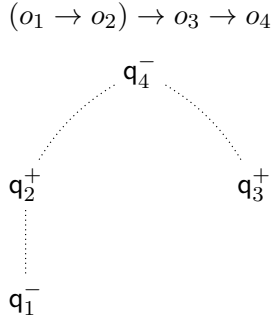
\begin{figure}
\[
(o_1 \to o_2) \to o_3 \to o_4
\]
\[
\xymatrix{
&\qu^-_4\\
\qu^+_2	\ar@{.}@/^/[ur]&&
\qu^+_3	\ar@{.}@/_/[ul]\\
\qu^-_1	\ar@{.}[u]
}
\]
\caption{Interpretation of $(o \to o) \to o \to o$ as an arena}
\label{fig:ex_arena}
\end{figure}
We show in Figure \ref{fig:ex_arena} the interpretation of the simple
type $(o \to o) \to o \to o$ as an arena -- the reader familiar with
Hyland-Ong games will recognize here the familiar arena for that type,
where each move corresponds to an atom occurrence.

\subsubsection{Mixed boards} Mixed boards are boards as in Definition
\ref{def:board}, except that moves are labeled by moves from an
underlying arena, and various conditions are satisfied to ensure the
links between the two\footnote{Mixed boards were developed in
\cite{DBLP:journals/lmcs/CastellanC24,hdr} to make explicit the link
between the interpretation of types in concurrent games and in standard
Hyland-Ong games.}. This leads to the following slightly bulky
definition -- here $\pred(-)$ denotes the unique predecessor of a
non-minimal move, exploiting that boards are \emph{forestial}:

\begin{defi}\label{def:mixed_board2}
A \textbf{mixed board} is $(A, \mar{A})$ with $A$ a $-$-board,
$\mar{A}$ an arena, with
\[
\lbl_A : \ev{A} \to \ev{\mar{A}}\,,
\qquad
\qquad
\ind_A : \ev{A} \to \mathbb{N} \uplus \{\grey{*}\}\,,
\]
with $\lbl_A$ a \textbf{label} function preserving polarities
and $\ind_A$ an \textbf{indexing} function such that $\ind_A(a) \in
\Ind(\lbl_A(a))$ for all $a \in A$, satisfying
the following additional conditions:
\[
\begin{array}{rl}
\text{\emph{rigid:}} & \text{$\lbl_A$ preserves and reflects
minimality, and
preserves $\imc$,}\\
\text{\emph{transparent:}} & \text{for any $x, y \in \conf{A}$ and
bijection $\theta : x \simeq y$,}\\
&\text{then $\theta \in \tilde{A}$ iff $\theta$ is an order-iso
preserving $\lbl_A$,}\\
\text{\emph{local conflict:}} & \text{if $a_1 \mconflict~~a_2$, we have
$\pred(a_1) = \pred(a_2)$ and $\ind(a_1) = \ind(a_2)$,}\\
\text{\emph{invariant conflict:}} & \text{if $a_1 \mconflict~~a_2$,
$\lbl(a_1) = \lbl(b_1)$, $\lbl(a_2) = \lbl(b_2)$,}\\
&\text{$\pred(b_1) = \pred(b_2)$ and $\ind(b_1) = \ind(b_2)$, then $b_1
\mconflict~~b_2$.}\\
\text{\emph{jointly injective:}} & \text{for $a_1, a_2 \in {A}$, if
$\lbl(a_1) = \lbl(a_2)$, $\ind(a_1) = \ind(a_2)$,} \\
&\text{and $\pred(a_1) = \pred(a_2)$, then $a_1 = a_2$.}\\
\text{\emph{wide:}} & \text{for any $a \in {A}$, $\mar{b} \in
{\mar{A}}$, $\lbl(a) \imc_{\mar{A}}
\mar{b}$, and $\grey{i} \in \Ind_A(\mar{b})$,}\\
&\text{there is $b \in {A}$ s.t. $\pred(b) = a$,
$\lbl(b) = \mar{b}$ and $\ind(b) = \grey{i}$,}\\ 
\text{\emph{$+$-transparent:}} & \text{for $\theta : x \sym_A y$,
$\theta \in \ptilde{A}$ iff for all $a^- \in x$, $\ind(\theta(a)) =
\ind(a)$,}\\
\text{\emph{$-$-transparent:}} & \text{for $\theta : x \sym_A y$,
$\theta \in \ntilde{A}$ iff for all $a^+ \in x$, $\ind(\theta(a)) =
\ind(a)$.}
\end{array}
\]

\noindent
A mixed board is \textbf{strict} if $A$ is strict and $\Ind(\mar{a}) =
\{\grey{*}\}$ for every $\mar{a} \in \mar{A}$ minimal.
\end{defi}

\noindent
This looks complicated, but those conditions are really simple
structural properties expressing that the board is an expansion of the
arena\footnote{It is not quite the case that the board is determined
(up to iso) by the arena, because the arena lacks the information about
which moves are related additively and multiplicatively -- it conflates
$A \with B$ with $A \tensor B$. In principle this could be added, but
this does not seem worth the even heavier definition.}.

The mixed board for the atom is $(o, \mar{o})$ with $\lbl : \qu^-
\mapsto \qu^-$ and $\ind : \qu^- \mapsto \grey{*}$ -- by abuse of
notations, we shall denote this mixed board by $o$.
The \textbf{tensor} of mixed boards has $\mar{A \tensor
B} = \mar{A} \parallel \mar{B}$, with components inherited. The
\textbf{with} of strict mixed boards $S$ and $T$ is defined likewise.
The \textbf{bang} of strict $S$ has $\mar{\oc S} = \oc \mar{S}$
(\emph{i.e.} with just $\Ind(\mar{s}) = \mathbb{N}$ for $\mar{s} \in
\mar{S}$ minimal and other components unchanged). The \textbf{linear
arrow} $A \lin S$ is extended to mixed boards in the obvious way.
However, note that only the objects of our category will be mixed
boards -- we do not give a mixed board construction for the hom-game. 

\subsection{Cartesian Morphisms} \label{subsec:carmor}
On mixed boards, we may now define cartesian morphisms.

\subsubsection{Structural maps.} From now on, fix a mixed board $A$.
Cartesian maps are motivated as
variations of symmetries that can additionally contract and weaken
resources. The condition \emph{transparent} characterises those as
order-isomorphisms which leave the label unchanged; by weakening
\emph{order-isomorphism} to simply \emph{forest morphism}, we obtain
the following notion:

\begin{defi}\label{def:str_map_game}
A \textbf{structural map} is a function $f : x \to y$, for $x, y \in
\conf{A}$,
satisfying
\[
\begin{array}{rl}
\text{\emph{min-preserving:}} & 
\text{for all $a \in \min(x)$, then $f\,a \in \min(y)$,}\\
\text{\emph{$\imc$-preserving:}} &
\text{for all $a \imc_A b$, then $f\,a \imc_A f\,b$,}\\
\text{\emph{label-preserving:}} &
\text{for all $a\in x$, $\lbl_A\,(f\,a) = \lbl_A\,a$.}
\end{array}
\]

\noindent
We write $f : x \strto y$ to indicate that $f : x \to y$ is a
structural map.
\end{defi}

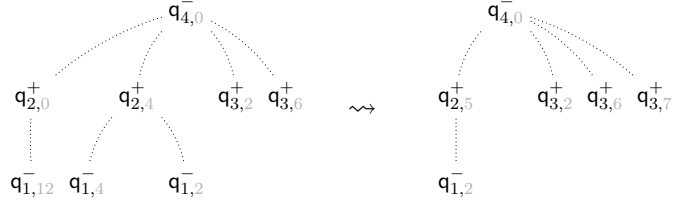
\begin{figure}
\[
\raisebox{40pt}{
\scalebox{.8}{
\xymatrix@R=20pt@C=0pt{
&&&\qu^-_{4,\grey{0}}\\
\qu^+_{2,\grey{0}}
        \ar@{.}@/^/[urrr]
&&\qu^+_{2,\grey{4}}
        \ar@{.}@/^/[ur]
&&\qu^+_{3,\grey{2}}
        \ar@{.}@/_/[ul]
&\qu^+_{3,\grey{6}}
        \ar@{.}@/_/[ull]\\
\qu^-_{1,\grey{12}} \ar@{.}@/^.1pc/[u]&
\qu^-_{1,\grey{4}} \ar@{.}@/^/[ur]
&&\qu^-_{1,\grey{2}}
        \ar@{.}@/_/[ul]
}}}
\quad\leadsto\quad
\raisebox{40pt}{
\scalebox{.8}{
\xymatrix@R=20pt@C=0pt{
&\qu^-_{4,\grey{0}}\\
\qu^+_{2,\grey{5}}
        \ar@{.}@/^/[ur]
&&\qu^+_{3,\grey{2}}
        \ar@{.}@/_/[ul]
&\qu^+_{3,\grey{6}}
        \ar@{.}@/_/[ull]
&\qu^+_{3, \grey{7}}
	\ar@{.}@/_/[ulll]\\
\qu^-_{1,\grey{2}}
        \ar@{.}[u]
}}}
\]
\caption{A structural map on $(o_1 \to o_2) \to o_3 \to o_4$}
\label{fig:ex_str_map}
\end{figure}

\noindent
In Figure \ref{fig:ex_str_map} we give an example of a structural map,
where $\qu^+_{3, \grey{2}}$ and $\qu^+_{3, \grey{6}}$ are sent to
themselves, and the other assignments are forced. Note that all copy
indices can be changed freely. The structural map contracts both
positive and negative moves, while $\qu^+_{3, \grey{7}}$ is not reached
-- it is regarded as \emph{weakened} by this structural map.

Structural maps form a category, and one can consider the associated
preorder with configurations as elements, and $x \leq y$ iff there is
some structural map $f : x \strto y$. However, this preorder is not
actually the one we need, because it is not compatible with the linear
arrow construction of preorders. Indeed, recall that in $\ScottL$, the
linear arrow was 
\[
A \lin B 
\quad = \quad
A^\op \times B
\]
contravariant on the left hand side, whereas structural maps on $A \lin
B$ are covariant on both sides. To recover the appropriate variance, we
must take polarities into account:

\begin{defi}\label{def:structural_maps}
For a structural map $f : x \strto y$, we define the conditions:
\[
\begin{array}{rl}
\text{\emph{$-$-total:}} &
\text{if $a^+ \in x$, for all $f\,a^+ \imc b^-$ in $y$, there
is $a^+ \imc c^-$ in $x$ s.t. $f\,c^- = b^-$;}\\
&\text{and for all $b^-$ minimal in $y$ there is $c^-$ in $x$ such that
$f\,c^- = b^-$.}\\ 
\text{\emph{$+$-total:}} &
\text{if $a^- \in x$, for all $f\,a^- \imc b^+$,
there is $a^- \imc c^+$ in $x$ s.t. $f\,c^+ = b^+$,}\\
\text{\emph{$-$-preserving:}} &
\text{if $a^- \in x$, $\ind_A\,(f\,a) = \ind_A\,a$,}\\
\text{\emph{$+$-preserving:}} &
\text{if $a^+ \in x$, $\ind_A\,(f\,a) = \ind_A\,a$,}
\end{array}
\]
we call a structural map \textbf{positive} iff it is
\emph{$-$-preserving} and \emph{$-$-total}; we call it
\textbf{negative} iff it is \emph{$+$-preserving} and \emph{$+$-total}.
For these notions, we use notations
$f : x \pstrto y$ and $f : x \nstrto y$.
\end{defi}

\noindent
Intuitively, positive structural maps can only contract positive moves:
\emph{$-$-preserving} ensures that they cannot contract negative moves
(as the copy index is preserved), and \emph{$-$-total} entails that
they cannot weaken negative moves: negative extensions must have a
pre-image. Dually, negative structural maps can only contract and
weaken negative moves.

Structural maps generalize symmetries, in a way compatible with
polarities: 

\begin{lem}\label{lem:sym_str}
Any symmetry $\theta : x \sym_A y$ is also a structural map
$\theta : x \strto y$. 
Moreover, $\theta$ is positive (resp. negative) as a symmetry iff it is
positive (resp. negative) as a structural map.
\end{lem}
\begin{proof}
For the first part, note that by \emph{transparent}, $\theta$ is an
order-isomorphism preserving $\lbl_A$. It is thus evident from the
definition that it is also a structural map.
For the second part, if $\theta : x \sym_A^+ y$, then by
\emph{$+$-transparent} it is \emph{$-$-preserving}. It is also
\emph{$-$-total} because it is an order-iso. The other implication
proceeds similarly, and the negative case is symmetric.
\end{proof}

\subsubsection{Cartesian morphisms.} We are now in position to define
our \emph{cartesian morphisms}, which take positive structural maps
covariantly and negative structural maps contravariantly.

\begin{defi}\label{def:carmor}
A \textbf{cartesian morphism} $\chi : x \carmor y$ is any
composite relation:
\[
x 
\quad = \quad
x_1 
\quad \pstrto \quad
x_2
\quad \nstrot \quad
x_3
\quad
\dots
\quad
x_{n-2}
\quad \pstrto \quad
x_{n-1}
\quad \nstrot \quad
x_n
\quad = \quad
y
\]
where $x_1, \ldots, x_n \in \conf{A}$.
\end{defi}

\noindent
A cartesian morphism $\chi : x \carmor y$ is a relation between $x$ and
$y$, \emph{i.e.} $\chi \subseteq x \times y$, but it is in general not
functional in either direction. This, of course, is unavoidable: basic
contractions are functional, but they are taken covariantly or
contravariantly.
\begin{figure}
\[
\raisebox{40pt}{
\scalebox{.8}{
\xymatrix@R=20pt@C=0pt{
&&&\qu^-_{4,\grey{0}}\\
\qu^+_{2,\grey{0}}
        \ar@{.}@/^/[urrr]
&&\qu^+_{2,\grey{4}}
        \ar@{.}@/^/[ur]
&&\qu^+_{3,\grey{2}}
        \ar@{.}@/_/[ul]\\
\qu^-_{1,\grey{12}} \ar@{.}@/^.1pc/[u]&
\qu^-_{1,\grey{4}} \ar@{.}@/^/[ur]
&&\qu^-_{1,\grey{2}}
        \ar@{.}@/_/[ul]
}}}
\quad\nstrot\quad
\raisebox{40pt}{
\scalebox{.8}{
\xymatrix@R=20pt@C=-5pt{
&&&&&\qu^-_{4,\grey{0}}\\
&\qu^+_{2,\grey{0}}
        \ar@{.}@/^/[urrrr]
&&&\qu^+_{2,\grey{4}}
        \ar@{.}@/^/[ur]
&&\qu^+_{3,\grey{2}}
        \ar@{.}@/_/[ul]\\
\qu^-_{1,\grey{12}} 
	\ar@{.}@/^/[ur]&&
\qu^-_{1,\grey{2}}
	\ar@{.}@/_/[ul]&
\qu^-_{1,\grey{12}} \ar@{.}@/^/[ur]
&&\qu^-_{1,\grey{2}}
        \ar@{.}@/_/[ul]
}}}
\quad\pstrto\quad
\raisebox{40pt}{
\scalebox{.8}{
\xymatrix@R=20pt@C=0pt{
&&\qu^-_{4,\grey{0}}\\
&\qu^+_{2,\grey{0}}
        \ar@{.}@/^/[ur]
&&\qu^+_{3,\grey{2}}
        \ar@{.}@/_/[ul]
&\qu^+_{3,\grey{6}}
        \ar@{.}@/_/[ull]\\
\qu^-_{1,\grey{12}} \ar@{.}@/^/[ur]
&&\qu^-_{1,\grey{2}}
        \ar@{.}@/_/[ul]
}}}
\]
\caption{A cartesian morphism}
\label{fig:ex_carmor}
\end{figure}
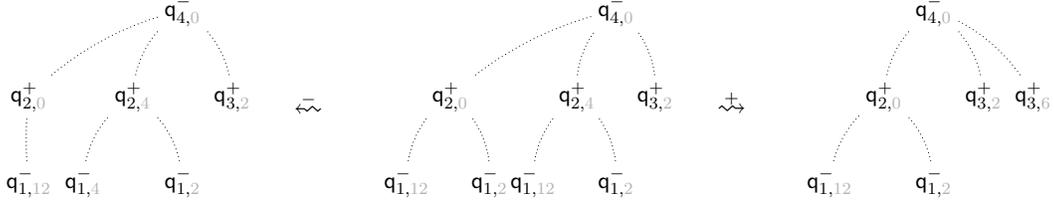
We give in Figure \ref{fig:ex_carmor} an example of a cartesian
morphism (we do not give the exact definition of structural maps to
avoid overloading the picture, but they are almost uniquely defined).
Note that there is also a cartesian morphism from the rightmost diagram
to the leftmost one, so that these two are considered equivalent,
qualitatively.

The definition of cartesian morphism obviously yields a category,
but the first property we aim for is that every cartesian
morphism factors uniquely as the notation $\carmor$ suggests. But this
is not a trivial fact, establishing it requires building up a number of
prerequisites. 

We start with this easy property of structural maps:

\begin{lem}\label{lem:strmap_pred}
Consider $f : x \strto y$ a structural map, and $f(a) = b$ with
both non-minimal.

Then, $f(\pred(a)) = \pred(b)$ as well.
\end{lem}
\begin{proof}
Immediate from \emph{$\imc$-preserving} and the fact that the game $A$
is \emph{forestial}. 
\end{proof}

\noindent
Next we show a uniqueness property of intermediate witnesses for
cartesian morphisms:

\begin{lem}\label{lem:linking_unique_wit}
Consider $\chi : x \carmor y$ a cartesian morphism, obtained through
the chain:
\[
x =
x_1 
\quad \stackrel{f^+_1}{\strto} \quad
y_1
\quad \stackrel{f^-_1}{\strot} \quad
x_2
\quad
\dots
\quad
x_n
\quad \stackrel{f^+_n}{\strto} \quad
y_n
\quad \stackrel{f^-_n}{\strot} \quad
x_{n+1}
= y\,.
\]

\noindent
Then, for any $(a_1, a_{n+1}) \in \chi$, there is a unique sequence of
witnesses
\[
\xymatrix@R=0pt{
x_1     \ar@{}[r]|\strto^{f^+_1}&
y_1     \ar@{}[r]|\strot^{f^-_1}&
x_2     &\dots&
x_n     \ar@{}|\strto[r]^{f^+_n}&
y_n     \ar@{}|\strot[r]^{f^-_n}&
x_{n+1}\\
\rotatebox{90}{$\in$}&
\rotatebox{90}{$\in$}&
\rotatebox{90}{$\in$}&\dots&
\rotatebox{90}{$\in$}&
\rotatebox{90}{$\in$}&
\rotatebox{90}{$\in$}\\
a_1&b_1&a_2&\dots&a_n&b_n&a_{n+1}
}
\]
such that for every $1\leq i \leq n$, $(a_i, b_i) \in f^+_i$ and
$(a_{i+1}, b_i) \in f^-_i$.
\end{lem}
\begin{proof}
Existence is obvious by definition of relational composition, we prove
uniqueness.
Note that since structural maps preserve minimality and $\imc$, all
moves in such as sequence must have the same \emph{depth}, where the
\textbf{depth} of $a$ minimal is $0$, and the depth of $b$ where $a
\imc b$ is $\depth(b) = \depth(a) + 1$. It follows that $\chi$
preserves depth as well. Thus, we prove the uniqueness of the chain
above by induction on $\depth(a_1) = \depth(a_{n+1})$.

Assume the depth is $0$. For any $1 \leq i \leq n-1$, if $a_{i+1} \in
\min(x_{n+1})$ is fixed, necessarily negative since $A$ is negative,
then there is a unique $b_i \in y_n$ such that $(a_{i+1}, b_i) \in
f^-_i$ since $f^-_i$ is a function. Now, in turn, we show that
there is a unique $a_i \in \min(x_i)$ such that $(a_i, b_i) \in
f^+_i$. It must be minimal, or contradict the minimality of $b_i$
since $f^+_i$ is $\imc$-preserving. Consider then another $a'_i$ such
that $(a'_i, b_i) \in f^+_i$ also; necessarily also minimal. Then,
$\lbl(a'_i) = \lbl(b_i)$ since $f^+_i$ is \emph{label-preserving}.
Additionally, $\ind(a'_i) = \ind(b_i)$ since $f^+_i$ is
\emph{$-$-preserving} -- as all elements in this chain are negative.
But then, $a'_i$ and $a_i$ are both minimal, with same label and copy
index, thus $a'_i = a_i$ by \emph{jointly injective}. 

Now, consider $(a_1, a_{n+1}) \in \chi$ of depth $d+1$. Since $A$ is
forestial, there are unique $c_1 \imc a_1$ and $c_{n+1} \imc a_{n+1}$.
By Lemma \ref{lem:strmap_pred}, we have $(c_1, c_{n+1}) \in \chi$ as
well, and by induction hypothesis, there is a unique sequence of
witnesses
\[
\xymatrix@R=0pt{
x_1     \ar@{}|\strto[r]^{f^+_1}&
y_1     \ar@{}|\strot[r]^{f^-_1}&
x_2     &\dots&
x_n     \ar@{}|\strto[r]^{f^+_n}&
y_n     \ar@{}|\strot[r]^{f^-_n}&
x_{n+1}\\
\rotatebox{90}{$\in$}&
\rotatebox{90}{$\in$}&
\rotatebox{90}{$\in$}&\dots&
\rotatebox{90}{$\in$}&
\rotatebox{90}{$\in$}&
\rotatebox{90}{$\in$}\\
c_1&d_1&c_2&\dots&c_n&d_n&c_{n+1}
}
\]
for this. Now for negative extensions, we reason as above from
right to left: if $c_{i+1} \imc a_{i+1}$ negative, there is a
unique $d_i \imc b_i$ such that $(a_{i+1}, b_i) \in f^-_i$ since it
is a function; and then there is a unique $c_i \imc a_i$ s.t.
$(a_i, b_i) \in f^+_i$ by \emph{$-$-total}, uniqueness being by
\emph{label-preserving} and \emph{$-$-preserving} for $f^+_i$, along
with \emph{jointly injective} for $A$. For positive extensions, the
reasoning is symmetric from left to right, concluding the proof.
\end{proof}

\noindent
From this unique witness property, we derive the following lifting
property:

\begin{lem}\label{lem:main_linkings}
For any structural map  $\chi : x \carmor y$, we have the following
lifting properties:
\[
\begin{array}{rl}
\text{\emph{(1)}} & 
\text{if $b_2^- \in \min(y)$, then there is a unique $b_1^- \in
\min(x)$ such
that $(b_1^-, b_2^-) \in \chi$,}\\
\text{\emph{(2a)}} &
\text{If $(a_1, a_2) \in \chi$, then for all $a_2 \imc_A b_2^-$ in
$y$,}\\
&\text{there is a unique $a_1 \imc b_1^-$ in $x$ such that $(b_1, b_2)
\in
\chi$,}\\
\text{\emph{(2b)}} &
\text{If $(a_1, a_2) \in \chi$, then for all $a_1 \imc_A b_1^+$ in
$x$,}\\
&\text{there is a unique $a_2 \imc b_2^+$ in $y$ such that $(b_1, b_2)
\in
\chi$.}
\end{array}
\]
\end{lem}
\begin{proof}
Since $A$ is negative, there is only one case for transporting minimal
events.

By definition, $\chi$ is obtained as a composition of 
structural maps:
\[
x =
x_1 
\quad \stackrel{f^+_1}{\strto} \quad
y_1
\quad \stackrel{f^-_1}{\strot} \quad
x_2
\quad
\dots
\quad
x_n
\quad \stackrel{f^+_n}{\strto} \quad
y_n
\quad \stackrel{f^-_n}{\strot} \quad
x_{n+1}
= y\,.
\]

\emph{(1)} For $1\leq i \leq n$, for any $a_{i+1}^- \in x_{n+1}$
minimal, then there is a unique $b_i^- \in y_n$ such that $(a_{i+1}^-,
b_i^-) \in f^-_i$, because it is a function; and $b_i^-$ is minimal
since $f^-_i$ preserves minimality. Then, there is some $a_i^- \in
x_i$ such that $(a^-_i, b^-_i) \in f^+_i$ by \emph{$-$-total}. Again
it must be minimal: a predecessor would yield a predecessor for $b^-_i$
by \emph{$\imc$-preserving}. It is unique since $f^+_i$ is
\emph{label-preserving} and \emph{$-$-preserving} and by \emph{jointly
injective} for $A$. Iterating this, we get a unique sequence of
witnesses terminating in $a_{n+1}$, and in particular unique $(a_1,
a_{n+1}) \in \chi$. 

\emph{(2a)} Now consider $(a_1, a_{n+1}) \in \chi$, and $a_{n+1} \imc
c_{n+1}^-$. By Lemma \ref{lem:linking_unique_wit}, there is
\[
\xymatrix@R=0pt{
x_1     \ar@{}|\strto[r]^{f^+_1}&
y_1     \ar@{}|\strot[r]^{f^-_1}&
x_2     &\dots&
x_n     \ar@{}|\strto[r]^{f^+_n}&
y_n     \ar@{}|\strot[r]^{f^-_n}&
x_{n+1}\\
\rotatebox{90}{$\in$}&
\rotatebox{90}{$\in$}&
\rotatebox{90}{$\in$}&\dots&
\rotatebox{90}{$\in$}&
\rotatebox{90}{$\in$}&
\rotatebox{90}{$\in$}\\
a_1&b_1&a_2&\dots&a_n&b_n&a_{n+1}
}
\]
a unique sequence of witnesses. Note that by Lemma
\ref{lem:strmap_pred}, if there is indeed some $a_1 \imc c_1^-$ such
that $(c_1^-, c_{n+1}^-) \in \chi$, then its sequence of witnesses
(unique by Lemma \ref{lem:linking_unique_wit}) must be above the
(unique) sequence of witnesses for $(a_1, a_{n+1})$. Hence it suffices
to show that there is a unique sequence of witnesses above the above
ending in $c_{n+1}^-$, and this is what we shall do. Thus assume
$a_{i+1} \imc c_{i+1}^-$. As just above, there is a unique $(c_{i+1}^-,
d_i^-) \in f^-_i$ because it is a function, and we do have $b_i \imc
d_i^-$ by \emph{$\imc$-preserving}. Then, there is a unique $(c_i^-,
d_i^-) \in f^+_i$ by \emph{$-$-total}, \emph{$-$-preserving},
\emph{label-preserving}, \emph{$\imc$-preserving} for $f^+_i$ and
\emph{jointly injective} for $A$. Iterating this we get a unique
sequence of witnesses terminating in $c^-_{n+1}$ above the unique
sequence witnessing $(a_1, a_{n+1})$; and in particular unique $(c_1^-,
c_{n+1}^-)$ as required.

\emph{(2b)} The reasoning is symmetric.
\end{proof}

\noindent
Now, we are almost ready to prove our factorization result. As we will
need to build structural maps gradually, we need to cover intermediate
cases where total constraints are not satisfied, and shall therefore
need the following notions:

\begin{defi}
A \textbf{partial positive map}, written $f : x \ppstrto y$, is a
structural map satisfying \emph{$-$-preserving} (but not
\emph{$-$-total}). Likewise, a \textbf{partial negative map}, written
$f : x \pnstrto y$, is a structural map satisfying \emph{$+$-preserving}
(but not \emph{$+$-total}).  
\end{defi}

\noindent
These satisfy the following lifting lemmas:

\begin{lem}\label{lem:lift_neg}
Consider $f : x \ppstrto y$ a partial positive structural map. Then:\\
\indent\emph{(1)} for $y \vdash_A b_2^-$ minimal, there is a unique $x
\vdash_A b_1^-$ minimal such that
\[
f \uplus \{(b_1^-, b_2^-)\} : x\uplus \{b_1^-\} \ppstrto
y \uplus \{b_2^-\}\,,
\]
\indent\emph{(2)}
for $a \in x$, for $y \vdash_A b_2^-$ with $f\,a \imc_A
b_2^-$, there is a unique $x \vdash_A b_1^-$ with $a \imc_A b_1^-$ s.t.
\[
f \uplus \{(b_1^-, b_2^-)\} : x\uplus \{b_1^-\} \ppstrto
y \uplus \{b_2^-\}
\]
\end{lem}
\begin{proof}
We detail \emph{(2)}, as \emph{(1)} is similar but simpler.

\emph{Existence.} First, we note that $f$ is
\emph{label-preserving}, so that $\lbl\,a = \lbl\,(f\,a)$. Thus, by
\emph{rigid}, $\lbl\,a \imc_{\mar{A}} \lbl\,b_2^-$. Thus by
\emph{wide}, there is indeed some $b_1^-$ such that $a \imc_A b_1^-$,
$\lbl(b_1^-) = \lbl(b_2^-)$ and $\ind(b_1^-) = \ind(b_2^-)$. Now, we
must justify that $x \vdash_A b_1^-$. Clearly, $x \uplus \{b_1^-\}$ is
down-closed. If it was inconsistent, there would be some $b \in x$ such
that $b \mconflict~b_1^-$, in which case $a \imc b$ and $\ind(a) =
\ind(b)$ by \emph{local conflict}. Since $A$ is alternating, $b$ is
negative as well, and $f$ is \emph{$-$-preserving}, so that by
\emph{invariant conflict} we have $f b \mconflict~b_2^-$ as well,
contradiction.

\emph{Uniqueness.} Straightforward by \emph{jointly injective} and that
$f$ is positive.
\end{proof}

\begin{lem}\label{lem:lift_pos}
Consider $f : x \pnstrto y$ a negative structural map.
Then
for all $a \in x$, for all $y \vdash_A b_2^+$ with $f\,a \imc_A
b_2^+$,
there is a unique $x \vdash_A b_1^+$ with $a \imc_A b_1^+$ such that
\[
f \uplus \{(b_1^+, b_2^+)\} : x\uplus \{b_1^+\} \pnstrto y \uplus \{b_2^+\}
\]
\end{lem}
\begin{proof}
Same as for Lemma \ref{lem:lift_neg}.
\end{proof}

\noindent
Now, we are finally in position to prove our factorization result. For
the proof, we introduce the following notation: if $R \subseteq A
\times B$ is a relation from $A$ to $B$, we write $R^\perp \subseteq B
\times A$ for the reverse relation. We will also apply this to
functions, regarded as functional relations.

\begin{lem}\label{lem:factorization}
Consider $A$ a mixed board, and $\chi : x \carmor z$ a cartesian
morphism.

Then, there are unique $y \in \conf{A}$, $\chi_- : y
\nstrto x$ and $\chi_+ : y \pstrto z$, such that
\[
\xymatrix{
x        \ar@{<->}|{(-+)}[rr]^{\chi}&&
z\\
&y      \ar@{->}[ul]|{(-)}^{\chi_-}
        \ar@{->}[ur]|{(+)}_{\chi_+}
        \ar@{}[u]|{\rotatebox{90}{$\subseteq$}}
}
\]
where the bottom path is composed relationally, \emph{i.e.} we ask
$\chi_+ \circ \chi_-^\perp \subseteq \chi$. 

Finally, the inclusion is actually an equality. 
\end{lem}
\begin{proof}
First, remark that we can find $y, \chi_-$ and $\chi_+$ such that
\[
\xymatrix{
x        \ar@{<->}|{(-+)}[rr]^{\chi}&&
z\\
&y      \ar@{->}|{(-p)}[ul]|{}^{\chi_-}
        \ar@{->}|{(+p)}[ur]|{}_{\chi_+}
        \ar@{}[u]|{\rotatebox{90}{$\subseteq$}}
}
\]
with $\chi_- : y \pnstrto x$ and $\chi_+ : y \ppstrto z$ -- indeed, it
suffices to take $y$ and $\chi_-, \chi_+$ to be empty. Let us call such
data a \emph{solution}; solutions are partially ordered by
componentwise inclusion. There is an upper bound to the size of $y$,
because it is a tree whose depth is bounded by both that of $x$ and
$z$, and its width is bounded by the maximum of those of $x$ and $z$
(by \emph{$-$-preserving} and \emph{$+$-preserving} of $\chi_+$ and
$\chi_-$). Hence, there is a solution $y, \chi_-, \chi_+$ where $y$ has
maximal size. We show that it satisfies the conditions of the lemma.

We must first check that $\chi_+$ and $\chi_-$ are not partial.
First, we show that $\chi_+$ is \emph{$-$-total}. Thus take $a^+ \in
y$, and $a_2^+ = \chi_+\,a^+ \imc b^-_2$ in $z$. By Lemma
\ref{lem:lift_neg}, there is a unique $y \vdash b^-$ with $a^+ \imc
b^-$ and 
$\chi_+ \uplus \{(b^-, b_2^-)\} : y \uplus \{b^-\} \ppstrto z$.
To extend $\chi_-$ accordingly, write $a_1^+ =
\chi_-\,a^+$. By hypothesis, we have $(a_1^+, a_2^+) \in \chi$. Hence,
by Lemma \ref{lem:main_linkings}, there is a (unique) $a_1^+ \imc
b_1^-$ such that $(b_1^-, b_2^-) \in \chi$. We form
\[
\chi_- \uplus \{(b^-, b_1^-)\} : y \uplus \{b^-\} \pnstrto z
\]
which is clearly a partial negative structural map. The
required inclusion is still satisfied, so this contradicts the
maximality of the solution $y, \chi_-, \chi_+$ -- the case where $b^-$
is minimal is similar but simpler. The proof that $\chi_-$ is $+$-total
is symmetric.

For \emph{uniqueness}, consider we have
\[
\xymatrix{
&y'     \ar@{->}|{(-)}[dl]_{\xi_-}
        \ar@{->}|{(+)}[dr]^{\xi^+}
        \ar@{}[d]|{\rotatebox{270}{$\subseteq$}}\\
x       \ar@{<->}|{(-+)}[rr]^{\chi}&&
z\\
&y      \ar@{->}[ul]|{(-)}^{\chi_-}
        \ar@{->}[ur]|{(+)}_{\chi_+}
        \ar@{}[u]|{\rotatebox{90}{$\subseteq$}}
}
\]
and show $y' \subseteq y$, and that for every $a \in y'$, we have
$\chi_-\,a = \xi_-\,a$ and $\chi_+\,a = \xi_+\,a$.

Consider $a \in y'$ minimal not satisfying this property. If it is
minimal in $A$
then it is in particular negative. Then write $a' = \xi^+\,a$; by
definition of structural maps we have $a'$ minimal with $\lbl\,a' =
\lbl\,a$, by \emph{$-$-preserving} we have $\ind\,a' = \ind\,a$ -- so
that in fact $a = a'$ by \emph{jointly injective}. And by
\emph{$-$-total}, there is $a'' \in y$ such that $\chi_+\,a'' = a$, and
for the same reason $a'' = a$, so that $a \in y$. Moreover, we have
seen that $\chi_+\,a = \xi_+\,a = a$; but we also know that
$(\chi_-\,a, \chi_+\,a), (\xi_-\,a, \xi_+\,a) \in \chi$ by hypothesis.
But now, by Lemma \ref{lem:main_linkings}, there is a unique $b$ such
that $(b, a) \in \chi$ -- so $\chi_-\,a = \xi_-\,a = b$ as required.
Now if $a \in y'$ is not minimal in $A$, it has a unique predecessor $b
\imc_A a$, for which we know by induction that $b \in y$ and $\chi_-\,b
= \xi_-\,b$ and $\chi_+\,b = \xi_+\,b$. Based on that, if $a$ is
negative, we can directly replay the argument above; and the symmetric
argument if $a$ is positive.
In particular, this shows that $y' \subseteq y$ with $\xi_-, \chi_-,
\xi_+, \chi_+$ compatible with the inclusion. But the argument is
symmetric, so we also have $y \subseteq y'$ -- which concludes
uniqueness.

Finally, we must show that the inclusion between the two sides of the
diagram is actually an equality. Consider $(b_1, b_2) \in \chi$
minimal (note that they must have the same depth) such that $(b_1,
b_2) \not \in \chi_+ \circ \chi_-^\perp$. If $(b_1, b_2)$ are
minimal, then they are negative $(b_1^-, b_2^-)$. By \emph{$-$-total}
for $\chi_+$, there is some $b^- \in y$ such that $\chi_+\,b^- =
b_2^-$. Writing $b'_1 = \chi_-\,b^-$, we must have $(b'_1, b_2) \in
\chi$, but by Lemma \ref{lem:main_linkings} this implies $b'_1 =
b_1^-$, so $(b^-_1, b^-_2) \in \chi_+ \circ \chi_-^\perp$ after
all, contradiction. So now if they are not minimal, there are $c_1 \imc
b_1$ and $c_2 \imc b_2$ with $(c_1, c_2) \in \chi$ (by Lemma
\ref{lem:strmap_pred}) and $(c_1, c_2) \in \chi_+ \circ \chi_-^\perp$
(by minimality of $(b_1, b_2)$), so there is $c \in y$ such that
$\chi_+\,c = c_2$ and $\chi_-\,c = c_1$. Now we distinguish by cases.
If $b_1, b_2$ are negative, then exploiting $c$, the reasoning is the
same as for $b_1, b_2$ minimal above, using the same lemmas -- and if
$b_1, b_2$ are positive, then the reasoning is symmetric. 
\end{proof}

\noindent
From this lemma, it follows in particular that as announced, every
cartesian morphism $\chi : x \carmor y$ can be written uniquely as the
relational composition $x \nstrot \cdot \pstrto y$.

\subsection{Reconstructing the Preorder} Now, in this section we
reconstruct a preorder from these cartesian morphisms, and show how
this is compatible with all the constructions on preorders involved in
the relative Seely category structure. Fix here a mixed board $A$.

\subsubsection{Uncolored case.} From the relational collapse, we know
that we must equip the set $\coll{A}$ of \emph{positions} of $A$
defined in \eqref{eq:def_pos} with a preorder derived from cartesian
morphisms. The obvious route is to start is to define it on
\emph{configurations}: for $x, y \in \conf{A}$, we set 
\[
x \pre y
\qquad\Leftrightarrow\qquad
\exists \chi : y \carmor x\,,
\]
noting the inversion in directions. This is compatible with symmetry:

\begin{lem}\label{lem:pre_inv_sym}
Consider $x, x', y, y' \in \conf{A}$ such that $x \sym_A x'$ and $y
\sym_A y'$. 

Then, $x \pre y$ iff $x' \pre y'$.
\end{lem}
\begin{proof}
Fix a cartesian morphism $\chi : y \carmor x$ which by Lemma
\ref{lem:factorization} factors as $\chi_+ \circ \chi_-^\perp$ for
$\chi_- : z \nstrto y$ and $\chi_+ : z \pstrto x$. Fix also symmetries
$\theta : x \sym_A x'$ and $\vartheta : y \sym_A y'$, by Lemma
\ref{lem:sym_factor} they factor as $\theta = \theta_+ \circ \theta_-$
and $\vartheta = \vartheta_+ \circ \vartheta_-$ for $\theta_+,
\vartheta_+$ positive symmetries and $\theta_-, \vartheta_-$ negative
symmetries. Now, with all this data we can form
\[
x' 
\quad
\stackrel{\theta_+}{\pstrot}
\quad
\cdot
\quad
\stackrel{\theta_-^{-1}}{\nstrto}
\quad
x
\quad
\stackrel{\chi_+}{\pstrot}
\quad
\cdot
\quad
\stackrel{\chi_-}{\nstrto}
\quad
y
\quad
\stackrel{\vartheta_-}{\nstrto}
\quad
\cdot
\quad
\stackrel{\vartheta_+^{-1}}{\pstrot}
\quad
y'
\]
converting symmetries to structural maps by Lemma \ref{lem:sym_str}; so
that $x' \pre y'$.
\end{proof}

\noindent
We write $\x \pre \y$ if for any $x \in \x$ and $y \in \y$, we
have $x \pre y$ -- by the lemma above, this does not depend on the
choice of $x$ and $y$, and immediately forms a preorder. Thus:

\begin{defi}
For any mixed board $A$, we set $\scoll{A} = (\coll{A}, \pre)$.
\end{defi}

\begin{figure}
\begin{minipage}{.48\linewidth}
\[
\scalebox{.9}{$
\begin{array}{rcrcl}
s^\tensor_{A, B} &:& \scoll{A} \times \scoll{B} &\bij& \scoll{A \tensor
B}\\
s^\one &:& \one &\bij& \scoll{\one}\\
s^{\oc}_S &:& \oc \scoll{S} &\bij& \scoll{\oc S}\\
s^\top &:& \emptyset &\bij& \scoll{\top}\\
s^\with_{S, T} &:& \scoll{S} + \scoll{T} &\bij& \scoll{S\with T}\\
s^\lin_{A, S} &:& \scoll{A}^\op \times \scoll{S} &\bij& \scoll{A\lin S}
\end{array}
$}
\]
\caption{Structural preorder-isos}
\label{fig:str_coll_iso}
\end{minipage}
\hfill
\begin{minipage}{.48\linewidth}
\[
\scalebox{.9}{$
\begin{array}{rcrcl}
s^\tensor_{A, B} &:& \cscoll{A}{\C} \times \cscoll{B}{\C} &\bij& \cscoll{A
\tensor B}{\C}\\
s^\one &:& \one &\bij& \cscoll{\one}{\C}\\
s^{\oc}_S &:& \oc \cscoll{S}{\C} &\bij& \cscoll{\oc S}{\C}\\
s^\top &:& \emptyset &\bij& \cscoll{\top}{\C}\\
s^\with_{S, T} &:& \cscoll{S}{\C} + \cscoll{T}{\C} &\bij& \cscoll{S\with
T}{\C}\\
s^\lin_{A, S} &:& \cscoll{A}{\C}^{\op} \times \cscoll{S}{\C} &\bij&
\cscoll{A\lin S}{\C}
\end{array}
$}
\]
\caption{Colored preorder-isos}
\label{fig:col_str_coll_iso}
\end{minipage}
\end{figure}

\noindent
This is compatible with all the constructions on preorders involved in
the relative Seely structure of $\ScottL$. More precisely, the
bijections of Figure \ref{fig:str_coll_bij} can be verified to be
compatible with the preorder, \emph{i.e.} to yield
preorder-isomorphisms as described in Figure \ref{fig:str_coll_iso}.
For most of them, the corresponding verification is straightforward; we
just detail two.

\begin{lem}
Fix $A, S$ mixed boards with $S$ strict, $\x_A \lin \x_S, \y_A
\lin \y_S \in \coll{A\lin S}$. 

Then, $\x_A \lin \x_S \pre \y_A \lin \y_S$ iff $\y_A \pre \x_A$
and $\x_S \pre \y_S$.
\end{lem}
\begin{proof}
By Lemma \ref{lem:pre_inv_sym}, it suffices to reason on
representatives. 

\emph{If.} Fix $y_A, x_A, x_S, y_S$ such that $y_A \pre x_A$ and $x_S
\pre y_S$. By definition, this means that there are $\chi_A : x_A
\carmor y_A$ and $\chi_S : y_S \carmor x_S$. By Lemma
\ref{lem:factorization}, there are $\chi_A^- : z_A \nstrto x_A$ and
$\chi_A^+ : z_A \pstrto y_A$, with also $\chi_S^- : z_S \nstrto y_S$
and $\chi_S^+ : z_S \pstrto x_S$. We may then form 
\[ 
\chi_A^+ \lin \chi_S^- ~:~ z_A \lin z_S ~\nstrto~ y_A \lin
y_S\,, \qquad \chi_A^- \lin \chi_S^+ ~:~ z_A \lin z_S ~\pstrto~ x_A 
\lin x_S 
\] 
defined in the obvious way -- observe the inversion of
polarities, to match that $A$ is dualized on the left hand side.
Altogether, we have $x_A \lin x_S \pre y_A \lin y_S$ as required.  

\emph{Only if.} Assuming $x_A \lin x_S \pre y_A \lin y_S$, we have
structural maps
\[
\chi_{A\lin S}^- ~:~ z_A \lin z_S ~\nstrto~ y_A \lin y_S\,,
\qquad
\xi_{A\lin S}^+ ~:~ z_A \lin z_S ~\pstrto~ x_A \lin x_S\,.
\]

First, as $\chi_{A\lin S}^-$ preserves labels, it decomposes uniquely
into $\chi_A \lin \chi_S$ for $\chi_A : z_A \strto y_A$ and $\chi_S :
z_S \strto y_S$ structural maps. Furthermore, the negativity of
$\chi_{A\lin S}^-$ ensures that $\chi_A$ is positive and $\chi_S$ is
negative. From the same reasoning on $\xi_{A\lin S}^+$, we get $\xi_A^-
: z_A \nstrto x_A$ and $\xi_S^+ : z_S \pstrto x_S$, which may be
directly assembled to witness $y_A \pre x_A$ and $x_S \pre y_S$. 
\end{proof}

\noindent
This precisely ensures that the bijection $s^\lin_{A, S}$ extends to an
isomorphism of preorders as in Figure \ref{fig:str_coll_iso} -- note
how the contravariant on the left hand side matches the dualization of
the moves in $A$ in the construction of $A \lin S$. The other
noteworthy case is the exponential:

\begin{lem}\label{lem:carac_pre}
Consider $S$ a strict mixed board, and $\x = [\x_i \mid i \in I], \y =
[\y_j \mid j \in J] \in \coll{\oc S}$. 

Then, $\x \pre \y$ iff for all $i \in I$, there is $j\in J$ such that
$\x_i \pre \y_j$.
\end{lem}
\begin{proof}
By Lemma \ref{lem:pre_inv_sym}, it suffices to reason on
representatives.

\emph{If.} Working with representatives, we have $I, J \subseteq_f \N$.
Assume that for all $i \in I$, there is $j \in J$ such that $x_i \pre
y_j$. Fix a function $f : I \to J$ such that for all $i \in I$, we have
$x_i \pre y_{f(i)}$. So, we have $\chi_i^+ : z_i \pstrto x_i$ and
$\chi_i^- : z_i \nstrto y_{f(i)}$. Form $z = \famc{z_i \mid i \in I}$.
Setting
\[
\begin{array}{rcrcl}
\chi^+ &:& \famc{z_i \mid i \in I} &\to& \famc{x_i \mid i \in I}\\
&& (i, s) &\mapsto& (i, \chi_i^+(s))\,,
\end{array}
\]
it is a positive structural map. Critically, it
leaves $i$ unchanged, as required by \emph{$-$-preserving} and it
reaches all minimal moves in $x$, as required by \emph{$-$-total}.
Likewise, setting
\[
\begin{array}{rcrcl}
\chi^- &:& \famc{z_i \mid i \in I} &\to& \famc{y_j \mid j \in J}\\
&& (i, s) &\mapsto& (f(i), \chi_i^-(s))\,,
\end{array}
\]
it follows that $\chi^-$ is \emph{$+$-preserving} and \emph{$+$-total},
so a positive structural map. Altogether, we have constructed a witness
$z$ and structural maps ensuring that $x \pre y$ as required.

\emph{Only if.} Assuming $x \pre y$, we have some $z = \famc{z_k \mid k
\in K}$ with maps
\[
\chi^+ : \famc{z_k \mid k \in K} \pstrto \famc{x_i \mid i \in I}\,,
\qquad
\chi^- : \famc{z_k \mid k \in K} \nstrto \famc{y_j \mid j \in J}\,.
\]

Now, any $(k, s) \in \famc{z_k \mid k \in K}$ minimal must be negative
and $\ind(k,s) = k$, so that $\chi^+(k, s) = (k, s')$ by
\emph{$-$-preserving}. In fact, for every $(k, s) \in \famc{z_k \mid k
\in K}$, there is
\[
(k, s_0) \imc_{\oc S} (k, s_1) \imc_{\oc S} \ldots \imc_{\oc S} (k, s)
\]
a (unique) sequence of justifiers with $s_0$ minimal. Since $\chi^+$ is
\emph{$\imc$-preserving}, we must have $(k, s'_0) = \chi^+(k, s_0)
<_{\oc S} \chi^+(k, s) = (k', s')$, but this entails that $k = k'$:
$\chi^+$ preserves the first component $k$; and also $K \subseteq I$.
In fact, this is an equality since $\chi^+$ is also \emph{$+$-total},
so that $z = \famc{z_i \mid i \in I}$. Additionally, it follows that
$\chi^+$ decomposes into the data, for each $i \in I$, of $\chi^+_i :
z_i \pstrto x_i$. Decomposing $\chi^-$ similarly, we get a function $f
: I \to J$ and $\chi^-_i : z_i \nstrto y_{f(i)}$. But altogether,
assembling all the data we get $x_i \pre y_{f(i)}$ for all $i \in I$ as
required. 
\end{proof}

\noindent
Thus, we have established that the construction of a
preorder from cartesian morphisms is compatible with the constructions
on objects involved in the relative Seely structure.

\subsubsection{Colored case.} \label{subsubsec:colored_preorder}
In the colored case, we must first extend cartesian morphisms:

\begin{defi}
Consider $A$ a mixed board, and $x, y \in \cconf{A}{\C}$ configurations
in colors.

A \textbf{structural map} $f : x \strto y$ is simply a structural map
in the previous sense, preserving colors, \emph{i.e.} $\lambda_y(f(a))
= \lambda_x(a)$ for all $a \in x$. It is \textbf{positive} (resp.
\textbf{negative}) if it is positive (resp. negative) as a plain
structural map. A \textbf{cartesian morphism} is defined as in
Definition \ref{def:carmor}, with positive and negative structural maps
preserving colors.
\end{defi}

\noindent
All the development of Section~\ref{subsec:carmor} adapts transparently
in the presence of colors, which are preserved everywhere. The induced
preorder is again compatible with (color-preserving) symmetries, so
that it yields a preorder on positions in colors; altogether extending
$\ccoll{A}{\C}$ into a preordered set $\cscoll{A}{\C}$. Finally, as
before the bijections matching the relative Seely constructions in
games and in the linear Scott model are obviously compatible with
colors, yielding the preorder-isomorphisms of Figure
\ref{fig:col_str_coll_iso}.

\section{Qualitative Collapse}
\label{sec:qualcoll}

We have built, from any mixed board $A$, a preorder
$\scoll{A}$ (or $\cscoll{A}{\C}$ in the colored case) in a way
compatible with the constructions involved in the relative Seely
structure of $\ScottL$. We shall now extend that to strategies,
as usual focusing first on the uncolored case.

\subsection{Introduction}\label{subsec:intro}
The basic idea for our collapse to the linear Scott model is simple: we
shall simply take the down-closure of the relational collapse of
\eqref{eq:coll_rel_strat}, \emph{i.e.}, without colors:
\begin{eqnarray}
\scoll{\sigma} 
\quad=\quad 
[\coll{\sigma}]_{\scoll{A}^{\op} \times \scoll{B}} 
\quad\in\quad
\ScottL[\scoll{A}, \scoll{B}]
\label{eq:scott_coll}
\end{eqnarray}
which means that $(\x_A, \x_B) \in \scoll{\sigma}$ provided we can find
$\y_A \in \scoll{A}, \y_B \in \scoll{B}$ with $\y_A \pre \x_A$ and
$\x_B \pre \y_B$ along with a witness $y^\sigma \in \confp{\sigma}$
such that $y^\sigma_A \in \y_A$ and $y^\sigma_B \in \y_B$.

It is obvious that this is indeed a morphism in $\ScottL$. We shall now
make our first steps towards proving functoriality, introducing the
main difficulties. First, we note:

\begin{lem}\label{lem:scoll_pres_id}
Consider $A$ a mixed board. Then, $\scoll{\cc_A} = \id_{\scoll{A}}$.
\end{lem}
\begin{proof}
This is straightforward: the latter comprises all
pairs $(\x_1, \x_2)$ with $\x_2 \pre \x_1$, while the former amounts to
all $(\y_1, \y_2)$ such that there exists $\y \in \scoll{A}$ with a
specific representative $y \in \y$ with $\y_2 \pre \y$ and $\y \pre
\y_1$ -- clearly, this concerns the same pairs.
\end{proof}

\noindent
We also easily have oplax functoriality:

\begin{lem}\label{lem:scoll_oplax}
Consider $A, B$ and $C$ mixed boards, and $\sigma : A \vdash B$, $\tau
: B \vdash C$.

Then, $\scoll{\tau \odot \sigma} \subseteq \scoll{\tau} \circ
\scoll{\sigma}$.
\end{lem}
\begin{proof}
Consider $(\x_A, \x_C) \in \scoll{\tau \odot \sigma}$. This means 
there are $\y_A \pre \x_A$ and $\x_C \pre \y_C$ with
$(\y_A, \y_C) \in \coll{\tau \odot \sigma}$, which we know is in
$\coll{\tau} \circ \coll{\sigma}$. But $\coll{\sigma} \subseteq
\scoll{\sigma}$ and likewise for $\tau$, hence $(\y_A, \y_C) \in
\scoll{\tau} \circ \scoll{\tau}$. But then, $(\x_A, \x_C) \in
\scoll{\tau} \circ \scoll{\tau}$ as well by down-closure.
\end{proof}

\noindent
Again, the final inequality $\scoll{\tau} \circ \scoll{\sigma}
\subseteq \scoll{\tau \odot \sigma}$ is more problematic, let us see
why. Consider $(\y_A, \y_C) \in \scoll{\tau} \circ \scoll{\sigma}$.
Unfolding definitions, there must be witnesses $x^\sigma \in
\confp{\sigma}$ and $x^\tau \in \confp{\tau}$ such that $x^\sigma_A \in
\x_A \pre \y_A$, $x^\tau_B \pre x^\sigma_B$, and $x^\tau_C \in \x_C$
with $\y_C \pre \x_C$. Recall that the situation in
Section~\ref{subsubsec:lax_rel}, concerning lax preservation of
composition for the relational collapse, was somewhat analogous, except
that we had $x^\sigma_B \sym_B x^\tau_B$ rather than $x^\sigma_B
\carmor x^\tau_B$ -- we needed to synchronize $x^\sigma$ and $x^\tau$
through a \emph{symmetry}, whereas we now want to synchronize them
through a \emph{cartesian morphism}. In the symmetry case, this was
handled by Proposition \ref{prop:sync_sym}, which we do not (yet!) have
for cartesian morphisms. 

Let us sum up below what we need, for fixed $\sigma : A \vdash B$ and
$\tau : B \vdash C$. What we have: witnesses $x^\sigma \in
\confp{\sigma}$ and $x^\tau \in \confp{\tau}$ along with
a cartesian morphism $\chi : x^\sigma_B \carmor x^\tau_B$.
What we want: some $y^\tau \odot y^\sigma \in \confp{\tau \odot
\sigma}$ such that $y^\sigma_A \pre x^\sigma_A$ and $x^\tau_C \pre
y^\tau_C$. In the sequel, we shall refer to this data as, respectively,
a \emph{cartesian (matching) problem}, and a \emph{solution} to the
problem.  
\begin{figure}
\begin{center}
\includegraphics[scale=.7]{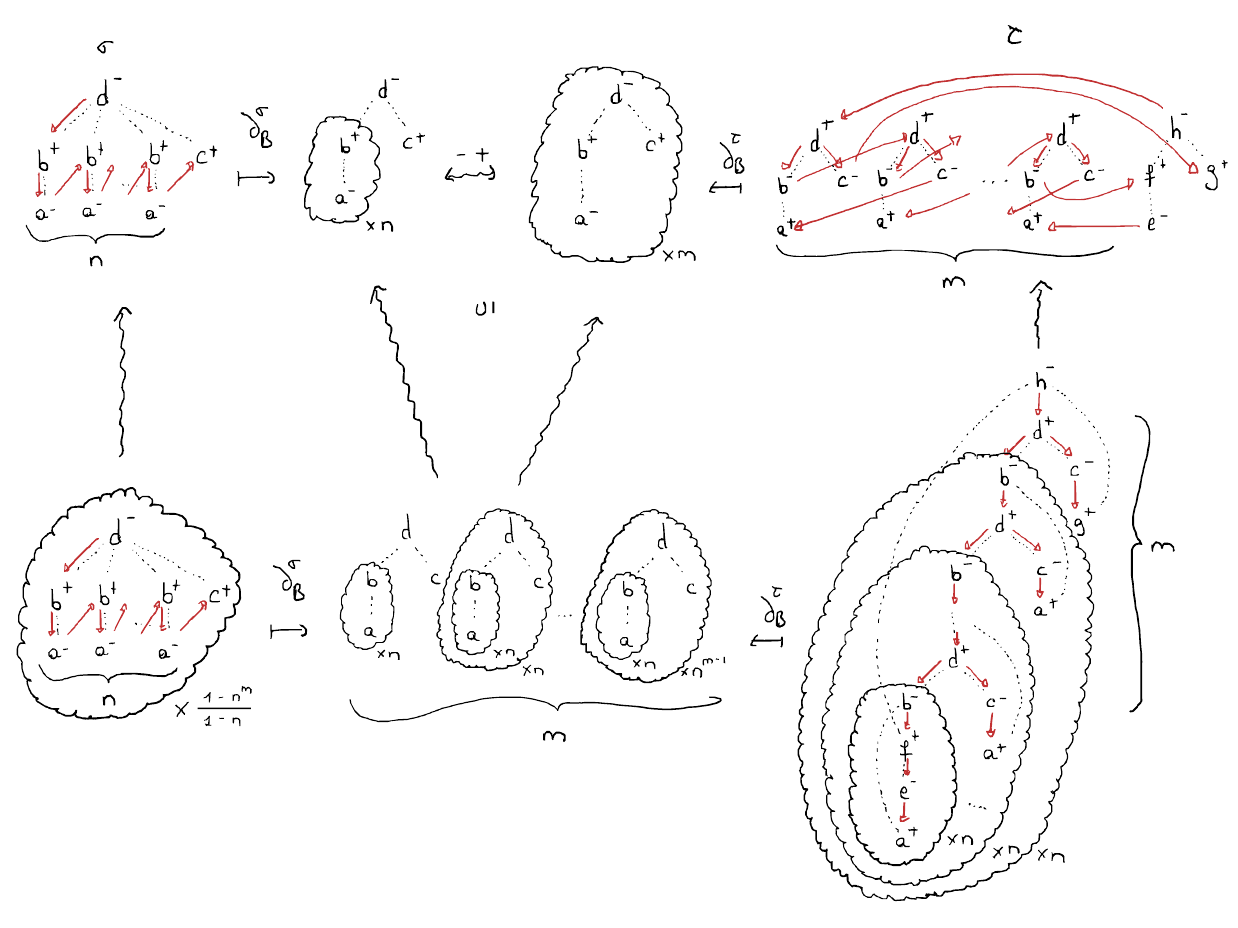}
\end{center}
\caption{Example resolution of a cartesian matching problem}
\label{fig:struct_pb}
\end{figure}
We call this a cartesian problem because both strategies are
actively trying to duplicate and erase each other, and a solution is a
situation where both strategies have reached a state where they have
all the resources they need, not more and not less. 

\begin{exa}
To illustrate this operation, we detail an example, using two
$\lambda$-terms
\[
\vdash M_\sigma = \lambda fx.\,\underbrace{f\,(\ldots
(f\,}_n\,x)\ldots) : (o \to o) \to o \to o\,,
\]
\[
g : (o \to o) \to o \to o \vdash M_\tau = \lambda x.\,
\underbrace{g\,(\ldots\,(g}_m\,x)\ldots) : (o \to o) \to o \to o\,.
\]

\noindent
The reader may recognize $M_\sigma$ the Church integer for $n$ and
$M_\tau$ that for $m$, though on different types.
Interpreting those (with the adequate promotion for $M_\sigma$), we obtain 
\[
\sigma \in \Dsinn[1, B]\,,
\qquad
\tau \in \Dsinn[B, C]
\]
strategies that we wish to compose -- here, $B = \oc \intr{(o \to o)
\to o \to o}$ and $C = \intr{(o\to o) \to o \to o}$. As events in $B$
in $C$ correspond to atoms in those types, we find it convenient to
write $B$ as $(a \to b) \to c \to d$ and $C$ as $(e \to f) \to g \to h$
to ease the correspondence.

With these notations, the upper part of Figure \ref{fig:struct_pb}
presents a cartesian problem involving $\sigma$ and $\tau$. On
the upper left part, we have the typical (unduplicated) configuration
of $\sigma$, iterating $n$ times its argument, displayed onto $B$ as
the configuration indicated (omitting copy indices). 
Likewise, on the upper right corner, we have the typical (unduplicated)
configuration of $\tau$, made larger than for $\sigma$ because of
$\eta$-expansion, displayed to $B$ as shown. There is a cartesian
morphism as shown, linking all pairs of moves with the same label. 

Resolving this problem involves performing all the necessary
duplications: $\tau$ makes $m$ copies of $\sigma$, but the first copy
of $\sigma$ makes $n$ copies of the $m-1$ remaining calls of $\sigma$,
and so on\ldots The solution appears in the bottom part of the diagram,
consisting in the indicated expansions of the configurations of
$\sigma$ and $\tau$, whose display on $B$ now match.
\end{exa}

\noindent
The example above illustrates that solving a cartesian problem
can involve an exponential blowup in the size of the configurations --
indeed, we know that the Church integer for $m$ applied to that for $n$
normalizes to the Church integer for $n^m$, witnessing the $n^m$ calls
to the event $f^+$ in the duplicated version of the configuration for
$\tau$ in the diagram. In general, the situation is far worse: the size
of the solution is not elementary in the size of the problem,
witnessing the usual bounds in the normalization of the simply-typed
$\lambda$-calculus\footnote{It is direct to extend Figure
\ref{fig:struct_pb} into an example of that, exploiting the explosion
of the application $n\,\ldots\,n$ of Church integers -- here $\tau$ is
$n$, and $\sigma$ is the tuple of $n$s typed appropriately.}. 

From this explosion, it is clear that the resolution of a cartesian
problem will be non-trivial. In particular, we rely on a
non-trivial termination argument, introduced next.

\subsection{Bounding Interactions} Here we provide an upper
bound on the size of solutions to cartesian problems -- this
relies on our earlier work on the size of interactions in
Hyland-Ong games
\cite{DBLP:conf/fossacs/Clairambault11,DBLP:conf/tlca/Clairambault13,DBLP:journals/corr/Clairambault15},
which we shall import into the realm of thin concurrent games.

\subsubsection{Structural maps in strategies.} In this endeavour, our
first step will be to formalize what it means for one configuration to
be an \emph{expansion} of another, as in the left and right hand sides
of Figure \ref{fig:struct_pb}. This involves adapting \emph{structural
morphisms} to strategies:

\begin{defi}
Consider $A, B$ mixed boards, $\sigma : A \vdash B$ a strategy,
$x^\sigma, y^\sigma \in \conf{\sigma}$.

A \textbf{partial structural map} is a function $f : x^\sigma \to
y^\sigma$ satisfying:
\[
\begin{array}{rl}
\text{\emph{min-preserving:}} & 
\text{for all $s \in \min(x^\sigma)$, then $f\,s \in \min(y^\sigma)$,}\\
\text{\emph{$\imc$-preserving:}} &
\text{for all $s \imc_\sigma t$, then $f\,s \imc_\sigma f\,t$,}\\
\text{\emph{valid:}} &
\text{$\pr_\sigma\,f$ has form $f_A \vdash f_B$ for $f_A : x^\sigma_A
\strto y^\sigma_A$ and $f_B : x^\sigma_B \strto y^\sigma_B$,}
\end{array}
\]
where $\pr_\sigma\,f$, the \textbf{display} of $f$ to $A\vdash B$, is
obtained as the composition
\[
x^\sigma_A \vdash x^\sigma_B
\quad \stackrel{\pr_\sigma^{-1}}{\bij} \quad
x^\sigma
\quad \stackrel{f}{\to} \quad
y^\sigma
\quad \stackrel{\pr_\sigma}{\bij} \quad
y^\sigma_A \vdash y^\sigma_B\,.
\]

\noindent
We write $f : x^\sigma \parstrto y^\sigma$.
It is a \textbf{structural map}, written $f : x^\sigma \strto y^\sigma$, if additionally
\[
\begin{array}{rl}
\text{\emph{total:}} &
\text{if $f\,s \imc_\sigma t^+$, there is a (necessarily) unique $s
\imc_\sigma u^+$ s.t. $f(u^+) = t^+$.}
\end{array}
\]
\end{defi}

\noindent
So in essence, a structural map between configurations of a strategy is
a forest-morphism which displays to a structural map in the game, in
the sense of Definition \ref{def:str_map_game}.

Morally, we wish that a structural map can only send an event of
$\sigma$ to one that is ``the same'', in the sense of the symmetry. 
For deterministic sequential innocent strategies this follows from the
definition above (which is indeed fine-tuned for deterministic
sequential innocent strategies, but would be poorly behaved beyond
that) -- we now introduce a lemma that expresses that.
For that, we recall from
\cite{hdr} the concept of a \textbf{grounded causal chain (gcc)} of a
strategy $\sigma$: it is a finite set $\rho = \{\rho_1, \ldots,
\rho_n\} \subseteq_f \ev{\sigma}$ that forms
\[
\rho_1 \imc_\sigma \ldots \imc_\sigma \rho_n
\]
a sequence with $\rho_1 \in \min(\sigma)$. In the general framework of \cite{hdr}, a
gcc may not be a configuration as it may not be down-closed. But here,
because strategies are sequential innocent, we have $\rho \in
\conf{\sigma}$ indeed -- it is then simply a \emph{branch} of $\sigma$.
We write $\gcc(\sigma)$ for the set of gccs of $\sigma$. Additionally,
if $x^\sigma \in \conf{\sigma}$, we write $\gcc(x^\sigma)$ for the set
of gccs within $x^\sigma$.  

\begin{lem}\label{lem:contr_br_sym}
Consider $A, B$ mixed boards, $\sigma : A \vdash B$ a strategy,
$x^\sigma, y^\sigma \in \conf{\sigma}$, and $f : x^\sigma \parstrto
y^\sigma$ a partial structural map.

Then, for any $\rho \in \gcc(x^\sigma)$, $f$ induces a symmetry
$f_{\restrict\rho} : \rho \sym_\sigma f\,\rho$.
\end{lem}
\begin{proof}
By induction on $\rho$. If $\rho$ is empty, then this is obvious.

Consider first $\rho \imc s_1^- \imc s_2^+ \in \gcc(x^\sigma)$. By induction
hypothesis, we have
\[
f : \rho \imc s_1^- \sym_\sigma f\,\rho \imc f\,s_1^-
\]
a symmetry, which extends on the left to $\rho \imc s_1^- \imc s_2^+$,
a configuration since $\sigma$ is deterministic sequential innocent.
Now by symmetry, there must be an extension 
\[
f \cup \{(s_2^+, t_2^+)\} : \rho \imc s_1^- \imc s_2^+ \sym_\sigma
f\,\rho \imc f\,s_1^- \imc t_2^+
\]
with in particular $f\,s_1^- \imc t_2^+$. But now we also have
$f\,s_1^- \imc f\,s_2^+$ since $f$ is rigid; and this implies $f\,s_2^+
= t_2^+$ since $\sigma$ is dsinn. So overall, we have, as required:
\[
f : \rho \imc s_1^- \imc s_2^+ \sym_\sigma
f\,\rho \imc f\,s_1^- \imc f\,s_2^+\,.
\]

Consider now $\rho \imc s_1^+ \imc s_2^- \in \gcc(x^\sigma)$. By induction
hypothesis,
\[
f : \rho \imc s_1^+ \sym_\sigma f\,\rho \imc f\,s_1^+
\]
is a symmetry. But also, $f\,s_1^+ \imc f\,s_2^-$ since $f$ preserves
$\imc$. But now, we argue that
\[
\pr_\sigma\,f : \pr_\sigma (\rho \imc s_1^+) \imc
\pr_\sigma\,s_2^-
\sym_{A\vdash B}
\pr_\sigma (\rho \imc s_1^+) \imc 
\pr_\sigma\,(f\,s_2^-)
\]
by \emph{transparent}, because as a structural map, $\pr_\sigma\,f$
preserves labels. It follows that
\[
f : \rho \imc s_1^+ \imc s_2^+ \sym_\sigma f\,\rho \imc f\,s_1^+ \imc
f\,s_2^-
\]
by \emph{$\sim$-receptive}, as required.

The last case is the
same where $s_2^-$ is minimal; which is similar but simpler.
\end{proof}

\subsubsection{Pointer structures.} In
\cite{DBLP:journals/apal/ClairambaultH10}, the present author together
with Harmer studied the termination of the simply-typed
$\lambda$-calculus, through the lense of game semantics. In particular,
they proved that any interaction between finite innocent strategies (in
the traditional sense of Hyland-Ong games
\cite{DBLP:journals/iandc/HylandO00}) must be finite. This was later
refined by Clairambault into a quantitative bound
\cite{DBLP:conf/fossacs/Clairambault11,DBLP:conf/tlca/Clairambault13,DBLP:journals/corr/Clairambault15},
that we shall import here into concurrent games.
As we rely on this result, we must first provide a reminder.

The result is more adequately phrased in terms of \emph{pointer
structures}:

\begin{defi}
A \textbf{pointer structure} is the data of a natural number $n\in \N$
together with 
\[
\phi : \{1, \ldots, n-1\} \to \{0, \ldots, n-2\}
\]
a \textbf{pointer} function, which is:
\[
\begin{array}{rl}
\text{\emph{contractive:}} & \text{for all $i \in \{1, \ldots, n-1\}$,
$\phi(i) < i$,}\\
\text{\emph{alternating:}} & \text{for all $i \in \{1, \ldots, n-1\}$,
$i$ is even iff $\phi(i)$ is odd.}
\end{array}
\]
\end{defi}

\noindent
Pointer structures are what remain from Hyland-Ong plays when forgetting
the identity of moves in arenas, and only remembering the
\emph{pointers}. We give an example below, pictured from left to right.
Instead of writing integers, we only write $\circ$ for even numbers
(reminiscent of Opponent moves) and $\bullet$ for odd numbers
(reminiscent of Player moves),
\[
\xymatrix{
\circ & 
\bullet	\ar@{-}[l]&
\circ	\ar@{-}[l]&
\bullet	\ar@{-}@/_1pc/[lll]&
\circ	\ar@{-}[l]&
\bullet	\ar@{-}@/^1pc/[lll]&
\circ	\ar@{-}@/^2pc/[lllll]
}
\]

\bigskip
\noindent and the function $\phi$ is indicated by following the edges from right to left. 

Not all these pointer structures can arise in interactions between
strategies; only those that satisfy an additional \emph{visibility}
condition. Given a pointer structure $\phi$, we define
\[
\scalebox{.85}{$
\begin{array}{rcrcl}
\pview{-} &:& \{0, \ldots, n-1\} &\to& \P(\{0, \ldots, n-1\})\\
 && 0 &\mapsto &\{0\}\\
 && 2k+1 &\mapsto& \{2k+1\} \cup \pview{2k}\\
 && 2k+2 &\mapsto& \{2k+2\} \cup \pview{\phi(2k+2)}
\end{array}$}
\hfill
\scalebox{.85}{$
\begin{array}{rcrcl}
\oview{-} &:& \{0, \ldots, n-1\} &\to& \P(\{0, \ldots, n-1\})\\
 && 0 &\mapsto &\{0\}\\
 && 2k+1 &\mapsto& \{2k+1\} \cup \oview{\phi(2k+1)}\\
 && 2k+2 &\mapsto& \{2k+2\} \cup \oview{2k+1}
\end{array}$}
\]
respectively called the \textbf{$P$-view} and the \textbf{$O$-view}.
The $P$-view captures the part of a play available to an innocent
strategy, while the $O$-view captures the part of a play available to
an innocent environment. A pointer structure is \textbf{visible} when
for all $1 \leq i \leq n-1$, we have $\phi(i) \in \pview{i}$ when $i$
is odd and $\phi(i) \in \oview{i}$ when $i$ is even -- visible pointer
structures are exactly those that may arise as an interaction between
innocent strategies.

Finally, in a pointer structure $\phi : \{1, \ldots, n-1\} \to \{0,
\ldots, n-2\}$, the \textbf{length} $|\phi|$ of $\phi$ is simply $n$. 
The \textbf{depth} of $1 \leq i
\leq n-1$ is the $k\in \N$ such that $\phi^k(i) = 0$; we extend this to
$0 \leq i \leq n-1$ by stating that the depth of $0$ is $0$. The
\textbf{depth} of a pointer structure $\phi$ is the maximal depth of $0
\leq i \leq n-1$. The \textbf{$P$-size} of $\phi$ is the minimal $N$
such that for all $0 \leq i \leq n-1$, we have $\# \pview{i} \leq
2N$, the \textbf{$O$-size} of $\phi$ is the minimal $N$ such that for
all $0 \leq i \leq n-1$, we have $\# \oview{i} \leq 2N+1$ -- the depth
roughly corresponds to the order of the type, the $P$-size and $O$-size
to the size of the two interacting strategies. 

Then, we have \cite[Theorem 4.17]{DBLP:journals/corr/Clairambault15}:
\begin{thm}\label{thm:main_bound}
Consider $\phi$ visible pointer structure of depth bounded by $d\geq 3$,
$P$-size bounded by $n\geq 1$ a $O$-size bounded by $p\geq 1$.
Then, 
writing $2_0(N) = N$ and $2_{d+1}(N) = 2^{2_d(N)}$, 
\[
|\phi| 
\quad\leq\quad
2_{d-3}\left(\frac{p^{n+1}-1}{p-1}-1\right)\,,
\]
additionally $|\phi| = 1$ if $d=0$, $|\phi| = 2$ if $d=1$, and
$|\phi|\leq 2n$ if $d=2$.
\end{thm}

\noindent
In \cite{DBLP:journals/corr/Clairambault15}, this bound is also shown
to be asymptotically tight. Though for the work here, this precise
bound does not matter: what matters is that it exists.

\subsubsection{The upper bound}
We now move back to the technical setting of this paper, for
interacting $\sigma : A \vdash B$, $\tau : B \vdash C$ with 
$x^\sigma \in \conf{\sigma}$, $x^\tau \in \conf{\tau}$ and a cartesian
morphism $\chi : x^\sigma_B \carmor x^\tau_B$, we want an
upper bound to the size of all solutions to the cartesian problem.

We call the \textbf{$\tau$-size} of $(x^\sigma,\chi, x^\tau)$ the
minimal $n$ s.t. every gcc of $x^\sigma_A \parallel x^\tau$ is smaller
than $2n$; its \textbf{$\sigma$-size} the minimal $p$ such that every
gcc of $x^\sigma \parallel x^\tau_C$ is smaller than $2p$; its
\textbf{depth} the minimal $d+2$ such that every gcc of $x^\tau_C$ is
smaller than $d+2$, every gcc of $x^\sigma_B, x^\tau_B$ is smaller than
$d+1$, and every gcc of $x^\sigma_A$ is smaller than $d$.  Finally, its
\textbf{branching degree} is the minimal $b$ such that (regarded as
trees), $x^\sigma$ and $x^\tau$ have branching degree smaller than $b$.

\begin{lem}\label{lem:bound_sol}
Consider $(x^\sigma \in \conf{\sigma}, \chi, x^\tau \in \conf{\tau})$
as above with $\tau$-size less than $n\geq 1$, $\sigma$-size less than
$p \geq 1$, depth less than $d\geq 3$ and branching degree less than
$b\geq 2$. 

Then, for any $y^\sigma \in \conf{\sigma}$ and
$y^\tau \in \conf{\tau}$ matching such that there are partial structural maps
$\chi^\sigma : y^\sigma \parstrto x^\sigma$, $\chi^\tau : y^\tau
\parstrto x^\tau$ with $\chi^\sigma_A : y^\sigma_A \pnstrto x^\sigma_A$
and $\chi^\tau_C : y^\tau_C \ppstrto x^\tau_C$, we have
\[
\# (y^\tau \inter y^\sigma) 
\quad\leq\quad
b^{2_{d-3}\left(\frac{p^{n+1}-1}{p-1}-1\right)}\,.
\]
\end{lem}
\begin{proof}
Firstly, because $\sigma$ and $\tau$ are deterministic sequential
innocent, it follows that $y^\tau \inter y^\sigma$ is a forest.
Consider one of its branches $\rho \in \gcc(y^\tau \inter y^\sigma)$,
written as
\[
\rho_0 \imc \rho_1 \imc \ldots \imc \rho_{l-1}
\]
for $l$ its length. For each $1\leq i \leq l-1$, if $\pr_{\tau \inter
\sigma}\,\rho_i$ is non-minimal in $A \parallel B \parallel C$, then
its unique immediate predecessor is some $\pr_{\tau \inter
\sigma}\,\rho_j$ for $j < i$, we set $\phi(i) = j$. If $\pr_{\tau
\inter \sigma}\,\rho_i$ is minimal in $A$, we must have
$(\rho_i)_\sigma$ defined. Because $y^\sigma$ is a forest, there is a
unique minimal $s \in y^\sigma$ such that $s <_\sigma (\rho_i)_\sigma$,
and it must correspond to some unique $\rho_j$ (such that
$(\rho_j)_\sigma = s$), with $j < i$ -- we set $\phi(i) = j$. If
$\pr_{\tau \inter \sigma}\,\rho_i$ is minimal in $B$ we set $\phi(i) =
0$, and it cannot be minimal in $C$ because $\rho_0$ is, and
$y^\sigma_C$ (thus $y^\sigma$) has a unique minimal event since $C$ is
strict\footnote{This is the definition of the
\emph{justifier} in an interaction between visible strategies
\cite[Section~10.2.2]{hdr}.}.

Then, it follows by \cite[Proposition 10.2.5]{hdr} that $\phi$ is a
visible pointer structure; that its $O$-views correspond to gccs
in $y^\sigma \parallel y^\tau_C$, and that its $P$-views correspond
to gccs in $y^\sigma_A \parallel y^\tau$. But because $\chi^\sigma$ and
$\chi^\tau$ and their displays $\chi^\sigma_A$, $\chi^\tau_C$ are
structural maps and hence forest morphisms, those are bounded
respectively by $2p$ and $2n$. Likewise, by construction its depth is
smaller than $d$. Hence, we may apply Theorem \ref{thm:main_bound} and
deduce that we have
\begin{eqnarray}
\# \rho 
&\leq&
2_{d-3}\left(\frac{p^{n+1}-1}{p-1}-1\right)\,.
\label{eqn:up_bound}
\end{eqnarray}

\noindent
So $y^\tau \inter y^\sigma$ is a forest whose depth is bounded by this
quantity. To deduce a bound on the size of $y^\tau \inter y^\sigma$, we
give a corresponding upper bound to the branching degree of that
forest. For that, consider $p \in y^\tau \inter y^\sigma$.
We prove that the branching at $p$ is bounded by $b$, reasoning by
cases on the polarity of $p$ and its component of occurrence.

If $p$ has polarity $\ell$ and occurs in $A$, then reasoning by cases
via \cite[Lemma 6.2.15]{hdr}, any $p \imc_{\tau \inter \sigma} p'$ must
satisfy that $p'$ has polarity $-$, $p'_\sigma$ is defined and
\[
\pr_{\tau \inter \sigma}(p_\sigma) \imc_{A\parallel B \parallel C}
\pr_{\tau \inter \sigma}(p'_\sigma)
\]
so that in particular, $p_\sigma \imc_\sigma p'_\sigma$ by \cite[Lemma
6.1.16]{hdr}. Since $\chi^\sigma$ is rigid,
\[
\chi(p_\sigma) \imc_\sigma \chi(p'_\sigma)
\]
as well. Hence, this defines a map from successors of $p$ in $y^\tau
\inter y^\sigma$ to successors of $\chi(p_\sigma)$ in $x^\sigma$. We
show this map is injective; for that, consider $p \imc_{\tau \inter
\sigma} p', p''$ with $\chi(p'_\sigma) = \chi(p''_\sigma)$. In
particular, $\pr_\sigma(\chi(p'_\sigma)) =
\pr_\sigma(\chi(p''_\sigma))$ so they have the same label, index and
predecessor. By \emph{rigidity}, \emph{label-preserving} and
\emph{negative}, it follows that $\pr_\sigma(p'_\sigma)$ and
$\pr_\sigma(p''_\sigma)$ have the same predecessor, label and copy
index, so that $\pr_\sigma(p'_\sigma) = \pr_\sigma(p''_\sigma)$ by
\emph{jointly injective} \cite[Definition 12.1.1]{hdr}. By
\emph{receptivity} of $\sigma$, it follows that $p'_\sigma =
p''_\sigma$, so that $p' = p''$ also by \emph{local injectivity} of
$\pr_\sigma$. It follows that the set of causal successors of $p$ in
$y^\tau \inter y^\sigma$ has cardinal less or equal than $B$. The case
where $p$ has polarity $\err$ and occurs in $C$ is symmetric.

If $p$ has polarity $-$, say \emph{e.g.} that it occurs in $C$. Then,
reasoning by cases based on \cite[Lemma 6.2.15]{hdr}, any $p \imc_{\tau
\inter \sigma} p'$ in $y^\tau \inter y^\sigma$ must have polarity
$\err$ and satisfy $p_\tau \imc_\tau p'_\tau$. By \emph{sequential
innocence}, there can be at most one such $p'_\tau$, so there is at
most one such $p'$ in $\tau \inter \sigma$. If $p$ has polarity $\ell$
and occurs in $B$, then again reasoning by cases on \cite[Lemma
6.2.15]{hdr}, any $p \imc_{\tau \inter \sigma} p'$ in $y^\tau \inter
y^\sigma$ satisfies $p_\tau \imc_\tau p'_\tau$, and the same reasoning
applies. The final case where $p$ has polarity $\err$ and occurs in $B$
is symmetric. 

Altogether, $y^\tau \inter y^\sigma$ is a forest of depth bounded by
\eqref{eqn:up_bound} and branching bounded by $b$, giving the announced
upper bound on its size.
\end{proof}

\noindent
This upper bound could be improved for $d = 1, 2$, but again the
precise quantity does not really matter here; only that once $x^\sigma$
and $x^\tau$ are fixed, the solutions to the cartesian problems have a
bounded size. Note also the importance of the assumptions that
$\chi^\sigma_A$ is negative and $\chi^\tau_C$ positive: this is what
ensures that $y^\sigma$ and $y^\tau$ do not have more duplications by
the external Opponent than in $x^\sigma$ and $x^\tau$, meaning that the
upper bound to the branching degree of $x^\sigma, x^\tau$ transports to
$y^\sigma, y^\tau$ and to $y^\tau \inter y^\sigma$. Finally, we note
that as $\#y^\sigma \leq \# (y^\tau \inter y^\sigma)$ and likewise for
$\tau$, the same upper bound applies to $y^\sigma$ and $y^\tau$.

\subsection{Functorial Collapse}
\label{subsec:scottl_funct}
With this, we have finished introducing the essential ingredients to
show functoriality of the collapse to the linear Scott model. 

\subsubsection{Cartesian problems and how to solve them.}
First, the tools introduced just above now allow us to be more precise
about what we mean by \emph{cartesian problem}. 

\begin{defi}
Consider $\sigma : A \vdash B$ and $\tau : B \vdash C$.
A \textbf{cartesian (matching) problem} is the data of $x^\sigma \in
\conf{\sigma}$, $x^\tau \in \conf{\tau}$ and a cartesian morphism $\chi
: x^\sigma_B \carmor x^\tau_B$.

A \textbf{solution} to this problem is given by $y^\sigma \in
\conf{\sigma}, y^\tau \in
\conf{\tau}$, $\chi^\sigma, \chi^\tau$ such that:
\[
\xymatrix{
x^\sigma & x^\sigma_B
        \ar@{<~>}[rr]|{\chi}^{-+}&&
x^\tau_B&x^\tau\\
&y^\sigma
        \ar@{~>}[ul]^{\chi^\sigma}&
y_B     \ar@{~>}[ul]|{\chi^\sigma_B}
        \ar@{}[u]|{\rotatebox{90}{$\subseteq$}}
        \ar@{~>}[ur]|{\chi^\tau_B}&
y^\tau  \ar@{~>}[ur]_{\chi^\tau}
}
\]
with $\chi^\sigma_A : y^\sigma_A \nstrto x^\sigma_A$ and $\chi^\tau_C :
y^\tau_C \pstrto x^\tau_C$.
\end{defi}

\noindent
This captures the notion of cartesian matching problem introduced in
Section~\ref{subsec:intro}. First, indeed we have $y^\sigma_A \pre
x^\sigma_A$ and $x^\tau_C \pre y^\tau_C$, so that this solution will
provide the required witness for functoriality. But we have a little
bit more here: we know that $y^\sigma$ is an \emph{expansion} of
$x^\sigma$ and likewise $y^\tau$ is an expansion of $x^\tau$; this is
witnessed by specific structural morphisms $\chi^\sigma$ and
$\chi^\tau$ whose display is compatible with the cartesian morphism
$\chi$. 

Now, the next key proposition shows how to solve cartesian problems:

\begin{prop}\label{prop:main_link}
Consider $\sigma : A \vdash B$ and $\tau : B \vdash C$ deterministic
sequential innocent.

Then any cartesian problem for $\sigma, \tau$ has a unique
solution. 
\end{prop}
\begin{proof}
Consider $x^\sigma \in \conf{\sigma}, x^\tau \in \conf{\tau}$ and $\chi
: x^\sigma_B \carmor x^\tau_B$ a cartesian problem.

First, note we can find a \emph{partial solution}, \emph{i.e.}
$y^\sigma \in \conf{\sigma}$, $y^\tau \in \conf{\tau}$, $\chi^\sigma$
and $\chi^\tau$ s.t.:
\[
\xymatrix{
x^\sigma & x^\sigma_B
        \ar@{<~>}[rr]|{\chi}^{-+}&&
x^\tau_B&x^\tau\\
&y^\sigma
        \ar@{~>}[ul]^{\chi^\sigma}|p&
y_B     \ar@{~>}[ul]|{\chi^\sigma_B}
        \ar@{}[u]|{\rotatebox{90}{$\subseteq$}}
        \ar@{~>}[ur]|{\chi^\tau_B}&
y^\tau  \ar@{~>}[ur]_{\chi^\tau}|p
}
\]
with $\chi^\sigma_A : y^\sigma_A \pnstrto x^\sigma_A$ and
$\chi^\tau_C : y^\tau_C \ppstrto x^\tau_C$ -- indeed one can take
$y^\sigma = y^\tau = \emptyset$. Such partial solutions are partially
ordered by componentwise inclusion.  By Lemma \ref{lem:bound_sol},
there is a bound $N \in \mathbb{N}$ on the cardinal of $y^\tau \inter
y^\sigma$ for partial solutions; thus there is a partial solution of maximal
size. From now on, we fix a partial solution $y^\sigma, y^\tau,
\chi^\sigma, \chi^\tau$ of maximal size and prove that it actually is a
total solution as desired.

First, we prove that $\chi^\tau : y^\tau \pstrto x^\tau$ is
\emph{total}. Consider $\chi^\tau\,s \imc_\tau t^+$. Assuming there is
no $s \imc_\tau u^+$ in $y^\tau$ such that $\chi^\tau\,u^+ = t^+$, we
shall construct an extension of the solution, contradicting its
maximality.  We start with the easy case where $t^+$ occurs in $C$.
Since $\tau$ is a forest, $[s]_\tau$ is a gcc. By Lemma
\ref{lem:contr_br_sym}, this means that the restriction of $\chi^\tau$
to $[s]_\tau$ is $\theta : [s]_\tau \sym_\tau [\chi^\tau\,s]_\tau$ a
symmetry on $\tau$. Hence, there is $s
\imc_\tau u^+$ in $\tau$ such that $\theta \cup \{(u^+, t^+)\} \in
\tilde{\tau}$ is still a symmetry.
Extend $y^\tau$ with $u^+$ and $\chi^\tau(u^+) = t^+$, it is a direct
verification that this yields a solution, contradicting maximality.

Now, let us assume that $t^+$ occurs in $B$, \emph{i.e.} $\pr_\tau(t^+)
= (1, b')$. First, we update $y^\tau$ and $\chi^\tau$ as above. This
means that we have $\pr_\tau(u^+) = (1, b)$ with $b$ negative in $B$,
with $y_B \vdash_B b$, and $\chi^\tau_B(b) = b'$. By  \emph{receptive},
there is a unique $y^\sigma \vdash v^-$ such that $\pr_\sigma(v^-) =
(2, b)$. We must now extend $\chi^\sigma$ accordingly. For that, assume
\emph{w.l.o.g.} that $b$ is not minimal -- the minimal case is similar
but simpler. So, there is a (unique) $a \imc_B b$. By \emph{locally
injective}, there is a unique $n^+ \in y^\sigma$ such that
$\pr_\sigma(n^+) = (2, a)$. Write $m^+ = \chi^\sigma(n^+)$ with
$\pr_\sigma(m^+) = (2, c)$ with $c \in x^\sigma_B$ -- by definition, $c
= \chi^\sigma_B(a)$. If we also write $d = \chi^\tau_B(a)$, then we
have $(c, d) \in \chi_B$ by hypothesis, and by
\emph{$\imc$-preserving}, $d \imc_B b'$ negative. Thus, by Lemma
\ref{lem:main_linkings}, there is a unique $c \imc_B e^-$ in
$x^\sigma_B$ such that $(e^-, b') \in \chi_B$. By \emph{locally
injective}, there is a unique $k^- \in x^\sigma$ such that
$\pr_\sigma(k^-) = (2, e^-)$. Then, we may finally extend $\chi^\sigma$
with $v^- \mapsto e^-$, and verify the required
conditions. Symmetrically,
$\chi^\sigma : y^\sigma \pstrto x^\sigma$ is
\emph{total} as well.

Finally, it remains to prove that $\chi^\sigma_A$ and $\chi^\tau_C$ are
not partial; first we check that $\chi^\tau_C$ is \emph{$-$-total}. So
consider $a \in y^\tau_C$ and $\chi^\tau_C\,a \imc_C c^-$ -- the case
where $c^-$ is minimal is similar but simpler. If $c^-$ has no
predecessor for $\chi^\tau_C$, then Lemma \ref{lem:lift_neg} provides a
unique extension of $y^\tau_C$ which yields an extension of $y^\tau$ by
receptivity, contradicting maximality. Likewise, the
\emph{$+$-totality} of $\chi^\sigma_A$ follows from Lemma
\ref{lem:lift_pos}.

For \emph{uniqueness}, assume we have two solutions, \emph{i.e.}
\[
\xymatrix{
&v^\sigma
        \ar@{~>}[dl]_{\xi^\sigma}
&v_B    \ar@{~>}[dl]_{\xi^\sigma_B}
        \ar@{~>}[dr]^{\xi^\tau_B}
        \ar@{}[d]|{\rotatebox{270}{$\subseteq$}}
&v^\tau \ar@{~>}[dr]^{\xi^\tau}\\
x^\sigma & x^\sigma_B
        \ar@{<~>}[rr]|{\chi_B}^{+-}&&
x^\tau_B&x^\tau\\
&y^\sigma
        \ar@{~>}[ul]^{\chi^\sigma}&
y_B     \ar@{~>}[ul]|{\chi^\sigma_B}
        \ar@{}[u]|{\rotatebox{90}{$\subseteq$}}
        \ar@{~>}[ur]|{\chi^\tau_B}&
y^\tau  \ar@{~>}[ur]_{\chi^\tau}
}
\]
and consider \emph{e.g.} an event $p \in v^\tau \inter v^\sigma$ which
is minimal such that it is not in $y^\tau \inter y^\sigma$. Now we
reason by cases: if $p$ is positive for $\sigma$, consider its
immediate predecessor $q \imc_{\tau \inter \sigma} p$. Necessarily, $q
\in y^\tau \inter y^\sigma$ and $q_\sigma \imc_\sigma p_\sigma$. But
then since $\chi^\sigma$ is \emph{total} and $\sigma$ is sequential
innocent, we have $p_\sigma \in y^\sigma$ as well. But since
$y^\sigma_B = y^\tau_B$ we must have some $m \in y^\tau\inter y^\sigma$
such that $m_\sigma = p_\sigma$. But then it is a consequence of
sequential innocence of $\sigma$ and $\tau$ that $q \imc_{\tau \inter
\sigma} m$; and so that $m = q$.
If $p$ is positive for $\tau$, the situation is symmetric. If $p$ is
negative in $A$, then it must be also in $y^\tau \inter y^\sigma$
because
$\chi^\sigma_A$ is \emph{$-$-preserving} and \emph{$-$-total} and by
receptivity of $\sigma$. If $p$ is negative in $C$, then the situation
is symmetric.
\end{proof}

\noindent
Thus as illustrated in Figure \ref{fig:struct_pb}, one can always find
solutions to cartesian problems: two deterministic sequential innocent
strategies trying to duplicate and erase each other alongside a
cartesian morphism can always find a successful resolution, and this
resolution is unique if one adequately takes into account the copy
indices. 

\subsubsection{Functorial collapse.}
The property above is the key conceptual contribution of this work. The
functoriality of the collapse to the linear Scott model immediately
follows.

\begin{thm}\label{thm:main_collapse}
We have a $\sim$-functor
\[
\scoll{-} : \Dsinn \to \Scott
\]
\end{thm}
\begin{proof}
Preservation of identities was handled in Lemma \ref{lem:scoll_pres_id}
and oplax preservation of composition in Lemma \ref{lem:scoll_oplax}.
For lax preservation of composition, consider $(\x_A, \x_B) \in
\scoll{\sigma}$ and $(\x_B, \x_C) \in \scoll{\tau}$. By definition,
this means that taking some representatives $x_A \in \x_A, x_B \in
\x_B$ and $x_C \in \x_C$, there are witnesses $x^\sigma
\in\confp{\sigma}$ and $x^\tau \in \confp{\tau}$ such that
\[
x^\sigma_A \pre x_A\,,
\qquad
x_B \pre x^\sigma_B\,,
\qquad
x^\tau_B \pre x_B\,,
\qquad
x_C \pre x^\tau_C\,,
\]
meaning in particular that $x^\tau_B \pre x^\sigma_B$, which means that
there is a cartesian morphism $\chi : x^\sigma_B \carmor x^\tau_B$.
This forms a cartesian problem, which we can solve using Proposition
\ref{prop:main_link}: the solution consists in $y^\sigma \in \conf{\sigma}, y^\tau
\in \conf{\tau}$ matching, with $\chi^\sigma, \chi^\tau$ such that: 
\[
\xymatrix{
x^\sigma & x^\sigma_B
        \ar@{<~>}[rr]^{\chi}&&
x^\tau_B&x^\tau\\
&y^\sigma
        \ar@{~>}[ul]^{\chi^\sigma}&
y_B     \ar@{~>}[ul]|{\chi^\sigma_B}
        \ar@{}[u]|{\rotatebox{90}{$\subseteq$}}
        \ar@{~>}[ur]|{\chi^\tau_B}&
y^\tau  \ar@{~>}[ur]_{\chi^\tau}
}
\]
with $\chi^\sigma_A : y^\sigma_A \nstrto x^\sigma_A$ and
$\chi^\tau_C : y^\tau_C \pstrto x^\tau_C$. But then, we may form
$y^\tau \odot y^\sigma \in \confp{\tau \odot \sigma}$, with 
\[
(y^\tau \odot y^\sigma)_A \nstrto x^\sigma_A \pre x_A\,,
\qquad
x_C \pre x^\tau_C \pstrot y^\tau_C = (y^\tau \odot y^\sigma)_C
\]
yielding a witness for $(\x_A, \x_C) \in \scoll{\tau \odot \sigma}$ as
required.
\end{proof}

\subsection{Further Structure} The above takes care of the main
obstacle in linking thin concurrent games with the linear Scott model,
but there is still some work to do to conclude.

\subsubsection{A relative Seely functor} First, the functor of
Theorem \ref{thm:main_collapse} extends to a relative Seely functor. To
show this we must adjoin to it a number of structural isomorphisms,
which are simply (the down-closure of) those of Figure
\ref{fig:str_coll_iso}. All required conditions are straightforward,
save for the preservation of promotion which we detail here.

\begin{lem}
Consider $S, A$ mixed boards with $S$ strict, and $\sigma : \oc S
\vdash A$. 

Then, the following diagram commutes (with 
$t^\oc_S$ the down-closure of $s^\oc_S$):
\[
\xymatrix@C=40pt{
\oc \scoll{S} \ar[r]^{(\scoll{\sigma}\,\circ\,t_S^{\oc})^{\dagger}}
        \ar[d]_{t_S^{\oc}}&
\oc \scoll{T} \ar[d]^{t^{\oc}_T}\\
\scoll{\oc S} \ar[r]_{\scoll{\sigma^{\dagger}}}&
\scoll{\oc T}
}
\]
\end{lem}
\begin{proof}
First, as $(\scoll{\sigma}\,\circ\,t_S^{\oc})^{\dagger}$ and
$\scoll{\sigma^{\dagger}}$ are down-closed, it is equivalent to compose
them with the preorder-isomorphisms $s^\oc_S$ and $s^\oc_T$, which are
bijections between multisets (quotiented lists) and symmetry classes of
configurations of $\oc T$ (quotiented families), for which we use
similar notations, thus we will deal with these bijections implicitly
in this proof.

Thus, take $([\x_{A,l} \mid l \in L], [\y_{B,k} \mid k \in K]) \in
(\scoll{\sigma}\circ s^\oc_S)^{\dagger}$. This means that for all $k
\in K$, we have $([\x_{A,l} \mid l \in L], \y_{B,k}) \in
\scoll{\sigma}$. So there is $\famc{x_{A,l} \mid l \in L}$ and $y_{B,k}
\in \y_{B,k}$ such that  
\[
y_{B,k} ~\pre~ x^{\sigma, k}_B\,,
\qquad\qquad
\famc{x^{\sigma,k}_{A,j} \mid j \in J_k} ~\pre~ 
\famc{x_{A,l} \mid l \in L}
\]
for some $x^{\sigma, k} \in \confp{\sigma}$ displaying to
$\famc{x^{\sigma,k}_{A,j} \mid j \in J_k} \vdash x^{\sigma,k}_B$. 
We may form $\famc{x^{\sigma,k} \mid k \in K} \in
\confp{\sigma^\dagger}$ displaying by definition to 
$\famc{x^{\sigma,k}_{A,j} \mid \tuple{k,j} \in \Sigma_{k\in K} J_k}
\vdash \famc{x^{\sigma,k}_B \mid k \in K}$ and
\[
\famc{y_{B,k} \mid k\in K} ~\pre~ \famc{x^{\sigma,k}_B \mid k \in K}\,,
\quad
\famc{x^{\sigma,k}_{A,j} \mid \tuple{k,j} \in \Sigma_{k\in K} J_k}
~\pre~
\famc{x_{A,l} \mid l \in L}
\]
as needed. Reciprocally, take $([\x_{A,l} \mid l \in L], [\y_{B,k}
\mid k \in K]) \in \scoll{\sigma^\dagger}$. By definition, there are
\[
\famc{x_{A,l} \mid l \in L} ~\in~[\x_{A,l} \mid l \in L]\,,
\qquad
\famc{y_{B,k} \mid k \in K} ~\in~[\y_{B,k} \mid k \in K]
\]
representatives witnessed by $\sigma^\dagger$, \emph{i.e.} there is
$\famc{x^{\sigma,i} \mid i \in I} \in \confp{\sigma^{\dagger}}$, such
that 
\[
\famc{x^{\sigma, i}_{A, j} \mid j \in J_i}
~\pre~
\famc{x_{A,l} \mid l \in L}\,,
\qquad
\famc{y_{B,k} \mid k \in K} ~\pre~
\famc{x^{\sigma, i}_B \mid i \in I}
\]
where $\pr_{\sigma^\dagger}\,\famc{x^{\sigma,i} \mid i \in I} = 
\famc{x^{\sigma, i}_{A, j} \mid j \in J_i} \vdash \famc{x^{\sigma, i}_B
\mid i \in I}$. But by Lemma \ref{lem:carac_pre} (on representatives),
this means that for all $k \in K$ there is $i\in I$ such that $y_{B,k}
\pre x^{\sigma,i}_B$. Thus for all $k \in K$, there is $i \in I$ such
that $x^{\sigma, i}$ witnesses that $([\x_{A,l} \mid l \in L],
\y_{B,k}) \in \scoll{\sigma}^\dagger$. 
\end{proof}

\noindent
Altogether, we have the following theorem:

\begin{thm}
The above provide the components for a relative Seely functor:
\[
\scoll{-} : \Dsinn \to \ScottL\,.
\]
\end{thm}

\subsubsection{Collapse in colors} Finally, we must now extend this
with colors -- this essentially follows the pattern of
Section~\ref{subsec:colors}. Most ingredients are in place already: the
relational collapse of strategies in colors was introduced in
Section~\ref{subsubsec:color_coll}, and in particular in
\eqref{eq:color_coll}.  As without colors, this collapse is adapted to
target the linear Scott model by taking the down-closure for the
preorder with colors introduced in
Section~\ref{subsubsec:colored_preorder}.

Thus, we simply restate the definition in \eqref{eq:scott_coll} with
colors:
\begin{eqnarray}
\cscoll{\sigma}{\C}
&=& 
[\ccoll{\sigma}{\C}]_{\cscoll{A}{\C}^{\op} \times \cscoll{B}{\C}} 
\quad\in\quad
\ScottL[\cscoll{A}{\C}, \cscoll{B}{\C}]\,.
\label{eq:def_coll_scottl_col}
\end{eqnarray}

\noindent
With this definition, we can finally state our main theorem:

\begin{thm}\label{thm:final}
For every set $\C$ we have a relative Seely functor:
\[
\cscoll{-}{\C} \quad:\quad \Dsinn ~\to~ \ScottL\,.
\]
\end{thm}
\begin{proof}
Preservation of identity as in Lemma \ref{lem:scoll_pres_id}, via
Lemma \ref{lem:col_pres_cc}. For composition, as in
Lemma \ref{lem:scoll_oplax}, oplax functoriality follows from that for
the colorful relational collapse (Proposition
\ref{prop:color_rel_fun}). Now for lax functoriality, consider
$(\x_A, \x_C) \in \cscoll{\tau}{\C} \circ \cscoll{\sigma}{\C}$, so
there is a position in colors $\x_B$ such that $(\x_A, \x_B) \in
\cscoll{\sigma}{\C}$ and $(\x_B, \x_C) \in \cscoll{\tau}{\C}$.
Unfolding the definitions, this means that there are experiments
$(x^\sigma, \lambda^\sigma) \in \cconfp{\sigma}{\C}$, 
$(x^\tau, \lambda^\tau) \in \cconfp{\tau}{\C}$
s.t.
\[
(x^\sigma_A, \lambda^\sigma_A) \pre (x_A, \lambda_A)\,,
\qquad
(x^\sigma_B, \lambda^\sigma_B) \carmor (x^\tau_B, \lambda^\tau_B)\,,
\qquad
(x_C, \lambda_C) \pre (x^\tau_C, \lambda^\tau_C)
\]
providing us in particular with a cartesian morphism $\chi : x^\sigma_B
\carmor x^\tau_B$ which preserves colors, \emph{i.e.} it relates moves
with the same color. By Proposition \ref{prop:main_link}, this
cartesian problem has a solution, giving us 
$y^\sigma \in
\conf{\sigma}, y^\tau \in
\conf{\tau}$, $\chi^\sigma, \chi^\tau$ such that:
\[
\xymatrix{
x^\sigma & x^\sigma_B
        \ar@{<~>}[rr]|{\chi}^{-+}&&
x^\tau_B&x^\tau\\
&y^\sigma
        \ar@{~>}[ul]^{\chi^\sigma}&
y_B     \ar@{~>}[ul]|{\chi^\sigma_B}
        \ar@{}[u]|{\rotatebox{90}{$\subseteq$}}
        \ar@{~>}[ur]|{\chi^\tau_B}&
y^\tau  \ar@{~>}[ur]_{\chi^\tau}
}
\]
which we must enrich with colors. To turn $y^\sigma$ and $y^\tau$ into
experiments we simply set
\[
\mu^\sigma(s) = \lambda^\sigma(\chi^\sigma\,s)\,,
\qquad\qquad
\mu^\tau(t) = \lambda^\tau(\chi^\tau\,t)\,,
\]
as $\chi^\sigma$ and $\chi^\tau$ are forest morphisms it is
straightforward that those are valid experiments. Furthermore,
$\mu^\sigma$ and $\mu^\tau$ are matching because in the inclusion in
the diagram above and since $\chi$ preserves colors. Hence, by Lemma
\ref{lem:col_pres_comp}, we may form an experiment
\[
(y^\tau\odot y^\sigma, \mu^\tau \odot \mu^\sigma) \in \cconfp{\tau
\odot \sigma}{\C}\,,
\]
additionally from the definition, $\chi^\sigma_A$ and $\chi^\tau_C$
preserve colors, so we still have
\[
(y^\sigma_A, \mu^\sigma_A) \pre (x_A, \lambda_A)\,,
\qquad
\qquad
(x_C, \lambda_C) \pre (y^\tau_C, \mu^\tau_C)\,,
\]
concluding lax functoriality and the fact that we have a functor.
Finally, all structural isomorphisms required for a relative Seely
functor are (the down-closure of) those in Figure
\ref{fig:col_str_coll_iso}. The proof for the coherence laws is
essentially undisturbed by colors.
\end{proof}

\noindent
Finally, from this we deduce:

\begin{cor} \label{cor:mainrelcollscott}
Consider $\Gamma \vdash M : A$ a simply-typed $\lambda$-term.

Then, the diagram commutes in $\ScottL$, for each set $\C$:
\[
\xymatrix@C=60pt{
\oc \intr{\Gamma}_{\Dsinn_\oc}
        \ar[r]^{\cscoll{\intr{M}_{\Dsinn_\oc}}{\C}}
        \ar[d]_{\oc t_\Gamma}&
\intr{A}_{\Dsinn_\oc}
        \ar[d]^{t_A}\\
\oc \intr{\Gamma}_{\ScottL_{\oc}}^{\C}
        \ar[r]_{\intr{M}_{\ScottL_\oc}^{\C}}&
\intr{A}_{\ScottL_\oc}^{\C}
}
\]
with $t_\Gamma, t_A$ the down-closure of $s_\Gamma, s_A$:
\end{cor}

\noindent
We phrase the result with $t_\Gamma, t_A$ because they are indeed the
structural isomorphisms in $\ScottL$ given through the relative Seely
functor. Note however that $s_\Gamma$ and $s_A$ are order-isomorphisms
and $\cscoll{\intr{M}_{\Dsinn_\oc}}{\C}$ and
$\intr{M}_{\ScottL_\oc}^{\C}$ are down-closed, so that the diagram also
holds if $t_\Gamma, t_A$ are simply replaced with the
order-isomorphisms $s_\Gamma, s_A$ (this glosses over the fact that the
functorial action of $\oc$ is different in $\Rel$ and $\ScottL$.
Fortunately, it is easily verified that if $\varphi : A \iso B$ is an
order-isomorphism between preorders, then $\oc^{\ScottL}
[\varphi]_{A^\op \times B} = [\oc^\Rel \varphi]_{(\oc A)^\op \times \oc
B}$; here this ensures that $\oc^{\ScottL} t_\Gamma = [\oc^{\Rel} s_\Gamma]$). 

This concludes the link between thin concurrent games and the linear
Scott model -- this is really the main result of this paper. However,
because we also know their relationship with the relational model, we
are able to leverage it to study the direct link between the relational
model and the linear Scott model, in the spirit of Ehrhard
\cite{DBLP:journals/tcs/Ehrhard12}.

\subsection{Quantitative Collapse} As announced, the interpretation of
a simply-typed $\lambda$-term in $\ScottL_\oc$ is simply the
down-closure of its interpretation in $\Rel_\oc$:

\begin{thm}\label{thm:main}
Consider $\Gamma \vdash M : A$ a simply-typed $\lambda$-term.
Then, $\intr{M}_{\ScottL_\oc} = [\intr{M}]_{\Rel_\oc}]$.
\end{thm}
\begin{proof}
This is direct via the following calculations in $\Rel$:
\begin{eqnarray*}
\intr{M}_{\ScottL_\oc} &=& s_A \circ
\cscoll{\intr{M}_{\Dsinn_\oc}}{\C} \circ \oc s_\Gamma^{-1}\\
&=& s_A \circ [\ccoll{\intr{M}_{\Dsinn_\oc}}{\C}] \circ \oc
s_\Gamma^{-1}\\
&=& [s_A \circ \ccoll{\intr{M}_{\Dsinn_\oc}}{\C} \circ \oc
s_\Gamma^{-1}]\\
&=& [\intr{M}_{\Rel_\oc}]
\end{eqnarray*}
using Corollary \ref{cor:mainrelcollscott} (inlining the remark about
phrasing it with $s_\Gamma, s_A$); then by
definition of the linear Scott collapse with colors in
\eqref{eq:def_coll_scottl_col}; then using that $s_\Gamma, s_A$ are
order-isomorphisms and hence commute with down-closure; and finally
using Corollary \ref{cor:mainrelcollcolor}. 
\end{proof}

\noindent
In \cite{DBLP:journals/tcs/Ehrhard12}, Ehrhard shows that $\ScottL_\oc$
is the \emph{extensional collapse} of $\Rel_\oc$. Ehrhard presents this
as a relatively complex categorical statement relating two cartesian
closed categories $\C$ (in this case, $\Rel_\oc$) and $\E$ (in this
case, $\ScottL_\oc$), the former intensional and the latter
extensional; roughly speaking expressing that the interpretation of the
$\lambda$-calculus in $\E$ corresponds the interpretation in $\C$
modulo extensional equivalence, a notion constructed by induction on
types. We do not reprove this extensional collapse theorem here. In
fact, Ehrhard's statement does not seem to follow easily from our
results: in order to prove the extensional collapse, Ehrhard provides
a cartesian closed category $\mathbf{Ppl}_\oc$ together with
\[
\mathbf{Ppl}_\oc 
\quad\to\quad
e(\Rel_\oc, \ScottL_\oc)
\]
a cartesian closed functor, where $(\C, \E)$ is a category of
``heterogeneous logical relations'' glueing together, intuitively, an
intensional model and an extensional model. There, $\mathbf{Ppl}_\oc$
is a sort of hybrid between $\Rel_\oc$ and $\ScottL_\oc$: morphisms are
relations that are well-behaved with respect to the preorder, a
property phrased via biorthogonality. In particular, we have
\[
\Rel_\oc 
\quad\ot\quad
\mathbf{Ppl}_\oc
\quad\to\quad
\ScottL_\oc
\]
cartesian closed functors, so that $\mathbf{Ppl}_\oc$ can play the same
role as $\Dsinn_\oc$ in the proof of Theorem \ref{thm:main}. Thus we
could imagine reproving Ehrhard's result by similarly providing 
\[
\Dsinn_\oc \quad\to\quad e(\Rel_\oc, \ScottL_\oc)
\]
a cartesian closed functor; however the natural definition does not
seem to work. Strategies are somehow ``too intensional''; the problem
is that the fact that they behave well with respect to the preorder is
derived combinatorially rather than maintained as an invariant -- so
they behave well only against other strategies, and not arbitrary
elements in $\Rel_\oc$.

This is not a bug! Rather than the extensional collapse statement
\emph{per se}, it is the relationship between qualitative and
quantitative models implicit in Ehrhard's paper, \emph{i.e.} Theorem
\ref{thm:main}, that turned out to be more influential, and that we
reproduced here. 

\section{Conclusions}

In addition to reproving Ehrhard's result by different means, this
result is a stepping stone for other work in progress. 

The first one is an infinitary extension of Theorem \ref{thm:final},
\emph{i.e.} an alternative collapse to the linear Scott model giving
also account of infinitary executions, \emph{i.e.} infinite
configurations; we hope that such a result is key in a complete
semantic understanding of the decidability of higher-order
model-checking \cite{DBLP:conf/lics/Ong06}. Ehrhard's does not extend
to an infinitary version of Theorem \ref{thm:main} with respect to the
infinitary relational model \cite{DBLP:conf/fossacs/GrelloisM15}, and
actually, it seems that surprisingly, the infinitary version of Theorem
\ref{thm:main} does not hold at all in the presence of greatest fixed
point. In contrast, work in progress suggests that Theorem
\ref{thm:final} does extend.

The second one is a bicategorical extension: from the result of this
paper, it is not too hard to send strategies to cartesian distributors
\cite{ol:intdist}. Proving that this is (pseudo-)functorial is more
subtle; this could potentially give a bicategorical version of
Ehrhard's result, thanks to the connection we already have between thin
concurrent games and generalized species of structure
\cite{DBLP:conf/lics/ClairambaultOP23}.

\section*{Acknowledgment}
The author would like to thank Charles Grellois for intense black board
discussions which motivated this development in the first place; as
well as the anonymous reviewers of all earlier versions of this paper
for their useful feedback. 

This work was supported by the ANR project DyVerSe
(ANR-19-CE48-0010-01); by the Labex MiLyon (ANR-10-LABX-0070) of
Universit\'e de Lyon, within the program ``Investissements d'Avenir''
(ANR-11-IDEX-0007), operated by the French National Research Agency
(ANR); and by the PEPR integrated project EPiQ ANR-22-PETQ-0007 part of
Plan France 2030.

\bibliographystyle{alphaurl}
\bibliography{main}

\end{document}